\documentclass[aps,prd,nofootinbib,showpacs,amssymb,twocolumn]{revtex4}
\usepackage{amsmath,amsfonts}
\usepackage{epsfig,subfigure}
\usepackage{ifpdf}
\usepackage{hyperref}
\usepackage{amsmath}
\usepackage{graphicx}
\usepackage{color}
\usepackage{natbib}
\usepackage{multirow}

\newcommand\editremark[1]{ {\color{red} #1}}

\newcommand\hidetosubmit[1]{}
\renewcommand\hidetosubmit[1]{#1}
\newcommand\optional[1]{}
\newcommand\ForInternalReference[1]{}

\newcommand\unit[1]{\, {\rm #1}}
\newcommand\mc{ {{\cal M}_c}}

\newcommand\Y[1]{Y^{(#1)}}

\newcommand\qmstateproduct[2]{\left<#1|#2\right>}

\newcommand\phiref{{\phi_{\rm ref}{}}}

\newcommand\rhoRecovered{\rho_{\rm rec}}

\newcommand\fref{f_{\rm ref}}
\newcommand\fSampleRateUsed{4096}

\newcommand\abbrvEFM{COOKL}  %

\newcommand\citeMCMC{\cite{LIGO-CBC-S6-PE,2011PhRvD..83h2002D,2011PhRvD..84f2003C,gr-extensions-tests-Europeans2011,gwastro-mergers-PE-Aylott-LIGOATest,2011ApJ...739...99N,2012PhRvD..85j4045V}}

\newcommand\citeFisher{\cite{1995PhRvD..52..848P,2008PhRvD..77d2001V,2008CQGra..25r4007C,2010PhRvD..82l4065V,gwastro-mergers-HeeSuk-FisherMatrixWithAmplitudeCorrections}}

\begin{document}

\author{Richard O'Shaughnessy$^{1}$}
\email{oshaughn@gravity.phys.uwm.edu}

\author{Ben Farr$^{2}$}

\author{Evan Ochsner$^{1}$}

\author{Hee-Suk Cho$^{3}$}

\author{Chunglee Kim$^{4,5,6}$}\email{chunglee@astro.snu.ac.kr}

\author{Chang-Hwan Lee$^{3}$}

\affiliation{$^1$Center for Gravitation and Cosmology, University of Wisconsin-Milwaukee, Milwaukee, WI 53211, USA}

\affiliation{$^2$Center for Interdisciplinary Exploration and Research in Astrophysics (CIERA) \& Dept. of Physics and Astronomy, Northwestern University, 2145 Sheridan Rd, Evanston, IL 60208, USA.}

\affiliation{$^3$Department of Physics, Pusan National University, Busan 609-735, Korea}

\affiliation{$^4$ Astronomy Program, Department of Physics and Astronomy, Seoul National University, 1 Gwanak-ro,
  Gwanak-gu, Seoul 151-742, Korea}

\affiliation{$^5$Korea Institute of Science and Technology Information, Daejeon 305-806, Korea}

\affiliation{$^6$Department of Physics, West Virginia University, PO Box 6315, Morgantown, WV 26505, USA}

\title{Parameter Estimation of Gravitational Waves from Nonprecessing BH-NS Inspirals with higher harmonics:
  \\ Comparing MCMC posteriors to an Effective Fisher Matrix  }
\date{\today}

\begin{abstract}
Most calculations of the gravitational wave signal from merging compact binaries limit attention to the leading-order quadrupole when constructing models for
detection or parameter estimation.   Some studies have claimed that if additional ``higher harmonics'' are included
consistently in the gravitational wave signal and search model, binary parameters can be measured much more precisely.
Using  the \texttt{lalinference} Markov-chain Monte Carlo parameter estimation code, we construct posterior parameter
constraints associated with two distinct nonprecessing black hole-neutron star (BH-NS) binaries, each with and
without higher-order harmonics.    All simulations place a plausible signal into a three-detector network with Gaussian
noise.    Our  simulations suggest that higher harmonics provide little information,
principally allowing us to measure a previously unconstrained  angle associated with the source geometry well but otherwise
improving knowledge of all other parameters by a few percent for our loud fiducial signal ($\rho=20$).     Even at this
optimistic  signal amplitude, different noise realizations have a more significant
impact on parameter accuracy than higher harmonics.   
We compare our results with the ``effective Fisher matrix''
introduced previously as a method to obtain robust analytic
predictions for complicated  signals with multiple significant harmonics. We find generally good
agreement with these predictions;  confirm that intrinsic parameter
measurement accuracy is nearly independent of detector network
geometry; and show  that uncertainties in extrinsic and intrinsic parameters
can to a good approximation be separated.
For our fiducial  example, the individual masses can be determined to lie between $7.11-11.48 M_\odot$ and
$1.77-1.276M_\odot$  at greater than 99\% confidence, accounting
for unknown BH spin.  Assuming comparable control over waveform systematics,  measurements of BH-NS binaries can
constrain  the BH and perhaps NS mass distributions. 
Using analytic arguments to guide extrapolation, we anticipate higher harmonics should provide little new information about nonprecessing BH-NS
binaries, for the signal amplitudes expected for the first few detections.  
Though our study focused on one particular example -- higher harmomics -- any study of subdominant degrees of freedom
in gravitational wave astronomy can adopt the tools presented here ($V/V_{\rm prior}$ and $D_{KL}$) to assess whether new physics is accessible (e.g., modifications of gravity;  spin-orbit
misalignment) and if so precisely what  information those new parameters provide.   
\end{abstract} 


\maketitle

\begin{widetext}
\tableofcontents
\end{widetext}

\ForInternalReference{

- Richard's mega-runs NOT in table (default evidence calculation not viable because of prior),
   but do use them for supporting data for thermo integration.

 - threshold

(0) Evidence theory: negative volume at low $\beta$: contrast with density of states

(-2) Use 'post', not 'corr\_post' (latter for progress of run) and returns correlated samples

(-4) Sanity check: high frequency response function is nontrivial.  We aren't sensitive to it (should NOT be)...but
might want to note it as a subtlety in a footnote, as it could impact tidal physics

}

\section{Introduction}

Ground based gravitational wave detector networks (notably LIGO \cite{gw-detectors-LIGO-original} and Virgo
\cite{gw-detectors-VIRGO-original})  are sensitive to the relatively well understood signal from  the lowest-mass compact binaries
$M=m_1+m_2\le 16 M_\odot$ \cite{2003PhRvD..67j4025B,2004PhRvD..70j4003B,2004PhRvD..70f4028D,BCV:PTF,2005PhRvD..71b4039K,2005PhRvD..72h4027B,2006PhRvD..73l4012K,2007MNRAS.374..721T,2008PhRvD..78j4007H,gr-astro-eccentric-NR-2008,gw-astro-mergers-approximations-SpinningPNHigherHarmonics,gw-astro-PN-Comparison-AlessandraSathya2009}.  
Strong signals permit  high-precision constraints on binary parameters, particularly when the binary precesses.
When interpreting plausible  data from ground-based instruments via Bayesian inference, 
a detailed model for these constraints can require the simulation of tens of millions of binary
waveforms, to adequately sample the high-dimensional model space \citeMCMC.   
The results of these calculations is a model for a strongly correlated, often highly multimodal,  and generically
broadly-distributed posterior in 15 dimensions.   
A simple procedure to reliably  estimate the performance of these detailed calculations is critical, to interpret and
communicate their results; to validate the estimate  itself; and to allow the broader astrophysical community to predict
how informative future gravitational wave surveys will be.

One general  algorithm is widely used to estimate the performance of these detailed calculations: 
the Fisher information matrix \citeFisher.  
Derived as a locally Gaussian approximation to the posterior, the Fisher matrix approximation is
expected to work well when observations provide tight constraints on all parameters
\cite{gw-astro-Vallis-Fisher-2007}. 
When valid, this easily-understood and calculated approximation can be applied both to simple test problems and  to the
full multi-detector likelihood, by incorporating information about detector network geometry \cite{CutlerFlanagan:1994}.  
 Repeated theoretical and practical investigations, however, have suggested the Fisher
matrix must be calculated with care, to avoid numerical  pathologies (e.g., associated with the sampling rate of the
signal; see \citet{gwastro-mergers-HeeSuk-FisherMatrixWithAmplitudeCorrections}, henceforth denoted \abbrvEFM); used with care, as it becomes ill-conditioned
and numerically unstable in the presence of strong degeneracies; and applied with care, only to sources in the appropriate
``strong-signal'' limit \cite{gw-astro-Vallis-Fisher-2007}.   
\abbrvEFM{} implicitly proposed three
idealizations which, together, convert the Fisher matrix to a more robust low-dimensional calculation, particularly for
nonprecessing binaries.   
 First and foremost,  for strong sources with well-localized sky locations, \emph{ignore} the sky location parameters: these
(and other) extrinsic, geometrical parameters almost perfectly separate from intrinsic
 parameters.\footnote{Some sources, identified electromagnetically, may already have well-determined sky locations.}  
Second, for
nonprecessing sources, assume the network has nearly equal sensitivity to both polarizations: only for a handful of
source orientations will asymmetric preferential sensitivity to one polarization or another bias our
conclusions about \emph{intrinsic} parameters. 
Finally, this paper pioneered an ``effective Fisher matrix'' approach,  eliminating observationally irrelevant scales by
suitably smoothing the local ambiguity function.  
Using these approximations, \abbrvEFM{} provided concrete
predictions for the performance of Bayesian parameter estimation codes, for three selected  cases.   

In this work, we  compare these predictions against direct and comprehensive Bayesian parameter estimation methods
which, by systematically comparing all possible candidate signals to data, construct a posterior probability
distribution \citeMCMC.    
Despite the considerable idealizations involved, the predictions of  \abbrvEFM{} work surprisingly well.  
Moreover, by repeated concrete calculations, we corroborate their  conclusions about the information
communicated via higher harmonics from  BH-NS binaries.

Even for nonprecessing binaries inspiralling along a quasicircular orbit,  the gravitational wave signal from merging
compact binaries remains surprisingly complicated, as source multipoles contribute to the gravitational wave
signal, at several different harmonics of the orbital frequency.    
Though the gravitational mass quadrupole dominates, these remaining
``higher harmonics'' influence the accessible signal, contributing noticeably to the overall signal amplitude and
influencing parameter estimation accuracy
\cite{2007CQGra..24..155V,gwastro-mergers-HeeSuk-FisherMatrixWithAmplitudeCorrections}.  
Using concrete examples,  \abbrvEFM{} argued that while higher
harmonics do provide additional information about BH-NS binaries, for observationally plausible signal amplitudes that
information is principally \emph{geometric}, encoding the source orientation relative to the line of sight, 
and does not significantly improve the measurement of intrinsic parameters (masses and spins) for BH-NS binaries.

\subsection{Context and prior work}
In the context of ground-based detectors, some studies outside and within the LIGO scientific collaboration 
have compared the
predictions of the Fisher matrix to parameter recovery strategies, including maximum likelihood
\cite{1996PhRvD..53.3033B,2008CQGra..25r4007C,2008PhRvD..77d2001V,2010PhRvD..82l4065V} and detailed MCMC calculations.  
Usually, these studies investigate measurement error, using large-scale simulations to determine (for example) how often
the maximum-likelihood point lies within the predicted confidence intervals derived from a Fisher matrix or MCMC
prediction.  
In this paper,  we emphasize  a  related problem: modeling the \emph{shape} of the posterior probability
distribution.  
Specifically, we compare two distributions by first (implicitly) translating
both to a common point, each centered on their
maximum-likelihood value.  Since this shift is explicitly coordinate dependent, we perform the shift and comparison in a handful of
phenomenologically-motivated coordinate systems.   
Once translated, we evaluate the covariance matrix for each distribution, then compare them.  
We similarly compare the covariance matrix against predictions from \abbrvEFM{}.  
We do not modify our procedure to account for edge effects, e.g. from the upper bound on the mass ratio or black hole
spin.

To determine the shape of each posterior, as noted above, we employ the  \texttt{lalsimulation} and
 \texttt{lalinference} \cite{LIGO-CBC-S6-PE,gw-astro-PE-Raymond} code libraries developed by the LIGO Scientific Collaboration and Virgo collaboration.  
 The former generates gravitational waveforms using several state-of-the-art signal models and the latter uses 
 Bayesian inference techniques to analyze gravitational-wave detector data 
 for parameter estimation and model selection purposes.    

We adopt fiducial initial LIGO and Virgo detector models both for efficiency and for comparison with
the results in COOKL.  The next generation of
gravitational wave detector networks will be sensitive to lower frequencies, leading to significantly longer
signals.  Existing parameter estimation methods are currently prohibitively computationally-expensive for these long signals. However, these existing methods have demonstrated they can reliably estimate posterior probability distributions
for signals in the most recently operating generation of detectors \cite{LIGO-CBC-S6-PE}.   
Therefore, one of the key goals of this work is to demonstrate that the predictions of the effective Fisher matrix method
are a reasonable proxy for the parameter estimation capabilities of a full Bayesian inference analysis.
This will justify using the effective Fisher matrix approach to predict 
the parameter estimation capabilities of advanced detectors.
Of course, it will be critical to enable Bayesian parameter estimation pipelines to run on the longer signals of advanced
detectors, but such work is outside the scope of this paper.

As described in \cite{gwastro-mergers-HeeSuk-FisherMatrixWithAmplitudeCorrections}, signal models including both higher
harmonics or spin have been extensively applied to parameter estimation problems for ground-based
\cite{2007CQGra..24..155V,LIGO-CBC-S6-PE} and
space-based \cite{2008PhRvD..78f4005P,2011PhRvD..84b2002L} detectors.   In the absence of precession, higher harmonics
are known to break degeneracies and improve sky localization, particularly for LISA \cite{2006PhRvD..74l2001L,2009PhRvD..80f4027K,2011PhRvD..84b2002L}.  
Our results suggest that  higher harmonics provide little useful additional information about BH-NS binaries.

\subsection{Executive summary and outline}
In this work, we describe how well the parameters of the nonprecessing signals studied in
\cite{gwastro-mergers-HeeSuk-FisherMatrixWithAmplitudeCorrections} can be recovered when inserted into  a realistic
three-detector network, both with and without higher harmonics.  

Our work builds upon  a large, community-wide effort to construct and validate parameter estimation strategies for
ground-based gravitational-wave networks, culminating in the \texttt{lalinference} code \cite{LIGO-CBC-S6-PE} employed
in this work.  In turn, this code implements parameter estimation strategies that have previously been extensively applied to gravitational wave
parameter
estimation \citeMCMC{}. 

Using a fixed source sky location and distance, we generate random noise for each detector; insert a signal into each
detector; and systematically compare the set of detector data with all plausible compact binaries in a broad prior range, 
allowing both the BH and NS to have spin.
\footnote{We effectively treat both objects as black holes, a priori allowing spins up to the Kerr limit on both bodies.
Due to the mass ratio $\simeq 7$, the spin of the smaller body is suppressed by a factor 
$\simeq 49$ relative to the larger body and has a negligible effect on our predictions, prior or posterior. }
We explicitly use identical noise realizations to compare signals with and without higher harmonics, to
isolate their effect on the posterior distribution and evaluate   what additional information higher harmonics provide. 

For the masses considered here, higher harmonics principally provide additional constraints on the source
\emph{geometry}, but not its intrinsic properties. In particular, they do not significantly improve the 
measurement of the masses and spins of our fiducial binary, but instead break a degeneracy between the polarization angle
and orbital phase of the binary at some reference point, thereby improving the measurement of these parameters.

We use several techniques to compare the posteriors of simulations with different noise realizations to one another 
and to effective Fisher matrix predictions.  These include a \emph{prior volume ratio}, 
$V/V_{\rm prior}$, as a measure of amount of information being extracted from the signal, 
a specialized version of the \emph{KL-divergence}, $D_{KL}$, to compare the similarity of two distributions, and the
\emph{mutual information}, $I(A,B)$, as a measure of the correlations between two subsets of parameters.
From the prior volume, we find that the higher harmonics provide more information 
than would an equivalent increase in SNR, but that it is still rather modest, 
the change in volume being comparable to the differences between different noise realizations. 
From $D_{KL}$ and inspection of marginalized posteriors, we find that simulations with different noise realizations
produce posteriors that are similar to one another 
and to the highly idealized effective Fisher matrix predictions provided in \abbrvEFM{}.
From the mutual information, we find that after marginalizing over time and polarization angle, there is 
a strong separation into intrinsic and extrinsic parameters with only very weak correlations between the two.

We also make two observations about the computational cost of Bayesian parameter estimation 
using amplitude-corrected waveforms. First, to perform calculations which include the low frequency portions of the higher harmonics, 
one needs to generate waveforms at lower orbital frequencies than for restricted waveforms.
Second, to fully resolve the highest harmonics during the late inspiral, 
one needs to use a higher sample rate to avoid aliasing. 
These effects mean that, to properly include amplitude corrections, one ostensibly needs much longer waveforms
with many more samples, and this could increase the computational cost of parameter estimation 
by one or two orders of magnitude.
Fortunately, as discussed briefly in Sec.~\ref{subsec:waveforms},
and in more detail in Appendix~\ref{ap:SampleRate}, we find that doing parameter 
estimation on amplitude-corrected waveforms with sample rates and low frequency limits 
appropriate for restricted waveforms has a negligible effect on any parameter estimation results.

There are, of course, some limitations to this work. For one thing, the computational cost of waveform generation 
limits our MCMC computations to producing $N_{\rm eff} \simeq 10^{4}$ independent samples. 
This introduces an error $\propto 1/\sqrt{N_{\rm eff}}$ in our confidence intervals and other results. 
Additionally, we have only used a single post-Newtonian waveform model (SpinTaylorT4), 
while it is well-known there are systematic biases
between different PN models~\cite{gw-astro-PN-Comparison-AlessandraSathya2009,Nitz:2013mxa,Harry:2013tca}. 
While these errors can change specific quantitative results, they will not affect our qualitative conclusions that higher
harmonics will provide a small improvement to the measurement of some extrinsic parameters 
but essentially no improvement to the measurement of intrinsic parameters for BH-NS binaries, 
and that the effective Fisher matrix method of COOKL provides a reasonably accurate prediction
of the results of full Bayesian parameter estimation methods. Lastly, this work is limited to studying two 
physical binary systems rather than fully exploring a larger parameter space. 
However, by verifying this these cases in detail, we argue that the effective Fisher matrix approach
can be used to efficiently explore the importance of physical effects such as spins, higher harmonics and tidal effects
across any and all regions of parameter space.

Our paper is organized as follows.
In Sec.~\ref{sec:Waveforms} we describe the waveform model, the specific BH-NS configurations 
and detector network being studied, and issues related to choices of parameter space coordinates.
In Sec.~\ref{sec:PEMethods} we review parameter estimation via Bayes' theorem for gravitational wave data analysis,
we describe the {\tt lalinference\_mcmc} code that is our Bayesian parameter estimation pipeline, 
and we introduce the quantitative techniques we will use to compare our results. 
In Sec.~\ref{sec:results} we examine the results of our simulations in detail, emphasizing the similarities between
various results, illustrating the effect of higher harmonics, and showing how intrinsic and extrinsic parameters 
decouple after marginalizing over time and polarization phase.  
In Sec.~\ref{sec:results:Compare}, we compare our simulations to the predictions of the
effective Fisher matrix and argue the latter is an inexpensive proxy for the former.
In Sec.~\ref{sec:implications}, we argue that the binary configurations we consider are, in a sense, ``typical'', and we briefly discuss some implications that can be inferred for other regions of parameter space.
Lastly, our conclusions can be found in Sec.~\ref{sec:conclusions} 
and we relegate several technical discussions to the appendices.

\begin{table*}[!]
\begin{tabular}{l|ccccccc|cc|rr }
type &$m_1$&$m_2$& $\iota$& $\phi$& $\phiref$ & $\psi$ &$\chi$& $M_{\rm c}$  & $\eta$ & $f_{\rm MECO}$ \\
  & $M_\odot$ & $M_\odot$ & & &  & &  & $M_\odot$ & & Hz \\
\hline
no spin&10&1.4&$\pi/4$& 0& $\pi/2$ &2.228  &0.0  & 2.994 & 0.1077 &  $559$ \\
aligned spin&10&1.4&$\pi/4$& 0 & $\pi/2$ & 2.228   &1.0  & 2.994 & 0.1077 &  $1926$ 
\end{tabular}
\caption{\label{tab1}{\bf Fiducial source parameters for the non-spinning and aligned-spin binaries.} We adopt the chirp
  mass $M_{\rm c}$ and symmetric mass ratio $\eta$ instead of individual masses.   
The orbital phase $\phiref$ is defined at 100 Hz.  
The (constant) orbital angular momentum direction is specified by the polar angles $(\iota,\psi)$, where the propagation
direction $\hat{n}$ is the reference axis (i.e., $\cos \iota  = \hat{L} \cdot \hat{n}$).  The black hole's spin is
parameterized by  $\chi = S_1/m_1^2$.  
The post-Newtonian signals used in the text terminate at a gravitational-wave frequency 
(of the second harmonic) $f_{\rm MECO}$, the smaller of the  ``minimum
energy circular orbit'' (hence the acronym) and the frequency at which $\dot{\omega}<0$. 
The values shown are derived from the same kind of \texttt{lalsimulation} output used in our simulations, 
albeit estimated from data evaluated at a $32$ kHz sampling rate for this table, rather than the 4kHz sampling rate
adopted for our MCMC calculations. 
}
\end{table*}

\begin{table}
\begin{tabular}{llllll}
$d$ & $t$ & $DEC$ & $RA$ & $\Delta t_{LH} $ & $\Delta t_{VH} $ \\
Mpc & s &  &  & ms & ms  \\
\hline
  23.1  &
 894383679.0 &
0.5747   &
  0.6485 &
  $- 3.93 $ &
 $ 5.98 $ \\
\end{tabular}
\caption{\label{tab:SourceGeometryOnSky}\textbf{Source location}: Source geocenter event time and sky location.  For a
  sense of scale, this table also provides  the
  time differences between different detector sites,  implied by that sky location and event time.
}
\end{table}

\section{Waveforms and coordinate choices}
\label{sec:Waveforms}

\subsection{Compact binary waveform model} \label{subsec:waveforms}
Following \cite{gwastro-mergers-HeeSuk-FisherMatrixWithAmplitudeCorrections},  we construct the post Newtonian (pN) gravitational wave signal from a BH-NS binary using the
\texttt{lalsimulation} SpinTaylorT4 code \cite{lal}.  
Based on previous implementations \cite{2003PhRvD..67j4025B,2004PhRvD..70j4003B},  this time-domain code solves the orbital dynamics of
an adiabatic, quasicircular inspiralling binary using the ``TaylorT4'' method \cite{gw-astro-PN-Comparison-AlessandraSathya2009} for the phase evolution and
(orbit-averaged) precession equations for the angular momenta \cite{1995PhRvD..52..821K}.   
The orbital phase and frequency evolution includes non-spinning corrections to 3.5pN order and 
spin corrections to 2.5pN order.\footnote{3PN and 3.5PN spin-orbit corrections
  have recently been calculated~\cite{Bohe:2013cla} and implemented into {\tt lalsimulation}. 
  They are not included in this work to avoid subtleties related to waveform termination 
  and because this work was already underway.}
 The precession equations are given to 2pN order.   
 
At each time, the gravitational wave signal measured by a distant observer is  
constructed from the orbital phase, orbital frequency and the orientations of the spins and orbital plane.    
The leading-order (``restricted'') expression  contains only the dominant second harmonic of the
orbital phase.    A more complete expression includes ``higher harmonics'' or ``amplitude corrections''. 
For quasi-circular, precessing binaries,  an expression for the signal including higher harmonics is implemented in 
\texttt{lalsimulation}\footnote{Precessing amplitude corrections were recently extended to 2PN order 
	in~\cite{Buonanno:2012rv}, but they have not yet been implemented into {\tt lalsimulation}.}
 up to  1.5pN order  
\cite{gw-astro-mergers-approximations-SpinningPNHigherHarmonics,1995PhRvD..52..821K,WillWiseman:1996}.   

We evolve the orbital dynamics of the binary by specifying ``initial'' conditions at $100\unit{Hz}$, 
then integrating the evolution equations forwards and backwards in time.   
At high frequency, this binary evolution is terminated prior to merger, either when it reaches the ``minimum energy circular orbit'', 
or when the orbital frequency ceases to increase monotonically. The stopping frequencies of our injected signals 
(which stop due to the minimum energy condition) are shown in Table \ref{tab1}. 
At low frequency, we investigated the effect of several different starting conditions, ensuring that we include the entirety of every harmonic
above the lower gravitational-wave frequency cutoff of our detectors, which we take to be $30$ Hz in this work.  
 In our preferred simulations,  indicated by stars (*) in Table \ref{tab:RunSummaryAndResults}, the data
contains the gravitational wave signal from a binary starting with a fixed initial \emph{orbital} frequency of $6\unit{Hz}$,\footnote{\label{note:TimeShift}At the time of writing, the  \texttt{lal} library has a bug: roughly speaking,  the injected sky location is associated
  with the first timesample in the waveform.  For long
  inspiral waveforms such as those injected from an initial  6 Hz orbital frequency,  the sky location can be slightly  offset from the intended
  source location.  We are confident this offset and related errors have no significant
  effect on our principal results.  In Figure \ref{fig:CompareGeometry}, as in all other calculations in this work, we
  have not corrected this offset when it appears.   }  
low enough to guarantee all higher harmonics employed here do not begin near the detector's sensitive band.  
The parameter estimation strategy fixes the initial  \emph{orbital} frequency at $15$ Hz 
in the absence of higher harmonics (only the second harmonic is present) or $6$ Hz if higher harmonics 
(up to the fifth harmonic) are present.  
Our simulations presented in the main text sample the signal at $\fSampleRateUsed\unit{Hz}$. 
From Table \ref{tab1}, an attentive reader may deduce that this sample rate is not sufficient to resolve 
the higher harmonics of the spin-aligned binary during the very late inspiral. 
However, the aliased portion of the signal, being higher order amplitude corrections, is suppressed by one or more powers of $v$, and the detectors have poor sensitivity at these high frequencies. As such, this aliasing has a  negligible impact on our results.
To further justify our undersampling, in Appendix \ref{ap:SampleRate}, we compare our results in the main text to results that use a higher sampling rate to eliminate aliasing. In a similar spirit, Appendix \ref{ap:SampleRate} 
also shows how a higher choice of lower frequency cutoff, while leaving out a portion of the higher harmonic signal, would drastically decrease the computational cost of our MCMC simulations without changing the results
in any significant way. 

We note that existing models of compact binary coalescences are imperfect, and there is extensive ongoing 
research to compute further PN corrections and construct inspiral-merger-ringdown waveforms 
that leverage information from numerical relativity simulations. In Appendix \ref{ap:SampleRate}, we briefly investigate
how systematics from different waveform models could affect our results. Much more extensive studies of
systematic errors from different waveform models can be found 
in~\cite{gw-astro-PN-Comparison-AlessandraSathya2009,Nitz:2013mxa,Harry:2013tca}, among others.

\subsection{Fiducial binary and detector network}

Following COOKL, we investigate two fiducial nonprecessing binaries (one non-spinning, one spin-aligned) 
along a single line of sight and at a specific sky location; see
Table \ref{tab1} for the intrinsic parameters and Table \ref{tab:SourceGeometryOnSky} for the extrinsic parameters.
For our MCMC simulations, we consider a three-detector network consisting of the first generation 
Hanford (H1) and Livingston (L1) LIGO detectors and the Virgo detector (V1). 
For each detector we use an analytic estimate of the design (one-sided) power spectral density (PSD):
\begin{widetext}
\begin{eqnarray} 
S_{\rm H1}(f) = S_{\rm L1}(f) &=& 9 \times10^{-46}\,\left[ \left( 4.49 \frac{f}{150} \right)^{-56} 
	+ 0.16 \left( \frac{f}{150} \right)^{-4.52} + 0.52 + 0.32 \left( \frac{f}{150} \right)^2 \right]\ , \label{eq:LIGOPSD}\\
S_{\rm V1}(f) &=& 10.2 \times 10^{-46} \left[ \left( 7.87 \frac{f}{500} \right)^{-4.8} + \frac{6}{17}\,\frac{500}{f} 
	+ \left( \frac{f}{500} \right)^2\right]\ . \label{eq:VIRGOPSD}
\end{eqnarray} 
\end{widetext}
The  \texttt{lalinference} code adopts these widely-used  choices as fiducial analytic  models for gaussian noise in the initial LIGO and Virgo
detectors;\footnote{The LIGO noise model was first described by \cite{2001PhRvD..63d4023D} and is available as
  \href{https://www.lsc-group.phys.uwm.edu/daswg/projects/lal/nightly/docs/html/_l_a_l_l_i_g_o_i_psd_8c.html}{lalinspiral:XLALLIGOIPsd}.
The Virgo noise model is available as
\href{https://www.lsc-group.phys.uwm.edu/daswg/projects/lal/nightly/docs/html/group___l_a_l_noise_models__h.html}{lalinspiral:LALVirgoPSD}. };
both correspond favorably to the best reported initial detector performance \cite{2010arXiv1003.2481T,arXiv:1203.2674,2010CQGra..27q3001A}.
We assume all detectors have no sensitivity below $30$ Hz and follow the above formulae from $30$ Hz to the 
Nyquist frequency.
No degeneracies exist: the source location corresponds to a delay of several milliseconds between each detector pair, larger than the timing
uncertainty in each interferometer.  The source sky location and orientation produce a comparable signal amplitude in
each interferometer; for example, for the nonspinning event, the individual detector SNRs are roughly 11.4 (H1), 14.2 (L1),
and 8.66 (V1), depending on noise realization.

In each MCMC simulation, we generate a random, synthetic noise realization for each instrument such 
that their PSDs match Eqs.~(\ref{eq:LIGOPSD}-\ref{eq:VIRGOPSD}). We label different noise realizations with the value
of a random seed used in their generation.
To isolate the effects of different noise realizations from the more interesting effects of spin and higher harmonics, 
we use identical noise realizations for simulations with different injected signals. In addition, we also 
do noiseless simulations - that is, the data stream being analyzed is just a signal with no synthetic noise - as 
another baseline for comparison.
When higher harmonics and/or spins were present in the signal, we performed parameter estimation with a
signal model that included them; when absent, our signal model omitted them.

We also emphasize that the effective Fisher matrix results predicted in COOKL and here assume
an idealized network of two co-located detectors with equal sensitivity to plus and cross polarizations and a 
PSD equal to $S_{\rm H1}$. Therefore, the rather good agreement between the two methods is especially 
encouraging considering one uses a simple, idealized network and the other a fully realistic one.

\subsection{The importance of coordinate choices} \label{subsec:coordinates}

In general, the waveform from a quasicircular compact binary is parameterized 
by its component masses ($m_1,m_2$), 
the distance from binary to observer $d$,
the time of arrival $t$,
the orbital phase at some fiducial point in the evolution $\phi_{ref}$,
the orientation of its orbital angular momentum $\hat{L}_N$ relative to the line of sight
(which can be described by the inclination and polarization angles ($\iota, \psi$)), 
the spin angular momenta of each body $\vec{S}_{1,2}$,
and the binary's location on the observer's sky (DEC, RA).   
When convenient, we reparametrize the component masses in terms of the
total mass ($M=m_1+m_2$), the  symmetric mass ratio $\eta=m_1 m_2 / M^2$, 
and the chirp mass $M_{\rm c}=M \eta^{3/5}$.  
We describe the BH spin with the parameter $\chi$, such that $\vec{S}_{\rm BH} = m_{\rm BH}^2 \chi \hat{L}_N$.
The allowed range of this parameter is $-1 \leq \chi \leq 1$ due to the Kerr limit and the fact that the spin
could be aligned or anti-aligned with the orbital angular momentum.

Our previous study COOKL demonstrated that subtle coordinate choices can have a dramatic effect on the ambiguity
function.    
For good coordinate choices, nearby points in parameter space should result in waveforms that appear as similar 
as possible to the detectors. The orbital phase constant and the orientations of the spin 
and the orbital angular momenta must be specified at some reference point during the binary's evolution.
Waveforms that have similar orientations and phasing \emph{where the detector is most sensitive} 
will appear more similar to each other than waveforms that happen to coincide at some much higher or lower 
frequency not around peak sensitivity.\footnote{For the non-precessing binaries considered here,
	the reference point is only relevant for the orbital phase, as the angular momenta have constant orientations.
}
For this reason, we choose $\phi_{\rm ref}$ to be the orbital phase at $100$ Hz, which is approximately 
the peak of the SNR integrand for an inspiral-only waveform in initial LIGO.

In addition to choosing coordinates describing the loudest portion of the waveform, it is also 
desirable to choose coordinates for which a parameter space metric is nearly flat.
There is a well-known method for metric-based placement of non-spinning inspiral templates based on the work 
of Owen and Sathyaprakash~\cite{Owen:1995tm,Owen:1998dk,2007PhRvD..76j2004C}. 
They find that the overlap (or ``distance'') ${\mathcal O}\left(h(\lambda, h(\lambda + \delta \lambda)\right)$ 
between two nearby waveforms with parameter separation $\delta\lambda$ is determined by a metric
whose definition coincides with that of the Fisher matrix
\begin{eqnarray}
{\mathcal O}\left(h(\lambda, h(\lambda + \delta \lambda)\right) &=& 1 - g_{ij}(\lambda) \delta \lambda ^i \delta \lambda ^j , \\
g_{ij}(\lambda) = - \frac{1}{2} \frac{\partial^2 \mathcal{O}}{\partial \delta \lambda ^i \partial
\delta \lambda ^j} &\simeq& \left(\frac{\partial h(\lambda)}{\partial \lambda ^i} \bigg|
\frac{\partial h(\lambda)}{\partial \lambda ^j}\right) = \Gamma_{ij} .
\end{eqnarray}

To lay a template bank covering the mass plane, one typically uses so-called ``chirp time'' 
parameters~\cite{Owen:1998dk} 
\begin{eqnarray}
\tau_0 &=& \frac{5}{256}\frac{(\pi \fref)^{-8/3}}{M^{5/3}\eta} = \frac{5}{256}
\frac{(\pi \fref)^{-8/3}}{\mc^{5/3}} , \label{eq:tau0} \\
\tau_3 
 &=& \frac{\pi}{4} \frac{(\pi M \fref)^{-2/3}}{2\pi \fref \eta} 
 = \frac{1}{8 \fref(\pi \fref \mc)^{2/3} \eta^{3/5}} .
\end{eqnarray}
These are the leading and 1.5pN coefficients for the pN prediction of the time it will take an adiabatic, quasicircular
inspiral to evolve from $f_{\rm ref}$ to coalescence. In this work, we use $f_{\rm ref} = 100$ Hz, which is the point at which
we define our extrinsic parameters.
The advantage of this parameterization is that the metric becomes very nearly flat in these coordinates. Since the metric 
is equivalent to the Fisher matrix, the Fisher matrix will be nearly constant in an appreciable region 
around the injected parameters, and the posterior probability distribution will be nearly Gaussian in this parameterization.
We therefore re-evaluated our analytic and simulation results in several coordinate systems. 
We find that the posteriors have surprisingly simple form in the coordinates 
$(\mc,1/\eta^2,\chi)$ and give results in this parameterization as well. 
For the high-amplitude signal explored in this work, the posterior distribution is tightly confined: coordinate-induced
nongaussianities and noise-realization-dependent parameter errors rarely occur.  For weaker signals with broader and
more noise-realization-dependent posteriors, our experience
suggests alternative coordinates will significantly improve the resemblence between Fisher matrix estimates and MCMC posteriors.

Recently,~\cite{gwastro-mergers-TemplateBank-AlignedSpin-Andy2012} have developed a generalized 
template metric approach which can be used to place spin-aligned templates. Rather than using the chirp times as 
coordinates, they use an 8-dimensional space of $\psi_k$, the pN coefficients of the 
stationary-phase approximation pN inspiral waveform. They then use a principal component analysis to find the 
dominant eigendirections (of which they only need two) in this space and lay template along those directions.
In principle, it might also be interesting to 
display our results in a coordinate system of three or more principal eigendirections for the spin-aligned space 
as in~\cite{gwastro-mergers-TemplateBank-AlignedSpin-Andy2012}, but finding such coordinates and
relating them to the physically interesting parameters is well outside the scope and focus of this paper.

\ForInternalReference{

\editremark{issue of not being invariant?} : parallel transport is not well defined; method is measure-dependent.
May matter eventually, but not so much at the accuracy we need.

}

\section{Parameter estimation methods}
\label{sec:PEMethods}

\subsection{Bayes' theorem for GW parameter estimation}
\label{subsec:Bayes}

We begin this section by reviewing the basics of how Bayes' theorem can be applied to quantify how much 
support any stretch of gravitational wave data $\{d\}$ provides for the hypotheses 
$H_1$, that a signal of a specific form is present,
or $H_0$, that the data contains only noise, and to estimate the likely parameters if a signal is present.
This is largely to clarify notation and terminology, 
and we refer the reader to \cite{CutlerFlanagan:1994,2003itil.book.....M,LIGO-CBC-S6-PE} 
and references therein for more information.

For sufficiently short time intervals, gravitational-wave detector data in the absence of a signal can be approximated
as a Gaussian, stationary, random process characterized by a power spectrum $S_h$,
which we take to be either of Eqs.~(\ref{eq:LIGOPSD}-\ref{eq:VIRGOPSD}).
In the limit of a long, continuous time duration, the relation describing the noise is:
\begin{eqnarray}
\left<n^*(f)n(f)\right> = \frac{1}{2} S_h(|f|) \delta(f-f')\ ,
\end{eqnarray}
The power spectrum can also be used to define an inner product between any two signals (such as a data stream and a template waveform) for a single detector:
\begin{eqnarray} \label{eq:ip}
\qmstateproduct{a}{b} \equiv 2 \int_{-\infty}^{\infty} df \frac{a^*(f)b(f)}{S_h(|f|)}\ .
\end{eqnarray}
Note that this defines a complex-valued inner product, while most of the gravitational-wave data analysis
is written in terms of a real-valued inner product.
In fact, as in COOKL, we use the complex-valued inner product for our effective Fisher matrix computations;
however, our MCMC runs use a real-valued inner product acting on real-valued signals $a(t),b(t)$, which is simply the real part of Eq.~(\ref{eq:ip}). 
All of the equations in this work involve inner products of real-valued signals with themselves, in which case the real- and 
complex-valued inner products coincide, so we will use a common notation for either inner product.
Also, while we have written the inner product as an integral over the entire real-valued frequency range, 
the discreteness of our signals and the low-frequency limits of our detectors mean that 
in practice the integration is over the frequencies
$[-f_{\rm Nyq}, -f_{\rm low}] \cup [f_{\rm low}, f_{\rm Nyq}]$.

By our assumptions, in the absence of a signal the noise will follow a Gaussian distribution 
such that louder noise realizations (as measured by the inner-product-induced norm) 
are less probable according to
\begin{eqnarray} \label{eq:noise_prob}
p(\{d\}|H_0) \propto \exp - \frac{\qmstateproduct{d}{d}}{2}\ .
\end{eqnarray}
where $d(t)$ is the timeseries in a single detector.  For a multidetector network, the posterior probability is the product of many such factors,
one for each detector.
Bayes' Theorem relates the  (``posterior'') probability distribution $p(\lambda|\{d\}H_1)$ 
to the conditional probability density or likelihood, $p(\{d\}|\lambda,H_1)$, 
of the data given the signal parameters $\lambda$; 
the prior probability $p(\lambda|H_1)$, describing
knowledge about the parameters within the model $H_1$ before the data is analyzed;
and the total probability of the observed data given our signal model hypothesis, $p(\{d\}|H_1)$:
\begin{align}
\label{eq:Bayes}
p(\vec{\lambda}|\{d\},H_1)&=\frac{p(\vec{\lambda}|H_1)p(\{d\}|\vec{\lambda},H_1)}{p(\{d\}|H_1)}\ ,  \\
 &=p(\vec{\lambda}|H_1) \frac{p(\{d\}|\vec{\lambda},H_1)/p(\{d\}|H_0)}{p(\{d\}|H_1)/p(\{d\}|H_0)}\ . \label{eq:Bayes2}
\end{align}
In the second line we have normalized by the probability of the null hypothesis $p(\{d\}|H_0)$ to eliminate sampling-dependent dimensional
factors present in $p(\{d\}|H_1)$ and the likelihood $p(\{d\}|\lambda,H_1)$.

The user is free to quantify their prior assumptions about the parameters, $p(\lambda|H_1)$ as they see fit.
We typically  use uniform priors over some broad range. 
The actual priors used for these runs are described in the next subsection.

For each detector, we assume the data takes the form
$d = h(\lambda) + n$ for some noise realization $n$, and we ask how likely it is to observe $d$.
Using Eq.~(\ref{eq:noise_prob}), we can compute the likelihood for a single detector \cite{CutlerFlanagan:1994}
\begin{eqnarray}
p(\{d\}|\lambda,H_1) \propto \exp   - \frac{\qmstateproduct{d-h(\lambda)}{d-h(\lambda)}}{2}\ .
\end{eqnarray}
and similarly for a multidetector network.  
In practice, with {\tt lalinference\_mcmc}, we work an expression which explicitly does not depend on the sampling rate
or length of data being examined, the  \emph{likelihood ratio} ${\cal L}$:
\begin{eqnarray}
\label{eq:def:Likelihood}
{\cal L} \equiv p(\{d\}|\vec{\lambda},H_1)/p(\{d\}|H_0) \\
 = \frac{e^{-\qmstateproduct{h(\lambda)-d}{h(\lambda)-d}/2}}{e^{-\qmstateproduct{d}{d}/2}}\ ,
\end{eqnarray}
where the former expression applies in general and the latter to a single detector.  

To quantify our overall confidence that a signal was present, with any allowed parameter values 
(and to properly normalize our posterior) we compute the evidence $Z$:\footnote{In this convention, 
	$Z$ is more properly called an evidence ratio, odds ratio or a Bayes factor 
	for the signal hypothesis versus the null hypothesis,  
	rather than the standard notion of evidence $\int d\lambda  p(\{d\}|\lambda H_1)p(\lambda|H_1)$.   
	Our expression has the distinct advantage of being a dimensionless quantity, independent of the 
	sampling rate or number of samples and fits naturally with our use of likelihood ratio.} 
\begin{align}
\label{eq:def:Z:Modified}
Z(d|H_1) &\equiv \frac{p(\{d\}|H_1)}{p(\{d\}|H_0)} = \int d\lambda
p(\vec{\lambda}|H_1)\frac{p(\{d\}|\vec{\lambda},H_1)}{p(\{d\}|H_0)} \nonumber\\
 & = \int d\lambda p(\vec{\lambda}|H_1) {\cal L}(\lambda|\{d\})\ .
\end{align}

For multidetector networks, the posterior probability or  likelihood of the network 
is simply the product of these quantities for each of the individual detectors 
due to the fact that probabilities interact multiplicatively.

\subsection{MCMC parameter estimation: {\tt lalinference$\_$mcmc}}
\label{sec:mcmc}

Several general  strategies have been developed to estimate posterior distributions $p(\lambda|\{d\},H_1)$ and evidence $Z$ given data 
\cite{2011RvMP...83..943V,book-BurnhamAnderson-ModelSelection,mm-stats-HandbookOfComputationalMethods,%
mm-stats-NestedSampling-Skilling2006,2009AIPC.1193..277S,2009MNRAS.398.1601F%
}.
Here we use the  \texttt{lalinference\_mcmc} code to estimate the posterior parameter distribution consistent with a candidate data
stream and a given noise model \cite{LIGO-CBC-S6-PE,gw-astro-PE-Raymond}.    
A detailed description of the \texttt{lalinference} code,  including its jump proposals and parallel tempering method, is far
beyond the scope of this paper.  In brief, the current \texttt{lalinference} code iteratively explores the parameter space, relying on detailed balance to
construct a sequence of samples $x_k$ for $k=1\ldots \infty$ that converges in distribution to the true posterior distribution
\cite{2011RvMP...83..943V}.  %
Though each element $x_k$ of the chain  is randomly distributed, neighboring elements are strongly correlated: the chain
``wanders'' through the posterior.   As a result, the whole chain contains fewer effectively  independent samples from
the posterior than would naively be supposed from its length.  To estimate the number of independent samples, we use the
correlation length, defined on a parameter-by-parameter basis as the smallest nonzero $s$ so
\begin{eqnarray}
1 + \sum_{k=1}^s 2 C(k) \le s
\end{eqnarray}
where $C(k) = \left<x_{q}x_{k+q}\right>/\left<x^2\right>$ is the autocorrelation function of the sequence.    In terms of this
number, the effective sample size $N_{\rm eff} \equiv N/s$ for $N$ the chain length.
As used in this work, the \texttt{lalinference\_mcmc} code terminated when roughly $N_{\rm eff}\simeq 10^4$ independent
samples were present in the posterior chain.\footnote{This termination condition was chosen to produce
  reasonably-well-determined confidence intervals in each parameter.    For example, to construct $90\%$ confidence intervals requires identifying a region with only $10\%$
probability; with $N_{eff}$ sample points, however, the systematic accuracy in assigning a probability $p$ to a region
is of order $p/\sqrt{N_{eff}}$. }   We additionally required the sequences $x_k$ of each individual parameter to
satisfy standard convergence criteria (e.g., the Gelman-Rubin R statistic).   

The procedure must start from some prior assumptions about the parameters, 
and these priors can have some influence on the details of the posterior.   We assume a source could lie
at any orientation and any distance within 100 Mpc,\footnote{\label{note:Dmax}The maximum distance adopted is  conventional for
  low-mass sources in \texttt{lalinferece\_mcmc}.  As described in Appendix \ref{ap:Thermo} near Eq. (\ref{eq:AverageLogL:DmaxIssue}), combined with the mass
  prior, this relatively small maximum distance allows enough
  rare distant and high-mass signals to significantly influence  averages over the prior, including the average log
  likelihood.  In particular, the evidence will depend  on this
  arbitrary choice,  in our opinion  nonphysically.   We strongly recommend subsequent
  calculations adopt a significantlly larger maximum distance, in significant excess of the horizon distance for all
  sources allowed by the mass prior, to insure comparable and astrophysically relevant evidence calculations.   A detection-weighted prior on the source distance and
  mass prior would also regularize the distance distribution.  Unfortunately, given the complexity of existing pipelines
  and the variability of noise,
  such a  prior would either be interpretation-dependent and ad-hoc or accurate,  calculated by Monte
  Carlo, and variable from  source to source, significantly complicating comparisons across sources.  }  
uniform in volume and angle, with  random masses
$(m_1,m_2)$  uniformly distributed in mass between \textbf{$1 M_\odot-30 M_\odot$ with $m_1+m_2\le 35 M_\odot$} and anywhere inside a
time window of length $\Delta T = 0.1\unit{s}$.  
When (aligned) spin is included, we allow both
objects' dimensionless spins $|S_k/m_k^2|$ to lie between $[0,1]$ uniformly, either aligned or antialigned with the
orbital angular momentum.   This two-spin model space includes one more parameter than the analytic predictions in
\abbrvEFM{}, which did not allow the smaller body to have
internal angular momentum.    Due to the high mass ratio, the smaller body's internal angular momentum ($|S_2|\lesssim
m_2^2 $) is expected to have  a relatively small effect on the radiated signal \cite{2011PhRvD..84h4037A,2012PhRvD..86h4017B}.  

We note that these priors are quite broad and uninformative, 
rather than being concentrated around the injected parameters.
The prior on time of arrival (uniform in a range of length $0.1$ s) may seem rather restrictive. 
In practice, however,  parameter estimation will usually be performed after  a search pipeline has  claimed a likely
detection; this time window is  broad compared to the typical time resolution 
of a compact binary inspiral search pipeline.

Once the run has produced the targeted number of effectively independent samples, 
we can extract a number of results from the computed posterior. 
For example, we can find the maximum log-likelihood and the parameter values at this peak.
We can produce one- or two-dimensional marginalized posteriors by integrating over the other parameters,
and using them to find confidence intervals for various parameters.
We can also use points near the maximum log-likelihood to compute approximate covariance and Fisher
matrices that describe this region.
All of these results can be compared between different runs and also with similar
quantities produced via the effective Fisher matrix. The rest of this section describes
several analytic techniques we use to facilitate such comparisons.

\subsection{Prior volume ratio}
\label{subsec:priorvolume}

Motivated by a locally Gaussian approximation, we define a characteristic parameter volume fraction:
\begin{eqnarray}
\label{eq:def:VoverVprior}
V/V_{prior} \equiv \frac{Z(d|H_1)}{\text{max}_\lambda {\cal L}(\lambda|\{d\})}\ .
\end{eqnarray}
In the limit that the posterior can be approximated by a Gaussian of the form ${\cal L}(\lambda)p(\lambda) \propto \exp (-
\delta\lambda_a\Gamma_{ab}\delta \lambda_b/2)$ in the neighborhood of $\lambda_*$, where the local maximum's location
and shape is dominated
by the likelihood ${\cal L}$ and not the prior $p$, then the prior volume ratio is simply
\begin{eqnarray} \label{eq:VoverVpriorFisher}
V/V_{prior}  = \int d\lambda \frac{p(\lambda|H) {\cal L}}{\text{max}_\lambda {\cal L}} \simeq 
	\frac{\sqrt{|\Gamma|}\,}{(2\pi)^{N/2}}\ p(\lambda_*)\ ,
\end{eqnarray}
where $|\Gamma|$ represents the determinant and $N$ the number of parameters.
Because the definition is explicitly the product of the prior times a function $\le 1$ with support concentrated in
high-probability regions, the volume fraction  characterizes the fraction of a priori plausible signals 
that are consistent with the data $\{d\}$.  
We can define a recovered network SNR from the max log-likelihood (ratio) as:
\begin{eqnarray}
\label{eq:def:rhoRecovered}
\rhoRecovered \equiv \sqrt{2 \text{max}_\lambda \ln  {\cal L}(\lambda|\{d\})} \ .
\end{eqnarray}

To characterize how rapidly the prior volume ratio changes with $\rho_{\rm rec}$, 
we define a characteristic ``effective dimension'' as
\cite{PSconstraints3-MassDistributionMethods-NearbyUniverse}
\begin{eqnarray}
\label{eq:def:Deff}
D_{\rm eff,rec}(\rhoRecovered) \equiv - \frac{d\ln (V/V_{\rm prior})}{d\ln \rhoRecovered} \ .
\end{eqnarray}
Intuitively, the effective dimension is the number of parameters which can be constrained relative to their prior range.
For example, suppose we had a Gaussian posterior whose width in every parameter was significantly
narrower than the prior. Then, the prior volume ratio would be given by Eq.~(\ref{eq:VoverVpriorFisher}), and we would
have $D_{\rm eff} = N$. Now, suppose we added one or more parameters which have absolutely 
no effect on the posterior.  Then, we would simply recover the prior on those parameters for any $\rho_{\rm rec}$.
We would have the same $D_{\rm eff}$ as before, 
even though $N$ has increased by adding unmeasurable parameters.

Our non-spinning waveform model has 9 parameters and the spin-aligned model has 11. In justifying our prior range
in Sec.~\ref{sec:mcmc}, we argued that the spin of the smaller body is strongly suppressed, and so we might expect
to be unable to constrain it significantly (this turns out to be the case). Furthermore, as will be discussed at length,
our restricted waveforms have a degeneracy between $\psi$ and $\phi_{\rm ref}$ such that only a certain combination
of the two can be measured; amplitude-corrected waveforms break this degeneracy and allow both angles
to be measured. For a sufficiently loud signal, we expect to be able to constrain 
all other parameters to some extent relative to their priors.
Therefore, we expect the following values for $D_{\rm eff}$ for our various waveform models:\footnote{Another
	simple way to predict the effective dimension $D_{\rm eff}$ is to compute a Fisher matrix and 
	simply count the number of eigenvalues that are smaller than the associated prior range.}
\begin{eqnarray}
\label{eq:DeffPredicted}
D_{\rm eff}(\rho \simeq 20) = \begin{cases}
8 & \text{Zero spin, no} \\
9 & \text{Zero spin, with} \\
9 & \text{Aligned spin, no} \\
10 & \text{Aligned spin, with}
\end{cases}
\end{eqnarray}
where ``no'' indicates a model without higher harmonics and ``with'' a model including higher harmonics.  
At lower SNR, certain parameters will become poorly measured or unmeasurable, and $D_{\rm eff}$ will drop.
We provide more details about the effective dimension in Appendix~\ref{ap:Thermo}. In particular, 
we compute $D_{\rm eff}$ via thermodynamic integration and find it's behavior changes around a network SNR 
$\rho_{\rm rec} \simeq 10$, indicating this is when parameter estimation begins to significantly degrade.

Both the recovered signal amplitude $\rhoRecovered$ and effective dimension $D_{\rm eff, rec}$ depend on the noise
realization.  
For example, for a fixed physical signal and random noise realizations, the recovered signal amplitude is a
$\chi^2$-distributed random variable, with $2 N_D$ degrees of freedom, where $N_D$ is the number of detectors,
normalized so $\left<\rho^2\right>=N_D$ in the absence of signal.  
To provide an invariant
measure of signal strength and the local density of states, we use ``noiseless'' data $\{d\}_0$
where the data $d=h(\lambda_*)$ for some parameters $\lambda_*$.  In particular, we define the intrinsic network
amplitude $\rho(\lambda_*)$ by  
\begin{eqnarray}
\label{eq:def:rho}
\rho(\lambda_*) \equiv \sqrt{2 \max_\lambda \ln {\cal L}(\lambda| d = h(\lambda_*))} \ .
\end{eqnarray}

\subsection{Comparing two distributions' shapes}
\label{sec:Methods:Metrics}

On physical grounds, we want to understand how tightly individual gravitational wave measurements will constrain
parameters.  In particular, we want to compare the \emph{shapes} of the nearly-Gaussian posterior probability
distributions, both to each other and to analytic approximations derived using the effective Fisher matrix.    Because
these distributions will be approximately Gaussian, we can compare shapes using a locally Gaussian
approximation.\footnote{In more general cases with less-Gaussian posteriors, similarity between two  one-dimensional
  posteriors is often quantified via a Kolmogorov-Smirnov (KS) test.  We adopt the KL-divergence method due to its more
  attractive scaling with dimension and its clear treatment of differing correlations.  }
Specifically, we compare two distributions with covariance matrices $\Sigma=K^{-1}$ and $\Sigma_*=K_*^{-1}$
by computing the quantity
\begin{eqnarray}
\label{eq:Dkl}
D_{KL}(K_*,K)&\equiv& \frac{1}{2}\left[\ln[ |K_*|/|K|]+ \text{Tr}[ K_*^{-1}(K -K_*)]\right] \ . \nonumber\\
\end{eqnarray}
As described in Appendix \ref{ap:Compare:Details}, this expression is a special case of a more general expression, the KL
divergence, which has been extensively applied to the theory and practice of Markov Chain Monte Carlo;  see
\cite{book-BurnhamAnderson-ModelSelection} and references therein.  
In this work, we will not exploit the statistical significance of $D_{KL}$, treating the expression above solely as a
phenomenological measure of distribution similarity.  
For one-dimensional distributions $K=1/\sigma^2$ and $K_*=1/\sigma_*^2$, this expression reduces to
\cite{PSconstraints3-MassDistributionMethods-NearbyUniverse}:
\begin{eqnarray}
\label{eq:Dkl:1d}
D_{KL}(\sigma_*,\sigma) &\equiv & \ln \frac{\sigma}{\sigma_*}   
- \frac{1}{2}
+ \frac{  \sigma_*^2}{2\sigma^2}  \\
&\simeq& (\ln \sigma/\sigma_*)^2 + O(\ln \sigma/\sigma_*)^3 \nonumber
\end{eqnarray}
where in the last line we take a limit of small $\ln (\sigma/\sigma_*)$.   
For multidimensional  distributions which share the same  principal axes, $D_{KL}(K_*,K)$ separates into a sum of one-dimensional
$D_{KL}$:
\begin{eqnarray*}
D_{KL}(K_*, K) = \sum_q  D_{KL}(\sigma_{q,*},\sigma_q)
\end{eqnarray*}
More generally, however, the quantity $D_{KL}$ in Eq. (\ref{eq:Dkl}) severely penalizes correlations with different
principal axes.  As a concrete example, if $K_*$ is a two-dimensional symmetric matrix, without loss of
generality of the form
\begin{eqnarray*}
K_* \equiv \begin{bmatrix}
\lambda_1 & 0 \\
 0 & \lambda_2
\end{bmatrix} \; ,
\end{eqnarray*}
and  $K=R K_* R^{-1}$ is a rotation of that matrix by an angle $\theta$, then 
\begin{eqnarray}
D_{KL}(K_*,K) = \frac{2 (\lambda_1-\lambda_2)^2\sin^2 \theta}{\lambda_1 \lambda_2}
\end{eqnarray}
As a result, in the physically common case $\lambda_1 \gg \lambda_2$,  a slightly misaligned error ellipsoid $K$ with
similar scales as $K_*$ can have a large $D_{KL} \simeq 2 \theta^2 \lambda_1/\lambda_2$. 
Due to the computational limits of our MCMC results, 
each parameter $x$ has only $N_{\rm eff}(x)$ independent elements.  Our
best estimate for the sample standard deviation, $\hat{\sigma}^2=\sum_k (x-\bar{x})^2/(N-1)$, has sampling error.
Assuming a Gaussian distribution,  this estimator has a relative mean-squared error\footnote{Briefly, 
on physical grounds, measurements can at best distinguish one ``natural'' parameter to a relative accuracy 
$1/\sqrt{N}$.  Because the standard deviation naturally enters
quadratically into the distribution and hence into all derived expressions, it can be measured to a relative accuracy
$1/n^{1/4}$.  Conversely,  the dramatic variation in $\sigma$ between data realizations and the slow convergence of those fluctuations
with $N$ is \emph{a feature of the coordinates
  used} to characterize the posterior.   As alternative coordinates make clear, evidence favoring one Gaussian
distribution over another
accumulates linearly with the number of samples  $N$. }
\begin{eqnarray}
\label{eq:ErrorInSigma}
\frac{\left<(\hat{\sigma}^2 - \left<\hat{\sigma}^2\right>)^2\right>^{1/4}}{\sigma} \simeq   \frac{2^{1/4}}{(N-1)^{1/4}} \; .
\end{eqnarray}
In particular, the standard deviation calculated from a relatively small number of effective samples $N_{\rm eff}\simeq
10^3-10^4$ can vary noticeably between different Markov chains, by tens of percent at least.
These  statistical fluctuations  from different MCMC realizations produce errors that add in quadrature with
the   statistical fluctuations associated with different noise realizations, described below.
These statistical fluctuations limit our ability to distinguish between two distributions with too-similar widths ($\ln
\sigma/\sigma_* \lesssim 1/\sqrt{N}$) and
hence too-small  $D_{KL}$ ($D_{KL} \lesssim 1/N$).   

Motivated by more detailed discussions (see, e.g., Eqs. (33-36) in
\citet{PSconstraints3-MassDistributionMethods-NearbyUniverse} and Appendix \ref{ap:Compare:Details}), we anticipate
that we choose between two hypotheses $H_1$ (a gaussian with covariance $K$) and $H_2$ (a gaussian with covariance
$K_*$) with $N_{\rm eff}$ samples if the two Gaussian distributions have KL divergence above an $N_{\rm eff}$-dependent threshold:
\begin{eqnarray}
\label{eq:DklDistinguishCriteria}
D_{KL} \gtrsim \frac{10}{N_{\rm eff}}\times d
\end{eqnarray}
where the $d$ in the numerator is the number of dimensions in $K$.  The factor 10 was  chosen via a Monte Carlo
over two one-dimensional gaussian distributions, to reduce the false alarm
probability to less than $10^{-4}$.
Similarly, two samples will have distinguishable width if $D_{KL}(\sigma_1,\sigma_2)$ is greater than roughly twice this threshold.

Though fully accounting for finite-MCMC-length effects, the above condition does not account for detector noise.   If we
construct the posterior for a special data realization -- exactly zero noise --  the best-fit parameters will be the
physical parameters $\lambda_*$  and the posterior will have a locally Gaussian shape set by $K_*$.   Each
noise realization shifts the best-fit point and changes the associated posterior's shape.  In the Gaussian limit, the
best-fit parameters $\lambda$ are consistent with a Gaussian defined by $K_*$, centered on the physical parameters; the
best-fit signal amplitude $\hat{\rho}$ differs from the physical signal amplitude $\rho$ by a random number of order
unity; and the posterior covariance $K$ therefore differs from $K_*$ by two effects: change in $\rho$ and $K$.  
In the first case, because the posterior scales as $K\propto \rho^{2}$ (i.e., $\sigma \propto 1/\rho$), fluctuations in
the signal amplitude directly produce fluctuations in $\sigma$ and $K$.  In the large-$\rho$ limit, we anticipate and
Monte Carlo simulations confirm that the  $D_{KL}$ between the intrinsic and  sample-estimated $K$ will
fluctuate.  Substituting in two proportional $d$-dimensional covariance matrices $K_*\rho_*^2$ and $K_* \rho^2$ into
Eq. (\ref{eq:Dkl}), we find that fluctuations in the scale factor ($\rho$) have relatively little effect:
\begin{align}
D_{KL}(K_*\rho_*^2,K_*\rho^2) &= -  \frac{1}{2}\ln (\rho/\rho_*)^{2d} + \frac{d}{2}((\rho/\rho_*)^2-1) \nonumber \\
 &
\label{eq:DklCriteria:FluctuateAmplitude}
\simeq  d \left( \frac{\delta\rho}{\rho_*} \right)^2
\end{align}
Each of the $N_D$ detectors contributes a comparable noise in the signal amplitude, so $\delta \rho$ is $\chi^2$
distributed with $\left<\delta \rho^2\right>=N_D$.    As a result, random fluctuations due to the noise realization
produce unavoidable 
changes in the posterior's shape relative to the noiseless posterior,
characterized by a typical $D_{KL}$ of order 
\begin{align}
\label{eq:DklCriteria:FluctuateAmplitude:Typical}
\left<D_{KL}(K_*\rho_*^2,K_*\rho^2)\right>& \simeq d N_D /\rho^2  \nonumber \\
 & \simeq 0.0675 (N_D/3)(d/9)(\rho/20)^{-2}
\end{align}
Any value of $D_{KL}$ comparable to or smaller than this expression suggests the two distributions have effectively indistinguishable shapes.  
This uncertainty adds (linearly) to the error expected from finite $N_{\rm eff}$; for $N_{\rm eff} \gtrsim 10^3$, this
expression is the dominant source of error.  

Conversely, to be confident two distributions have \emph{different} shapes, we want  $D_{KL}$ between those
two covariances to be several standard deviations away.   Again using a one-dimensional Gaussian Monte Carlo to select
the prefactor, we consider two distributions to be clearly distinguishable if
\begin{eqnarray}
\label{eq:DklCriteria:FluctuateAmplitude}
D_{KL}(K_*,K) \gtrsim \frac{4 N_D }{\rho^2} \times d
\end{eqnarray}
where the coefficient insures a probability less than $10^{-4}$.
Finally, our approach to diagnosing posterior differences is both qualitatively and quantitatively useful only when the
posterior is  locally Gaussian.   A gaussian with covariance $K^{-1}$ is only a good approximation  in a small region.
Roughly speaking, when error ellipsoids approximated by $K$ are sufficiently long and
narrow, the locally Gaussian approximation can break down, simply because the covariance $K^{-1}$ changes from point to point.  In the worst case, ubiquitous in the limit of low
signal amplitude,  the error ``ellipsoids'' are not ellipsoidal.   Less catastrophically, the principal axes and
eigenvalues of $K$ can vary rapidly over the signal space; as a result, when the best-fit parameters $\lambda$ are far
from the physical values, the posterior's covariance $K$ will differ significantly from the predicted (zero-noise)
limit.  
This second challenge can be mitigated or completely eliminated by adopting a different coordinate system
to parameterize the binary.  
For the systems explored in this work, a local gaussian approximation is effective, so alternative coordinate systems
only marginally improve our already good agreement.  

Strong nongaussianities can also arise from edge effects which restrict a posterior to be defined to a narrow range.
We do not modify our procedure to account for edge effects, e.g. from the upper bound on the mass ratio or black hole
spin.

\subsection{Quantifying multidimensional correlations}
\abbrvEFM{} suggested that the posterior
for nonprecessing binaries largely \emph{separates} into purely intrinsic and
purely extrinsic parameters, even in the presence of higher harmonics, when the posterior is marginalized over
polarization and event time.  If true, this powerful constraint implies posteriors can be understood using far fewer
correlations.
Motivated by information theory, we quantify the degree of correlation between  two subspaces $A,B$ using
the mutual information~\cite{2003itil.book.....M}.
For the case of a multi-dimensional Gaussian with subspaces $A$ and $B$, the mutual information
of the two subspaces is
\begin{eqnarray}
\label{eq:RelativeInfo}
I(A,B)  \equiv \frac{1}{2} \ln \frac{|\Sigma_A||\Sigma_B|}{|\Sigma|}\ ,
\end{eqnarray}
where $\Sigma_A$ and $\Sigma_B$ are projections of the 
full covariance matrix $\Sigma$ onto the $A,B$ subspaces, respectively.  
This expression provides a  way to evaluate the degree of correlation between two
parameter subspaces that does not change under local linear transformation.  

To illustrate how this measure of similarity compares with another often-used measure, 
the correlation coefficient, we apply Eq. (\ref{eq:RelativeInfo}) to a two-dimensional covariance matrix of the form
\begin{eqnarray}
\Sigma &=& \begin{bmatrix}
\sigma_A^2 & \sigma_A \sigma_B c_{AB} \\
\sigma_A \sigma_B c_{AB} & \sigma_B^2
\end{bmatrix} \\
I(A,B) &=& \frac{1}{2}\ln \frac{1}{1-c_{AB}^2}\ .
\end{eqnarray}
It is clear that $I(A,B)$ will be zero if $A$ and $B$ are completely uncorrelated, and larger values indicate a 
greater degree of correlation. The mutual information is unbounded and diverges in the case where $A$ and $B$
are perfectly correlated or anti-correlated.
Using a finite number of samples produces errors in $\Sigma$ and hence in $I$.   For our purposes, the information $I$
will indicate an identifiable correlation if
\begin{eqnarray}
\label{eq:Info:ErrorScale}
I \gtrsim \frac{10 [\text{dim}(A) + \text{dim}(B)]}{N_{\rm eff} } \ .
\end{eqnarray}
This rule of thumb was found empirically by performing a Monte Carlo with a simple two-dimensional toy model, 
finding a cutoff which ensures two uncorrelated subspaces are claimed as correlated with a false alarm probability of
$10^{-4}$, and scaling properly with $N_{\rm eff}$, ${\rm dim}(A)$ and ${\rm dim}(B)$.

\subsection{Bounded parameters and truncated distributions}

In the above discussion of numerical and analytic posteriors, we have implicitly assumed all parameter combinations are
allowed.  In fact, many parameters  are defined over a bounded nonperiodic domain, such as $\eta \in[0, 1/4],$ and
$|\chi|\le 1$ (but not $\phi\in[0,2\pi]$).    In the limit of high signal amplitude, the posterior will resemble a
truncated Gaussian $p(x)\theta_D(x)$ where $\theta_D(x)$ is a suitable step function to limit $x$ to the allowed parameter
region $D$ and $p(x)$ is a Gaussian over the parameters $x$.  
Many of our posterior distributions have this property.  For example, because we specify an extremal black hole spin
$\chi=1$, our posterior distributions cannot be Gaussian into the regime $\chi>1$.  

Even though the $D_{KL}$ expression was derived using Gaussian distributions, 
it provides an equally well-posed
scheme to compare any two covariances $\Sigma, \Sigma_*$.  
We therefore apply it unchanged when comparing any pair of predictions,
such as for distributions that are truncated due to bounded parameters.  

Truncation becomes more of an issue when comparing the predicted posterior distributions involving spin.  Because the
signal is highly degenerate, limits on
$\chi$  constrain the extent of the three-dimensional distribution in $\mc,\eta,\chi$.    By contrast, Fisher matrix
calculations of the kind described in \abbrvEFM{} do not include this prior.   As a result, due to truncation, even
if  the Gaussian $p(x)$ in $p(x)\theta_D(x)$ and the Fisher matrix prediction are identical, because the covariance of the
distribution $p(x)\theta_D$ differs from the covariance of $p(x)$, truncation introduces additional systematic
differences between the effective Fisher matrix approximation and the results of detailed calculations. 

We note that for our fiducial binaries $\eta = 0.1077$ is right in the middle of its domain, 
well away from the boundaries. As such, the bounds on $\eta$ do not affect our results. 
However, if we were to study, for example, 
an equal mass binary then the $\eta$ distribution would be truncated much like the spin distribution is in this case.

\section{Parameter estimation results}
\label{sec:results}

\begin{table*}
\begin{tabular}{lll|cc|cc|r}
Source & Harmonics & Seed & $\rho_{\rm inj}$ & $\rho_{\rm rec}$ & $\ln Z$ & $\ln V/V_{prior}$ & $N_{\rm eff}$\\ 
\hline
 {Zero spin} & {no} & - & 20.33 & 20.64 & 180. & -33.4 & 7517 \\
 {Zero spin} & {no} & {-*} & 20.33 & 20.64 & 179. & -34.2 & 9347 \\
 {Zero spin} & {no} & 1234 & 20.33 & 20.5 & 176. & -34.2 & 10697 \\
 {Zero spin} & {no} & {1234*} & 20.33 & 19.42 & 156. & -32.5 & 7501 \\
 {Zero spin} & {no} & 56789 & 20.33 & 20.34 & 172. & -34.8 & 10403 \\
 {Zero spin} & {no} & {56789*} & 20.33 & 21.65 & 207. & -27.3 & 10000 \\
\hline
 {Zero spin} & {with} & - & 21.03 & 21.33 & 191. & -36.3 & 8027 \\
 {Zero spin} & {with} & {-*} & 21.03 & 21.34 & 191. & -36.9 & 7348 \\
 {Zero spin} & {with} & 1234 & 21.03 & 21.76 & 200. & -37. & 7511 \\
 {Zero spin} & {with} & {1234*} & 21.03 & 20.38 & 170. & -36.9 & 6523 \\
 {Zero spin} & {with} & 56789 & 21.03 & 20.67 & 177. & -36.6 & 11358 \\
 {Zero spin} & {with} & {56789*} & 21.03 & 22.12 & 204. & -40.4 & 22843 \\
\hline
 {Aligned spin} & {no} & - & 22.34 & 22.67 & 222. & -34.6 & 9841 \\
 {Aligned spin} & {no} & {-*} & 22.34 & 22.66 & 223. & -34.2 & 10174 \\
 {Aligned spin} & {no} & 1234 & 22.34 & 22.81 & 225. & -35.1 & 8670 \\
 {Aligned spin} & {no} & {1234*} & 22.34 & 21.90 & 206. & -32.8 & 126040 \\
 {Aligned spin} & {no} & 56789 & 22.34 & 24.89 & 272. & -37.5 & 10508 \\
 {Aligned spin} & {no} & {56789*} & 22.34 & 22.62 & 220. & -35.5 & 10003 \\
\hline
 {Aligned spin} & {with} & - & 22.88 & 23.19 & 230. & -38.9 & 4289 \\
 {Aligned spin} & {with} & {-*} & 22.88 & 23.20 & 232. & -37. & 1737 \\
% {Aligned spin} & {with} & {+*} & 22.88 & 23.2 & 229. & -39.7 & 16529 \\
 {Aligned spin} & {with} & 1234 & 22.88 & 23.67 & 240. & -40.1 & 8569 \\
 {Aligned spin} & {with} & {1234*} & 22.88 & 22.65 & 217. & -38.9 & 10866 \\
 {Aligned spin} & {with} & 56789 & 22.88 & 25.4 & 279. & -43.3 & 10715 \\
 {Aligned spin} & {with} & {56789*} & 22.88 & 23.37 & 237. & -34.9 & 34921 \\
\end{tabular}

\caption{\label{tab:RunSummaryAndResults}\textbf{Simulation results}: Table of distinct simulations performed.  The
  first set of columns indicate which of the two fiducial binaries was used (zero spin vs aligned spin), 
  whether higher harmonics  were included (up to $1.5$PN in amplitude), and random seed choice used to generate noise 
  (a ``-'' means no noise was used; the asterisk indicates a different noise realization,  MCMC realization, and initial
  orbital  frequency).   
The two quantities $\rho_{\rm inj},\rho_{\rm rec}$ provide the
  injected and best-fit total signal amplitude in the network [Eqs. (\ref{eq:def:rhoRecovered},\ref{eq:def:rho})].    
The latter quantity depends on the noise realization of the network. 
[For zero noise and given infinitesimal time and frequency resolution, the injected and recovered amplitudes should
    agree.  Due to finite resolution, a different (continuously-timeshifted) template can slightly better fit a
    discretely-sampled signal, leading to $\rho_{rec}-\rho_{inj}>0$. ]
The columns for $\ln Z$ and $V/V_{\rm prior}$ provide the evidence [Eq. (\ref{eq:def:Z:Modified})] and volume fraction
[Eq. (\ref{eq:def:VoverVprior})]; the evidence, volume fraction, and signal amplitude are related by $\rho_{\rm rec}^2/2  =
\ln Z/(V/V_{\rm prior})$.  
For each binary, two indepedent MCMC posteriors were constructed  with zero noise (``-'' and ``-*''); the nonzero difference
between their $\ln Z$ and their  $\ln (V/V_{\rm prior})$ suggests how robust our estimates of these quantities are, as
described at length in Appendix \ref{ap:Compare:Details}.  
  Finally, $N_{\rm eff}$ is the effective number of independent samples in our calculations.
}
\end{table*}

For our fiducial binaries, one with and one without BH spin, we generate target inspiral signals  
with and without higher harmonics, using a
specific sky location and event time; see Tables \ref{tab1} and \ref{tab:SourceGeometryOnSky} for details.   Each signal
is injected into random Gaussian noise, then systematically compared with a source model from the same
waveform family (SpinTaylorT4) including \emph{identical physics}: 
higher harmonics if and only if the signal included higher harmonics; spin if and only if the BH had nonzero
spin.   
Table \ref{tab:RunSummaryAndResults} lists the specific simulations performed, providing the injected and recovered
signal-to-noise ratio.  
For each simulation, Table \ref{tab:ParameterErrors}  provides  the marginalized one-dimensional uncertainties in all parameters except
source inclination and distance.  
Finally, Table \ref{tab:RunComparisons:ToEachOtherAndTheory} uses $D_{KL}$ to compare the  simulations' intrinsic
 posterior distributions for  $(\mc,\eta,\chi)$    to one another and to our analytic estimates.  
In Figs.~\ref{fig:CompareAllIntrinsic:ZeroSpin}-\ref{fig:CompareGeometry}, 
we use selected two-dimensional posterior distributions derived from ``zero noise'' data to illustrate our principal conclusions.  
First, for the same binary, we consistently predict similarly-shaped distributions for intrinsic parameters
($\mc,\eta,\chi$).  
For most simulations, differences between the posteriors are qualitatively consistent with random
chance, limited by signal amplitude fluctuations (e.g., noise-realization-dependent differences in recovered $\rho$).  
As a concrete example, we have performed
multiple  zero-noise, zero-spin simulations  and found differences consistent with our expectations [e.g., as suggested by
Eq. (\ref{eq:DklDistinguishCriteria})]. 
Second, using apples-to-apples comparisons of the same binary in the same noise, 
we consistently find higher harmonics provide minimal additional information about intrinsic parameters.  
Instead, higher harmonics provide \emph{geometric} information.  
Finally, we confirm that for aligned-spin binaries, the intrinsic and extrinsic parameters nearly separate when marginalized
over event time.

\begin{table*}
\begin{tabular}{lll|ccc|lr|llll}
Source & Harmonics & Seed & 
  $\sigma_{\mc}$ & $\sigma_{\eta}$ & $\sigma_{\chi}$ & 
 $\sigma_{\psi_+}$ & $\sigma_{\psi_-}$ & $\sigma_{t}$ & $\sigma_{RA}$ & $\sigma_{DEC}$ & $\sigma_{A}$\\  
  & & &   
 $ \times 10^3$ & $ \times 10^3$ & &  
  &  &  ms & $\unit{deg}$ &  $\unit{deg}$ & $\unit{deg}^2$  
\\ \hline
 {Zero spin} & {no} & - & 2.13 & 1.40 & - & 0.095 & 1.8 & 0.32 & 0.48 & 0.73 & 1.0 \\
 {Zero spin} & {no} & {-*} & 2.15 & 1.39 & - & 0.095 & 1.8 & 0.31 & 0.49 & 0.74 & 1.1 \\
 {Zero spin} & {no} & 1234 & 2.26 & 1.40 & - & 0.098 & 1.8 & 0.28 & 0.48 & 0.69 & 0.96 \\
 {Zero spin} & {no} & {1234*} & 2.35 & 1.57 & - & 0.10 & 1.8 & 0.36 & 0.6 & 0.9 & 1.4 \\
 {Zero spin} & {no} & 56789 & 2.57 & 1.52 & - & 0.1 & 1.8 & 0.32 & 0.53 & 0.76 & 1.2 \\
 {Zero spin} & {no} & {56789*} & 2.11 & 1.43 & - & 0.1 & 1.7 & 0.43 & {\bf 2.9} & {\bf 1.3} & {\bf 6.8} \\
\hline
 {Zero spin} & {with} & - & 1.97 & 1.25 & - & 0.091 & 0.67 & 0.26 & 0.42 & 0.67 & 0.81 \\
 {Zero spin} & {with} & {-*} & 1.94 & 1.24 & - & 0.09 & 0.67 & 0.26 & 0.42 & 0.65 & 0.79 \\
 {Zero spin} & {with} & 1234 & 1.90 & 1.16 & - & 0.088 & 0.67 & 0.23 & 0.4 & 0.63 & 0.70 \\
 {Zero spin} & {with} & {1234*} & 2.12 & 1.35 & - & 0.099 & 0.55 & 0.28 & 0.49 & 0.75 & 0.98 \\
 {Zero spin} & {with} & 56789 & 2.34 & 1.33 & - & 0.095 & 0.75 & 0.24 & 0.43 & 0.69 & 0.87 \\
 {Zero spin} & {with} & {56789*} & 2.04 & 1.40 & - & 0.099 & 0.45 & 0.32 & 0.53 & 0.66 & 1.1 \\
\hline
 {Aligned spin} & {no} & - & 6.19 & 7.89 & 0.038 & 0.088 & 1.8 & 0.37 & 0.41 & 0.63 & 0.71 \\
 {Aligned spin} & {no} & {-*} & 6.33 & 8.28 & 0.039 & 0.087 & 1.8 & 0.4 & 0.4 & 0.62 & 0.70 \\
 {Aligned spin} & {no} & 1234 & 5.50 & 7.12 & 0.029 & 0.088 & 1.8 & 0.39 & 0.49 & 0.67 & 0.82 \\
 {Aligned spin} & {no} & {1234*} & 6.04 & 7.00 & 0.033 & 0.087 & 1.6 & 0.28 & 0.43 & 0.72 & 0.85 \\
 {Aligned spin} & {no} & 56789 & 4.70 & 4.34 & 0.021 & 0.077 & 1.8 & 0.26 & 0.37 & 0.57 & 0.59 \\
 {Aligned spin} & {no} & {56789*} & 6.83 & 8.54 & 0.050 & 0.095 & 1.8 & 0.3 & 0.4 & 0.62 & 0.68 \\
\hline
 {Aligned spin} & {with} & - & 5.26 & 5.94 & 0.035 & 0.087 & 0.63 & 0.24 & 0.33 & 0.55 & 0.50 \\
 {Aligned spin} & {with} & {-*} & 5.38 & 6.19 & 0.035 & 0.087 & 0.63 & 0.25 & 0.34 & 0.55 & 0.51 \\
 {Aligned spin} & {with} & 1234 & 4.73 & 5.05 & 0.030 & 0.082 & 0.62 & 0.21 & 0.36 & 0.59 & 0.55 \\
 {Aligned spin} & {with} & {1234*} & 4.84 & 4.87 & 0.027 & 0.089 & 0.64 & 0.21 & 0.32 & 0.56 & 0.47 \\
 {Aligned spin} & {with} & 56789 & 3.91 & 4.12 & 0.016 & 0.072 & 0.38 & 0.3 & 0.42 & 0.54 & 0.71 \\
 {Aligned spin} & {with} & {56789*} & 7.23 & 10.0 & 0.056 & 0.095 & 0.92 & 0.43 & 0.39 & 0.6 & 0.64 \\
 %% {Zero spin} & {no} & - & 2.15 & 1.39 & - & 0.095 & 1.8 & 0.31 & 0.49 & 0.74 & 1.1 \\
 %% {Zero spin} & {no} & {1234} & 2.35 & 1.57 & - & 0.10 & 1.8 & 0.36 & 0.6 & 0.9 & 1.4 \\
 %% {Zero spin} & {no} & {56789} & 2.11 & 1.43 & - & 0.1 & 1.9 & 0.43 & 2.9 & 1.3 & 6.8 \\
 %% {Zero spin} & {with} & - & 1.94 & 1.24 & - & 0.091 & 0.67 & 0.26 & 0.43 & 0.65 & 0.78 \\
 %% {Zero spin} & {with} & {1234} & 2.09 & 1.33 & - & 0.097 & 0.55 & 0.27 & 0.49 & 0.75 & 0.99 \\
 %% {Zero spin} & {with} & {56789} & 2.00 & 1.37 & - & 0.097 & 0.45 & 0.31 & 0.52 & 0.66 & 1.1 \\
 %% {Aligned spin} & {no} & - & 6.33 & 8.28 & 0.039 & 0.087 & 1.8 & 0.4 & 0.4 & 0.62 & 0.70 \\
 %% {Aligned spin} & {no} & {1234} & 6.04 & 7.00 & 0.033 & 0.087 & 1.9 & 0.28 & 0.43 & 0.72 & 0.85 \\
 %% {Aligned spin} & {no} & {56789} & 6.83 & 8.54 & 0.050 & 0.095 & 1.8 & 0.3 & 0.4 & 0.62 & 0.68 \\
 %% {Aligned spin} & {with} & - & 5.40 & 6.26 & 0.034 & 0.086 & 0.64 & 0.26 & 0.35 & 0.55 & 0.52 \\
 %% {Aligned spin} & {with} & {1234} & 4.79 & 4.91 & 0.028 & 0.086 & 0.61 & 0.21 & 0.31 & 0.58 & 0.47 \\
 %% {Aligned spin} & {with} & {56789} & 6.95 & 9.65 & 0.058 & 0.09 & 0.92 & 0.4 & 0.4 & 0.6 & 0.61 
\end{tabular}

\caption{\label{tab:ParameterErrors}\textbf{One-dimensional parameter errors}: Measurement accuracy $\sigma_x$ for
$x$, one of   several intrinsic ($\mc,\eta, \chi$) and extrinsic ($\psi_{\pm},t,RA,DEC$) parameters.    
The extrinsic parameters are the polarization and orbital phase combinations $\psi_{\pm}\equiv \psi\pm \phiref$ defined on
  $[-\pi/2,\pi/2]$, so $\sigma_{\psi}=1.81$ is consistent with a uniform distribution; the event time
  $t$; the sky position measured in RA and DEC; and the sky area $A$, estimated using the $2\times 2$ covariance matrix
  $\Sigma_{ab}$ on the sky via $\pi |\Sigma|$.  %
One  simulation  recovers nonzero probability at two antipodal sky locations, significantly increasing  $\Sigma$
 above the   ``natural'' value associated with each best-fit peak; impacted parameters are shown in boldface.   
 For selected cases, these measurement accuracies are demonstrated in Figure \ref{fig:CompareGeometry}.  
Comparing simulations with identical noise realizations with and without higher harmonics, this table suggests higher harmonics have minimal impact on most parameters, except for $\psi_-$.  
Comparing simulations with identical phyiscs but different noise realizations suggests that even for our optimistic
signal amplitude, the information provided by higher harmonics is small, comparable to fluctuations due to the noise
realization.  %
  All parameter accuracies shown qualitatively agree with simple numerical estimates derived from the Fisher matrix or
  timing [Section \ref{sec:sub:Geometric}], up to tens of percent relative fluctuations due to finite sample size
  [Eq. (\ref{eq:ErrorInSigma})]. 
}
\end{table*}

\ForInternalReference{
\begin{table*}
{\tiny
\begin{tabular}{llr|cc|lr|lll|llllll|r}
Source & Harmonics & Seed & 
$\rho_{\rm inj}$ & $ \rho_{\rm  rec}$ &  $\ln Z$ & $\ln V/V_{\rm prior}$ &
 $\sigma_{\mc}$ & $\sigma_{\eta}$ & $\sigma_{a}$ & 
\multicolumn{6}{c}{$D_{KL}$}
%$D_{KL}(\mc,\eta)$ & $D_{KL}(\tau_0,\tau_3)$ & $D_{KL}(\mc,\eta,a)$ & $D_{KL}(\tau_0,\tau_3,a_1)$ 
& $N_{\rm eff}$\\  
& & & &  & &  &  $ \times 10^3$ & $ \times 10^3$ &  &
  $(\mc,\eta)$ & $(\tau_0,\tau_3)$ & $(\mc,\eta^{-2})$& $(\mc,\eta,a_1)$ & $(\tau_0,\tau_3,a_1)$ & $(\mc,\eta^{-2},a_1)$ & \\
\hline
 {Zero spin} & {no} & - & 20.33 & 20.64 & 180. & -33.4 & 2.13 & 1.40 & - & 0 & 0 & 0 & - & - & - & 7517 \\
 {Zero spin} & {no} & 1234 & 20.33 & 20.5 & 177. & -33.3 & 2.26 & 1.40 & - & 0.026 & 0.035 & 0.048 & - & - & - & 8025 \\
 {Zero spin} & {no} & 56789 & 20.33 & 20.34 & 172. & -34.8 & 2.57 & 1.52 & - & 0.089 & 0.074 & 0.061 & - & - & - & 10403 \\
 {Zero spin} & {with} & - & 21.03 & 21.33 & 191. & -36.3 & 1.97 & 1.25 & - & 0.029 & 0.028 & 0.026 & - & - & - & 8027 \\
 {Zero spin} & {with} & 1234 & 21.03 & 21.76 & 200. & -37. & 1.90 & 1.16 & - & 0.087 & 0.076 & 0.068 & - & - & - & 7511 \\
 {Zero spin} & {with} & 56789 & 21.03 & 20.67 & 177. & -36.6 & 2.34 & 1.33 & - & 0.13 & 0.12 & 0.12 & - & - & - &
 11358 \\
 {Aligned spin} & {no} & - & 22.99 & 22.67 & 222. & -34.6 & 6.19 & 7.89 & 0.038 & 0 & 0 & 0 & 0 & 0 & 0 & 9841 \\
 {Aligned spin} & {no} & 1234 & 22.99 & 22.81 & 225. & -35.1 & 5.50 & 7.12 & 0.029 & 0.018 & 0.018 & 0.030 & 0.13 & 0.13 & 0.15 & 8670 \\
 {Aligned spin} & {no} & 56789 & 22.99 & 24.89 & 272. & -37.5 & 4.70 & 4.34 & 0.021 & 0.95 & 0.26 & 0.093 & 1.6 & 0.9 & 0.79 & 10508 \\
 {Aligned spin} & {with} & - & 23.54 & 23.18 & 231. & -37.8 & 5.15 & 5.83 & 0.035 & 0.16 & 0.12 & 0.10 & 0.19 & 0.14 & 0.12 & 3355 \\
 {Aligned spin} & {with} & 1234 & 23.54 & 23.67 & 241. & -39.5 & 4.74 & 4.98 & 0.030 & 0.45 & 0.27 & 0.17 & 0.56 & 0.35 & 0.24 & 5572 \\
 {Aligned spin} & {with} & 56789 & 23.54 & 25.41 & 280. & -43.2 & 3.93 & 4.13 & 0.017 & 1.1 & 0.43 & 0.44 & 2.8 & 2.3 & 2.4 & 12070 \\
\end{tabular}
}

\caption{\label{tab:RunSummaryAndResults:Original}\textbf{Simulation results}: Table of distinct simulations performed.  The
  first set of columns indicate which of the two fiducial binaries was used (zero spin vs aligned spin), 
  whether higher harmonics were included, and random seed choice used to generate noise 
  (a ``-'' or $+$ means no noise was used).  The starred simulations are used as representative examples in subsequent
  tables and figures.  
The two quantities $\rho_{\rm inj},\rho_{\rm rec}$ provide the
  injected and best-fit total signal amplitude in the network [Eqs. (\ref{eq:def:rhoRecovered},\ref{eq:def:rho})].    
The latter quantity depends on the noise realization of the network. 
The columns for $\ln Z$ and $V/V_{\rm prior}$ provide the evidence [Eq. (\ref{eq:def:Z:Modified})] and volume fraction
[Eq. (\ref{eq:def:VoverVprior})]; the evidence, volume fraction, and signal amplitude are related by $\rho_{\rm rec}^2/2  =
\ln Z/(V/V_{\rm prior})$.  
  The next three columns show the one-dimensional standard deviations in chirp mass ($\sigma_{\mc}$), symmetric mass
  ratio ($\sigma_\eta$), and BH dimensionless spin ($\sigma_a$).  
  These quantities fluctuate significantly, driven both by noise realization dependence and the large number of
  effective samples needed to accurately estimate their value [Eq. (\ref{eq:ErrorInSigma})]. 
The six quantities $D_{KL}$ are calculated from the
  two- and three-dimensional covariance matrices using Eq. (\ref{eq:Dkl}), 
  using the coordinate systems labeling the columns.    The two rows with zeros as  entries indicate the two reference
  choices, against which all nonspinning or spinning parameter estimation was compared.   
  For zero spin, the first three rows show differences consistent with noise fluctuations
  [Eq. (\ref{eq:DklCriteria:FluctuateAmplitude:Typical})];
  for aligned spin,  diffrences are more substantial and coordinate-system dependent, but not above the conditions
  needed to distinguish between distributions [Eq. (\ref{eq:DklCriteria:FluctuateAmplitude})].  
  Finally, $N_{\rm eff}$ is the effective number of independent samples in our calculations.
\editremark{FIXME:  Explain zero-noise SNR, or replace by zero noise value.}
\editremark{[Some $D_{KL}$ numbers are unclear which 2 rows are being compared. e.g. what is row 4 being compared to, row 1? Is row 5 compared to row 4 or row 2?]}
}
\end{table*}
}

\subsection{Intrinsic distributions agree} 
\begin{figure}
\includegraphics[width=\columnwidth]{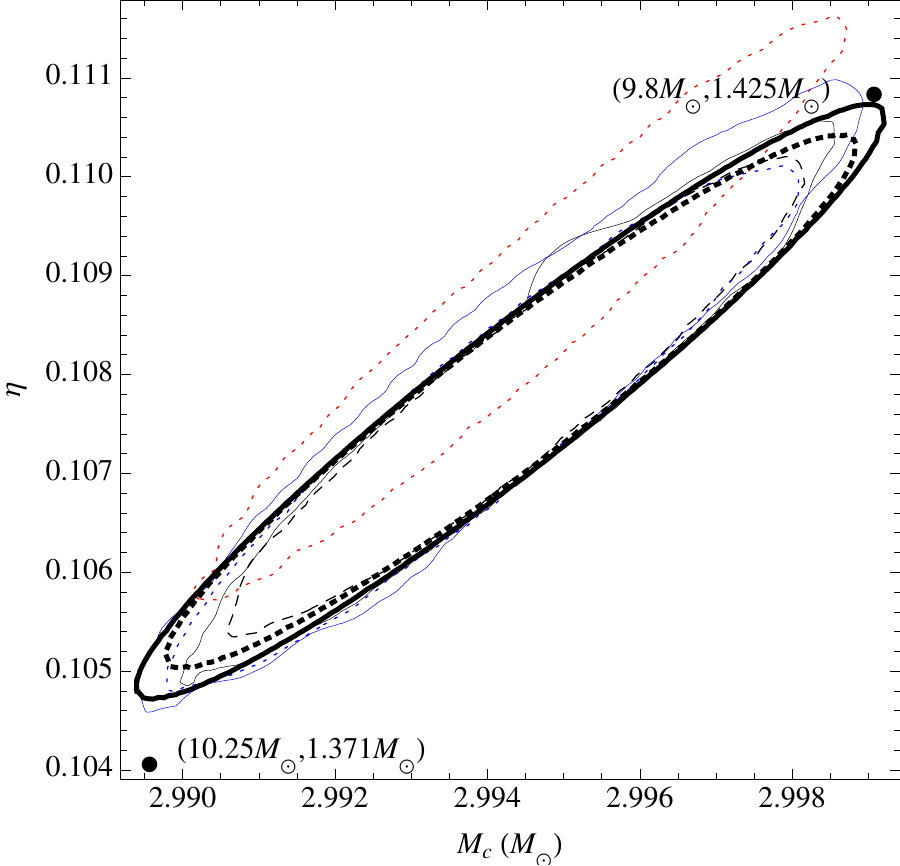}
\caption{\label{fig:CompareAllIntrinsic:ZeroSpin}\textbf{Chirp mass and mass ratio 90\% confidence interval: Zero spin}: Assuming our fiducial nonspinning signal
  is present in distinct realizations of Gaussian noise (colors, described below), the contour shows the  $90\%$ confidence intervals for
  $\mc,\eta$ derived from the half of our zero-spin calculations marked with ``*''.   Solid curves correspond to a signal without higher harmonics; dashed
  curves include higher harmonics; and colors denote specific noise realizations listed in Table \ref{tab:RunSummaryAndResults}, 
not all of which appear in this figure: 
zero noise (black); 1234 (blue); and 56789 (red). %
[To better distinguish between cases including higher harmonics, the zero noise case is shown  as a black dashed curve.]
  All posteriors have similar shape; differences between the estimated posteriors are consistent
  with finite sample and noise realization effects.    Higher harmonics do not improve our estimates of intrinsic parameters in any
  noticeable way.  
For comparison,  the thick black solid and dotted curves are  analytic estimates using the effective Fisher matrix
normalized to $\rho=20$, described in greater detail in
Section \ref{sec:results:Compare}.  
To help translate these results to an astrophysically relevant scale, the two black points and pairs indicate the chirp mass and mass ratios corresponding to
($m_1/M_\odot,m_2/M_\odot$).  %
}
\end{figure}

\begin{figure*}
\includegraphics[width=\textwidth]{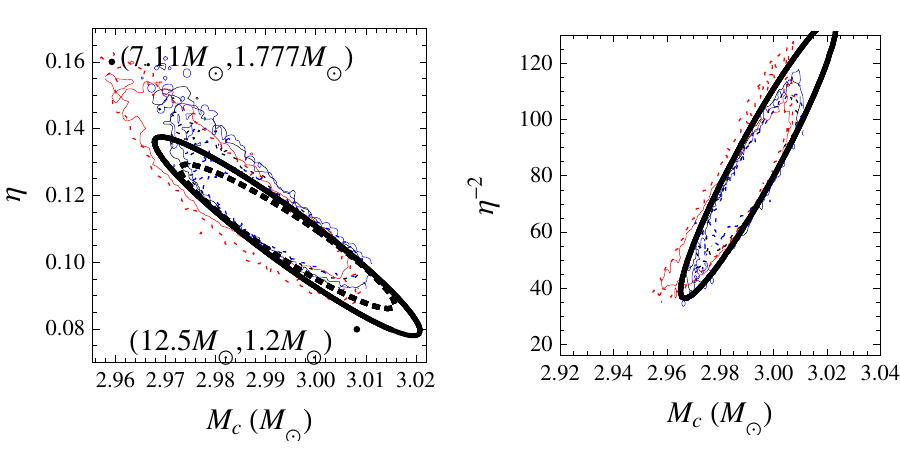}
\caption{\label{fig:CompareAllIntrinsic:AlignedSpin:Extreme}
  \textbf{99.9\% confidence intervals in mass plane for aligned-spin binary}: 
  For our fiducial aligned-spin signal injected into distinct realizations of Gaussian noise (colors, as described in the caption to Figure \ref{fig:CompareAllIntrinsic:ZeroSpin}), 
  the contours show the  $99.9\%$ confidence intervals 
  from each calculation in our various coordinates for the mass plane.   Contour styles are as described in Figure
  \ref{fig:CompareAllIntrinsic:ZeroSpin}.  This figure conveys three key points.  
  First, the similarity between the blue solid and dotted contours shows higher
  harmonics provide little additional information about intrinsic parameters.   
  Second,  measurements of spinning binaries can at best weakly distinguish the individual masses in BH-NS binaries.
For comparison, the  solid points and associated  $(m_1,m_2)$ pairs show where those points lie in the  $\mc,eta$ plane.
For our loud fiducial signal, the NS mass is constrained to lie well within the range allowed from prior expereince.
  Third, suitable coordinates can simplify all posterior probability distributions, independent of noise realization. 
}
\end{figure*}

\begin{figure*}
\includegraphics[width=\textwidth]{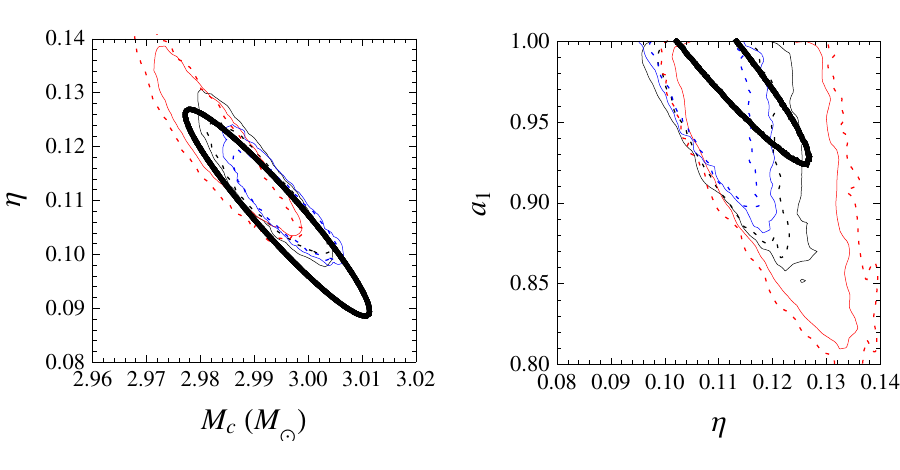}
\caption{\label{fig:CompareAllIntrinsic:AlignedSpin}\textbf{90\% confidence intervals for aligned-spin binary}: 
  For our fiducial aligned-spin signal injected into distinct realizations of Gaussian noise (colors, as described in the caption to Figure \ref{fig:CompareAllIntrinsic:ZeroSpin}), 
  the contour shows the  $90\%$ confidence intervals from each calculation 
  in the $\mc,\eta$ plane (left panel) and the $\eta,\chi$ plane (right panel).   Contour styles are as described in Figure
  \ref{fig:CompareAllIntrinsic:ZeroSpin}; as previously, the heavy black solid and dashed curves show revised analytic
  predictions using the \abbrvEFM{} method, provided in Table \ref{tab:Fisher:NoAndAlignedSpin}.   
}
\end{figure*}

As illustrated strikingly by Figure \ref{fig:CompareAllIntrinsic:ZeroSpin}, 
our zero-spin simulations consistently produce tightly-confined, highly-Gaussian, similarly-shaped posteriors in
$\mc,\eta$.    Small but nonzero differences in shape, size and position do exist between different noise realizations.  
Quantitatively, however, these shape differences have comparable magnitude 
to the effects expected when comparing different  noise fluctuations,
as seen by the corresponding column in Table  \ref{tab:RunSummaryAndResults}.  
Similarly, even for the the  $99.9\%$ confidence intervals in the left panel of Figure
\ref{fig:CompareAllIntrinsic:AlignedSpin:Extreme}, our aligned-spin simulations roughly agree.  
Over this relatively large region of parameter
space, particularly in mass ratio, the posterior distribution is  
not Gaussian in $\mc,\eta,\chi$ coordinates at this confidence level.    
Even in these coordinates, however, as illustrated using the 90\% confidence intervals in Figure \ref{fig:CompareAllIntrinsic:AlignedSpin},
most of the probability is adequately approximated by some locally-Gaussian approximation, times a cutoff at $\chi =1$.  
Strictly speaking, this cutoff forces the local distribution to be locally non-Gaussian and forces the local covariance to
be both noise-realization-dependent and different from our analytic calculations.  
In practice, ignoring these subtleties and treating the posterior as locally Gaussian, however, we
 find all posteriors have surprisingly similar one-, two-, and three-dimensional covariances $\Sigma$, as quantified in
 Table  \ref{tab:ParameterErrors}.    
Figure  \ref{fig:CompareAllIntrinsic:AlignedSpin:Extreme} 
also illustrates the value of well-chosen coordinates.  As
seen in the left panel of Figure  \ref{fig:CompareAllIntrinsic:AlignedSpin:Extreme}, using $(\mc,\eta,\chi)$ coordinates the
error contours are both weakly nonellipsoidal and have shape that  weakly depends on noise realization. 
As seen in the right panel of Figure  \ref{fig:CompareAllIntrinsic:AlignedSpin}, however, alternative coordinates
mitigate nongaussianity and reduce noise-realization-dependent effects. 
This improvement persists for low signal amplitudes, which have  broader posteriors than shown here. 

\begin{table*}
\begin{tabular}{lll|ll|ll}
Source & Harmonics & Seed & $D_{KL}(\mc,\eta)$ & $D_{KL}(\mc,\eta|\rm eff)$ & $D_{KL}(\mc,\eta,\chi)$ &
$D_{KL}(\mc,\eta,\chi|{\rm eff})$\\ 
\hline
 {Zero spin} & {no} & - & 0.00128 & 0.0411 & {} & {} \\
 {Zero spin} & {no} & {-*} & 0 & 0.0285 & {} & {} \\
 {Zero spin} & {no} & 1234 & 0.0157 & 0.00261 & {} & {} \\
 {Zero spin} & {no} & {1234*} & 0.0321 & 0.0462 & {} & {} \\
 {Zero spin} & {no} & 56789 & 0.0733 & 0.0193 & {} & {} \\
 {Zero spin} & {no} & {56789*} & 0.0209 & 0.0886 & {} & {} \\
\hline
 {Zero spin} & {with} & - & 0.0229 & 0.0299 & {} & {} \\
 {Zero spin} & {with} & {-*} & 0.0235 & 0.0339 & {} & {} \\
 {Zero spin} & {with} & 1234 & 0.0718 & 0.00941 & {} & {} \\
 {Zero spin} & {with} & {1234*} & 0.00202 & 0.0338 & {} & {} \\
 {Zero spin} & {with} & 56789 & 0.112 & 0.0288 & {} & {} \\
 {Zero spin} & {with} & {56789*} & 0.0305 & 0.122 & {} & {} \\
\hline
{Aligned spin} & {no} & - & 0.00391 & 0.0984 & 0.00427 & 2.29 \\
 {Aligned spin} & {no} & {-*} & 0 & 0.123 & 0 & 2.31 \\
 {Aligned spin} & {no} & 1234 & 0.0272 & 0.16 & 0.140 & 2.12 \\
 {Aligned spin} & {no} & {1234*} & 0.110 & 0.0959 & 0.163 & 2.07 \\
 {Aligned spin} & {no} & 56789 & 1.17 & 0.44 & 1.82 & 2.13 \\
 {Aligned spin} & {no} & {56789*} & 0.0112 & 0.0971 & 0.0833 & 2.51 \\
\hline
 {Aligned spin} & {with} & - & 0.216 & 0.0734 & 0.265 & 2.3 \\
 {Aligned spin} & {with} & {-*} & 1.45 & 0.0733 & 1.68 & 2.21 \\
 {Aligned spin} & {with} & 1234 & 0.512 & 0.0465 & 0.609 & 2.19 \\
 {Aligned spin} & {with} & {1234*} & 0.724 & 0.044 & 0.898 & 2.02 \\
 {Aligned spin} & {with} & 56789 & 1.31 & 0.0807 & 3.10 & 1.8 \\
 {Aligned spin} & {with} & {56789*} & 0.0661 & 0.511 & 0.163 & 3.12 \\

\end{tabular}

\caption{\label{tab:RunComparisons:ToEachOtherAndTheory}\textbf{Comparing simulations}: 
For each simulation, a comparison of that
  simulation's shape with either (a) some other similar simulation's shape (columns $D_{KL}(\mc,\eta)$ and
  $D_{KL}(\mc,\eta,\chi)$), either (zero spin, no, $-*$) or (aligned spin, no, $-*$), or (b) the corresponding effective Fisher matrix provided in Table
  \ref{tab:AlignedSpin:SensitiveToPN}.
  For zero spin, the first three rows show differences consistent with noise fluctuations
  [Eq. (\ref{eq:DklCriteria:FluctuateAmplitude:Typical})];
  for aligned spin,  diffrences are more substantial and coordinate-system dependent, but not above the conditions
  needed to distinguish between distributions [Eq. (\ref{eq:DklCriteria:FluctuateAmplitude})].  
}
\end{table*}

\subsection{Marginal information from higher harmonics is confined to source orientation }
Using apples-to-apples comparisons of the same source in the same data, we can explicitly confirm that higher harmonics
provide minimal new information about intrinsic parameters.  In fact, the differences between the zero-spin, zero-noise posterior in $\mc,\eta$
calculated with and without harmonics are at best comparable to  the fluctuations seen between different data
realizations; see Table \ref{tab:ParameterErrors} for the one-dimensional measurement errors, Table
\ref{tab:RunComparisons:ToEachOtherAndTheory} for comparisons between simulations using $D_{KL}$,   and Table
\ref{tab:RunSummaryAndResults} for a comparison using $V/V_{\rm prior}$.   

With aligned spin, higher harmonics seem to provide some additional information.  For example,  Figure
\ref{fig:CompareAllIntrinsic:AlignedSpin} shows the two-dimensional
posteriors in $\mc,\eta$ for three  starred  data realizations (black, red, blue) both with (dotted) and without (solid) higher
harmonics; each pair of contours  differ slightly in direction and extent.  
These distributions are manifestly similar: the presence of higher have less of an effect than a
change of noise realization (e.g., a change in $\rho$ of order unity).    Physically, though higher harmonics provide information, different
data realizations shift the error ellipsoids' positions, orientations, and scales so much that their marginal impact cannot
be easily isolated.  
In all cases, however, higher harmonics seem to provide minimal additional information about our two fiducial sources'
intrinsic parameters.

\begin{figure*}
\includegraphics[width=\columnwidth]{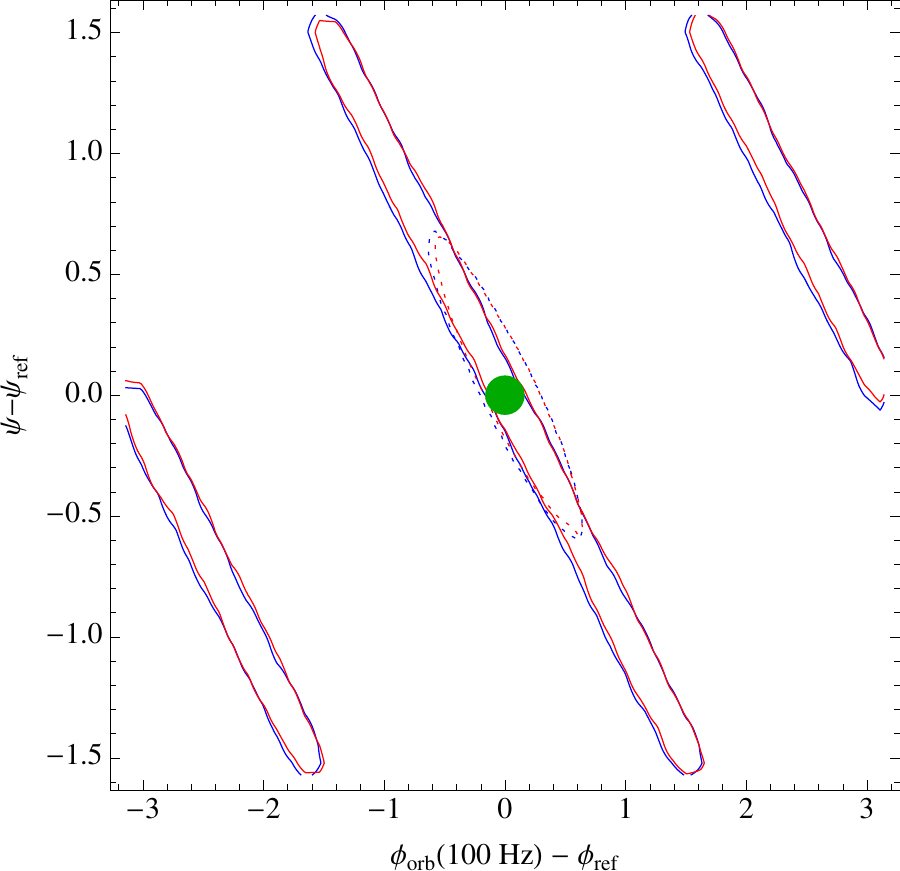}
\includegraphics[width=\columnwidth]{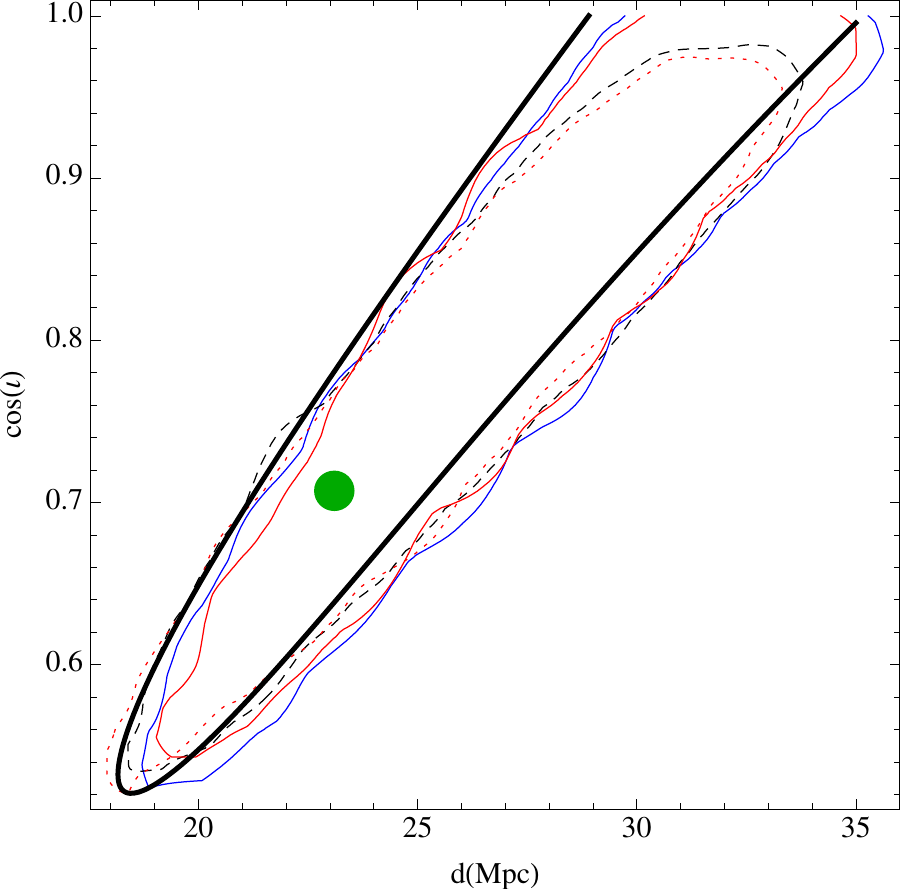}
\includegraphics[width=\columnwidth]{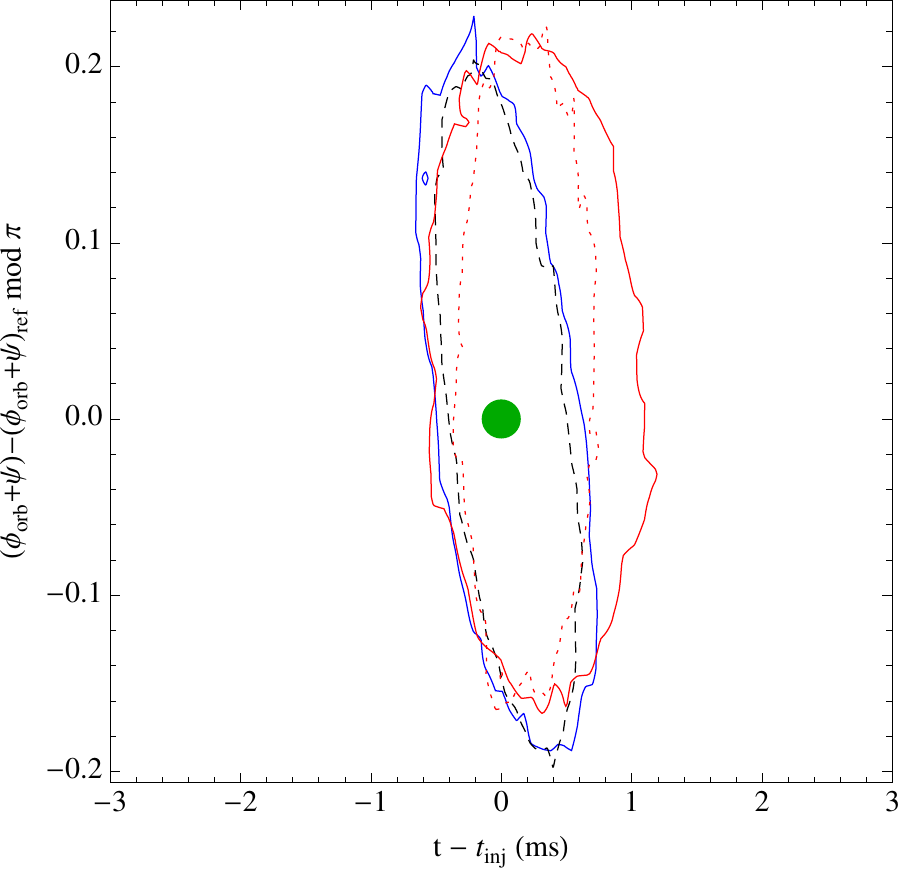}
\includegraphics[width=\columnwidth]{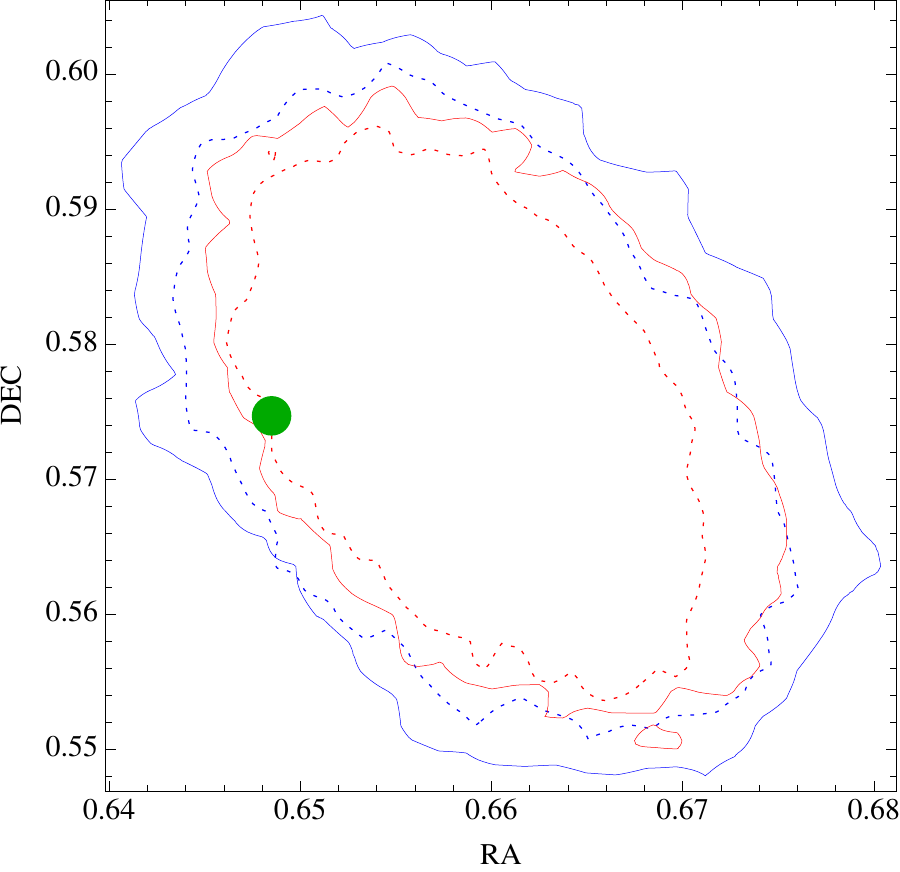}
\caption{\label{fig:CompareGeometry}\textbf{Higher harmonics provide information about geometry}: For zero-spin binaries (blue) and
  aligned-spin binaries (red), a comparison of 90\% confidence intervals derived with and without higher harmonics (dotted,
  solid, respectively).   Though the plotted results are obtained without noise, similar results are obtained with each
  noise realization
\emph{Top left panel}: Posterior in $\psi$ (the angle of $\hat{L}$  projected onto the plane of the sky) 
and $\phiref$ (the orbital phase at $100\unit{Hz}$).  In the absence of higher harmonics, $\phiref+\psi$ is
strongly constrained by observations  but $\phiref-\psi$ is unconstrained.  When higher harmonics are included, the
range of possible values of $\phiref-\psi$ is significantly reduced.   
\emph{Top right panel}: Posterior in $d$ (distance to the source) and $\cos \iota$ for $\iota$ the inclination.  
The heavy black curve shows an analytic approximation to the 90\% confidence interval
  [Eq. (\ref{eq:PosteriorDistanceIota:NoHarmonics})].   
Except for limiting cases ($\cos \iota \simeq \pm 1,0$) or special sky locations, higher harmonics provide relatively little new information
about the source orientation and distance.  
Our results are consistent with prior work  \cite{CutlerFlanagan:1994,gw-astro-ShortGRBSirens-Hughes2009,2011ApJ...739...99N,2013ApJ...767..124N,2012PhRvD..85j4045V,LIGO-2013-WhitePaper-CoordinatedEMObserving}.
\emph{Bottom left panel}: Posterior in $\psi+\phiref$ versus $t-t_{inj}$.    While higher harmonics improve our ability
to measure the previously unconstrainted $\psi-\phiref$, they do not significantly improve our ability to measure time
or the other phase ($\phiref+\psi$). 
\emph{Bottom right panel}: Posterior distribution for the source position on the plane of the sky.  Higher harmonics
slightly improve our ability to isolate the direction to the source, reducing the 90\% confidence interval sky area by
tens of percent. 
  The improved sky area is smaller only in part  because higher harmonics increase the signal amplitude and thus improve our
ability to identify the source's location.  While a source with higher harmonics has larger amplitude ($\rho^2$ larger
by about 3\%), the increased amplitude alone does not explain the significantly smaller sky area.   
The center of our recovered distributions and the actual source sky location (green dot) are offset by about
$0.01$ rad. This is due to an error in the injection routines, see Footnote \ref{note:TimeShift}.  
}
\end{figure*}

By contrast, as illustrated by Figure \ref{fig:CompareGeometry},  higher harmonics do provide geometric information,
improving our knowledge about the source position and orientation   relative to the line of sight.  
Higher harmonics are known to break almost-perfect degeneracies present in the leading-order gravitational wave 
signal \cite{2006PhRvD..74l2001L,2009PhRvD..80f4027K,2011PhRvD..84b2002L}.
This  signal can be represented in a compact complex form as
\begin{align} \label{eq:complexh}
h &= h_+ - i h_\times = - e^{-2i\psi} \frac{8 \mu v^2}{d_L} \sqrt{\frac{\pi}{5} }
\nonumber \\& \times
\left[   e^{-2i \Phi_{orb}}\Y{-2}_{22}(\iota,0)
   +    e^{2i \Phi_{orb}}\Y{-2}_{2-2}(\iota,0)
 \right]
\end{align}
where $\Y{s}_{lm}$ are spin-weighted spherical harmonics and we note that\footnote{The azimuthal argument of 
$\Y{-2}_{22}$ is degenerate with $\phiref$, so we can set it to zero without loss of generality.}:
\begin{eqnarray}
\Y{-2}_{22}(\iota,0) &=& \frac{1}{2}\sqrt{\frac{5}{\pi}} \cos^4 \left(\frac{\iota}{2}\right) = \frac{1}{8}\sqrt{\frac{5}{\pi}}  \left( 1 + \cos \iota \right)^2 , \\
\Y{-2}_{2-2}(\iota,0) &=& \frac{1}{2}\sqrt{\frac{5}{\pi}} \sin^4 \left(\frac{\iota}{2}\right) = \frac{1}{8}\sqrt{\frac{5}{\pi}}  \left( 1 - \cos \iota \right)^2. 
\end{eqnarray}    

For most orientations, either one or the other spin-weighted harmonic dominates this sum\footnote{Only when 
the binary is nearly edge-on, i.e. $\iota \simeq \pi/2$, are the two comparable}.  Our two sources have
$\iota=\pi/4$, so the $(2,2)$ mode dominates by a factor $\cos^4 (\pi/8) / \sin^4 (\pi/8) \simeq 34$. 
This means that to a good approximation the gravitational wave signal depends on 
$\psi$ and $\phiref$ principally through $\psi + \phiref$ and on $d_L$ and $\iota$ through $(1+\cos \iota)^2/d_L$.  
More generally, in the absence of higher harmonics, to a first approximation the distance, inclination, polarization,
and orbital phase enter into the signal via $e^{-2i(\psi \pm \phi_{\rm ref})} (1\pm \cos \iota)^2 /d_L$.  
This functional form explains the two-dimensional correlations between 
$(\psi,\phi_{\rm ref})$ and $(\iota,d_L)$ shown in Figure \ref{fig:CompareGeometry}.

When higher harmonics are included in the signal, the expression for $h$ in Eq.~(\ref{eq:complexh}) 
generalizes to a sum over many multipoles
$h_{lm}\Y{-2}_{lm}(\iota,0)$, each with a distinct angular dependence. Unlike the leading-order case, several
terms contribute to the overall amplitude for our fiducial case $\iota = \pi/4$ (and in general), 
with each harmonic having a different dependence on $\iota$ and $\phi_{\rm ref}$.
As a result, a signal including higher harmonics
communicates additional information about its orientation, as illustrated by the dotted curves in Figure
\ref{fig:CompareGeometry}.  
As with intrinsic parameters,  however, the amount of information we gain about source orientation 
seems to depend on the specific noise realization.

\ForInternalReference{
\begin{table*}
\begin{tabular}{llr|llllll|l}
Source & Harmonics & Seed & $\sigma_{\psi_+}$ & $\sigma_{\psi_-}$ & $\sigma_{t}(ms)$ & $\sigma_{RA}$ & $\sigma_{DEC}$
&${A}$  & $N_{eff}$\\ 
 &  &  &  && ms & $\unit{deg}$ &  $\unit{deg}$ & $\unit{deg}^2$  & \\ 
\hline
 {Zero spin} & {no} & 1234 & 0.0981 & 1.8 & 0.285 & 0.483 & 0.694 & 0.964 & 8025 \\
 {Zero spin} & {no} & 56789 & 0.102 & 1.76 & 0.324 & 0.529 & 0.756 & 1.18 & 10403 \\
 {Zero spin} & {no} & - & 0.0948 & 1.82 & 0.318 & 0.484 & 0.734 & 1.02 & 7517 \\
 {Zero spin} & {with} & 1234 & 0.088 & 0.666 & 0.225 & 0.396 & 0.626 & 0.702 & 7511 \\
 {Zero spin} & {with} & 56789 & 0.0946 & 0.772 & 0.242 & 0.427 & 0.693 & 0.869 & 11358 \\
 {Zero spin} & {with} & - & 0.0909 & 0.672 & 0.258 & 0.42 & 0.674 & 0.806 & 8027 \\
 {Aligned spin} & {no} & 1234 & 0.0879 & 1.82 & 0.393 & 0.486 & 0.669 & 0.824 & 8670 \\
 {Aligned spin} & {no} & 56789 & 0.0766 & 1.8 & 0.265 & 0.372 & 0.57 & 0.590 & 10508 \\
 {Aligned spin} & {no} & - & 0.0879 & 1.8 & 0.372 & 0.407 & 0.627 & 0.712 & 9841 \\
 {Aligned spin} & {with} & 1234 & 0.0832 & 1.08 & 0.209 & 0.371 & 0.569 & 0.562 & 4000 \\
 {Aligned spin} & {with} & 56789 & 0.0715 & 1.13 & 0.296 & 0.418 & 0.544 & 0.701 & 11599 \\
 {Aligned spin} & {with} & - & 0.0875 & 0.882 & 0.236 & 0.321 & 0.546 & 0.486 & 3893
\end{tabular}

\caption{\label{tab:Extrinsic}\textbf{Simulation results II: Extrinsic parameters}: Measurement accuracy $\sigma_x$ for several  extrinsic
  parameters $x$: the polarization and orbital phase combinations $\psi_{\pm}\equiv \psi\pm \phiref$ defined on
  $[-\pi/2,\pi/2]$, so $\sigma_{\psi}=1.81$ is consistent with a uniform distribution; the event time
  $t$; the sky position measured in RA and DEC; and the sky area $A$, estimated using the $2\times 2$ covariance matrix
  $\Sigma_{ab}$ on the sky via $\pi |\Sigma|$.  %
 For selected cases, these measurement accuracies are demonstrated in Figure \ref{fig:CompareGeometry}.  
 Higher harmonics have minimal impact on most parameters, except for $\psi_-$.  
 Our simulations also cannot resolve whether higher harmonics improve the evidence for a signal, principally due to
 finite computational resources (e.g., a finite number of sample points and noise realizations).  
  All parameter accuracies shown qualitatively agree with simple numerical estimates derived from the Fisher matrix or
  timing [Section \ref{sec:sub:Geometric}], up to tens of percent relative fluctuations due to finite sample size
  [Eq. (\ref{eq:ErrorInSigma})]. 
}
\end{table*}
}

\subsection{Bounding the relative impact of higher harmonics }

Including gravitational wave content beyond the leading-order quadrupole provides greater signal power and hence on
average provides stronger constraints on source physics.   
Using selected low-dimensional examples, we have used the correlation matrix $\Sigma$ to suggest that higher harmonics
provide relatively little new information about source physics.  
In this section, we argue that our information census has already identified all ways that higher harmonics can improve
our understanding of this source.

To define the relative impact of higher harmonics invariantly, we use  $V/V_{prior}$ as computed 
via Eq. (\ref{eq:def:VoverVprior}).\footnote{For sufficiently strong sources, 
we could assume the Fisher matrix provides a valid approximation to the posterior 
and compute the prior volume ratio from its determinant, as in Eq.~(\ref{eq:VoverVpriorFisher}).
However, we do not rely on this assumption and instead compute it directly from the evidence.
}  
This quantity is a ratio of characteristic (parameter) volumes: the volume consistent with observations and the
prior volume.  It is straightforward to calculate from our simulations and does not require any assumptions such as
approximate Gaussianity.  
By comparing measurements with and without higher harmonics, we can quantify their relative impact.  
The results of these comparison are summarized  in Tables \ref{tab:RunSummaryAndResults} and \ref{tab:ParameterErrors}.  

To provide a sense of scale, the numerical increase in evidence and decrease in $V/V_{\rm prior}$ can be estimated using the signal amplitude
$\rho$ and the expected number of measurable dimensions $D_{\rm eff}$.   A model with higher harmonics has a higher signal amplitude $\rho$ in any noise realization; for the
nonspinning model with zero noise and hence $8$ or $9$ parameters, $\rho$ is  21.03 versus 20.32,
respectively [Table \ref{tab:RunSummaryAndResults}].    The evidence scales as 
$Z\propto \rho^{-D_{\rm eff}}\exp (\rho^2/2)$
[Eqs. (\ref{eq:def:Deff},\ref{eq:def:VoverVprior},\ref{eq:def:rhoRecovered})]; for example,  $\Delta \ln Z \simeq 7.5$
between the nonspinning model with and
without higher harmonics, in zero noise.   This expression crudely explains  the large evidence differences between
scenarios with and without higher harmonics, up to systematic errors in our calculation of $\ln Z$ explained in Appendix
\ref{ap:Thermo}.  
Similarly, the prior volume scales as 
\begin{eqnarray}
\Delta \ln (V/V_{\rm prior}) \simeq
 -D_{\rm eff}\Delta \ln \rho \simeq -D_{\rm eff} \Delta \rho/\rho
\end{eqnarray}
This expression suggests that the volume fraction $V/V_{\rm prior}$ deceases relatively little because the amplitude
increases little; for example, this expression suggests $\Delta \ln (V/V_{\rm prior}) \simeq  0.3$ for the zero-spin
binary in zero noise.  
In fact,  higher harmonics have a much more significant effect on  $V/V_{\rm prior}$ than this estimate would suggest:
$\Delta \ln (V/V_{\rm prior}) \simeq -3$  (zero spin) or perhaps $-4$ (aligned spin).   
Higher harmonics provide more information than the increase in SNR would suggest 
by breaking degeneracies in the Fisher matrix.  
For this system, though, our experience with most one- and two-dimensional distributions [Table
  \ref{tab:ParameterErrors} and Figures \ref{fig:CompareAllIntrinsic:ZeroSpin},
  \ref{fig:CompareAllIntrinsic:AlignedSpin}, and \ref{fig:CompareGeometry}] suggests the broken degeneracy is between two largely uninteresting parameters 
(the polarization angle $\psi$ and the reference orbital phase $\phi_{\rm ref}$)
with small improvements in measurability distributed among the other parameters.
Because  $\ln (V/V_{prior})$ changes by less than 3 in our simulations, higher harmonics cannot improve
the product of uncertainty in parameters by more than a factor $e^{-3 }\simeq 20$.  
By contrast, for the systems simulated, 
higher harmonics improve our ability to measure one polarization combination (here, $\psi_-$), reducing
$\sigma_{\psi_-}$ by about a factor of 3 [Table \ref{tab:ParameterErrors}] -- roughly 1/3 of all of the available information content.   The remaining factor is distributed among small
changes in the remaining 8 parameter combinations, at the tens of percent level or less (i.e., set by $(\ln 20/3)/d$). 
These results strongly suggest higher harmonics have little global impact, bounding above the extent to which higher
harmonics can modify global correlations for these strong nonprecessing signals.  

Similar conclusions can be drawn from almost all of our simulations with noise: comparing simulations with the same
noise realization and physics with and without higher harmonics, usually   $\ln (V/V_{\rm prior})$ changes by less than
of order $3$.  
That said, a few pairs of simulations have claimed  prior volume differences that are
significantly larger than this value; most notably,  the zero spin 56789* realizations with and without noise have
$\Delta \ln (V/V_{\rm prior}) \simeq -27.3+40.4\simeq 13$.   
While these pairs of simulations have large prior volume differences, follow-up comparisons show no clear sign that higher harmonics have
any significant impact.  For example, these simulations have nearly regular one- and two-dimensional parameter distributions [Table
  \ref{tab:ParameterErrors} and Figures \ref{fig:CompareAllIntrinsic:ZeroSpin} and \ref{fig:CompareGeometry}].
Too, when the  evidence is calculated using a different method, described in Appendix \ref{ap:Thermo}, this 
discrepancy  disappears.    We suspect that direct evidence integration, the default method used in
\texttt{lalinference\_mcmc} and reported in Table   \ref{tab:ParameterErrors}, may behave pathologically for outlier
noise realizations, particularly when an extremely low-probability secondary maxima is under-resolved in set of
posterior samples.  
By contrast, in our experience, thermodynamic integration produces no extreme outliers, has $V/V_{\rm prior}$
consistently smaller for simulations with higher harmonics versus without, produces human-readable intermediate
output (i.e., the log-likelihood versus temperature), and has analytically-tractable limits;  see Appendix
\ref{ap:Thermo}.  
Unfortunately, in our experience, thermodynamic integration also produces significantly different  results for the
absolute value of evidence $Z$ and hence $V/V_{\rm prior}$.  
For the purposes of this work, however, the two methods qualitatively agree on evidence and $V/V_{\rm prior}$
\emph{differences} and qualitatively support the conclusions drawn above.   Our interpretation of these results is  that the change in $V/V_{\rm prior}$ is
smaller than the systematic error in the evidence ($\Delta \ln (V/V_{\rm prior})\lesssim 5$); a significant fraction of
that change is responsible for improving our understanding of one parameter; and the remainder is distributed across all
other parameters, improving their measurement accuracy by tens of percent at best. 

\subsection{Posteriors separate into intrinsic and extrinsic variables, after maximizing in time and phase }
\abbrvEFM{} claimed that the posterior
for nonprecessing binaries largely \emph{separates} into purely intrinsic and
purely extrinsic parameters, even in the presence of higher harmonics, when the posterior is marginalized over
polarization and event time.  
In Table \ref{tab:Separate}, we use mutual information  [Eq. (\ref{eq:RelativeInfo})] to quantify correlations between
intrinsic parameters ($\mc,\eta,a$) and extrinsic  parameters ($\ln L, \cos\iota,\phi_{orb},RA,DEC$ and
$t,\psi$).\footnote{From experience and following prior work, we change variables, eliminating distance in favor of  the
  signal amplitude $\rho$.
}
To provide a sense of scale, we expect numerical and sampling error  could introduce random values of $I \simeq
10^{-2} (N_{\rm eff}/10^4)^{-1} $ [Eq. (\ref{eq:Info:ErrorScale})].  By contrast, a strongly-coupled Fisher and covariance matrix
will have $I$ greater than or of order unity [Eq. (\ref{eq:RelativeInfo})].

When all intrinsic parameters are included in the covariance matrix, strong correlations exist between 
intrinsic and extrinsic parameters, as show by mutual information of order unity 
in the $I(A,B+C)$ and $I(A',B+C)$ columns.     
These correlations reflect the well-understood strong correlations between time,  orbital phase, polarization, and the
best-fitting intrinsic parameters.    In fact, as claimed in \abbrvEFM{}, most correlations between intrinsic and
extrinsic parameters are intimately tied to measuring time or phase.  Marginalizing over these variables vastly reduces
the correlations between intrinsic and the remaining extrinsic variables, both with and without spin,
as can be seen by examining the $I(A,B+\psi)$ (marginalize over time) and $I(A,B)$ columns 
(marginalize over time and phase).  

\begin{table*}
\begin{tabular}{lll|cccccc}
Source & Harmonics & Seed & $I(A,B+C)$ & $I(A,B+\psi)$ & $I(A,B)$ & $I(A',B+C)$ & $I(A',B+\psi)$ & $I(A',B)$\\ \hline
 {Zero spin} & {no} & - & 0.87 & 0.02 & 0.02 & - & - & - \\
 {Zero spin} & {with} & - & 0.76 & 0.02 & 0.02 & - & - & - \\
 {Zero spin} & {no} & {-*} & 0.86 & 0.01 & 0.01 & - & - & - \\
 {Zero spin} & {with} & {-*} & 0.78 & 0.06 & 0.02 & - & - & - \\
 {Zero spin} & {no} & 1234 & 0.78 & 0.01 & 0.01 & - & - & - \\
 {Zero spin} & {with} & 1234 & 0.68 & 0.04 & 0.04 & - & - & - \\
 {Zero spin} & {no} & {1234*} & 0.92 & 0.04 & 0.04 & - & - & - \\
 {Zero spin} & {with} & {1234*} & 1.61 & 0.68 & 0.05 & - & - & - \\ 
 {Zero spin} & {no} & 56789 & 0.83 & 0.01 & 0.01 & - & - & - \\
 {Zero spin} & {with} & 56789 & 0.69 & 0.04 & 0.04 & - & - & - \\
 {Zero spin} & {no} & {56789*} & 0.97 & 0.09 & 0.09 & - & - & - \\
 {Zero spin} & {with} & {56789*} & 1.06 & 0.25 & 0.16 & - & - & - \\
\hline
 {Aligned spin} & {no} & - & 0.84 & 0.08 & 0.06 & 0.94 & 0.06 & 0.06 \\
 {Aligned spin} & {with} & - & 1.08 & 0.52 & 0.04 & 0.75 & 0.09 & 0.05 \\
 {Aligned spin} & {no} & {-*} & 0.86 & 0.05 & 0.05 & 0.99 & 0.05 & 0.05 \\
 {Aligned spin} & {with} & {-*} & 1.13 & 0.53 & 0.04 & 0.79 & 0.09 & 0.05 \\
 {Aligned spin} & {no} & 1234 & 0.91 & 0.16 & 0.14 & 0.99 & 0.15 & 0.15 \\
 {Aligned spin} & {with} & 1234 & 1.58 & 1.11 & 0.09 & 0.77 & 0.21 & 0.06 \\
 {Aligned spin} & {no} & {1234*} & 0.75 & 0.06 & 0.06 & 0.84 & 0.06 & 0.06 \\
 {Aligned spin} & {with} & {1234*} & 1.56 & 1.06 & 0.10 & 0.77 & 0.20 & 0.06 \\
 {Aligned spin} & {no} & 56789 & 0.68 & 0.09 & 0.06 & 0.72 & 0.06 & 0.06 \\
 {Aligned spin} & {with} & 56789 & 2.24 & 1.50 & 0.23 & 1.31 & 0.56 & 0.15 \\
 {Aligned spin} & {no} & {56789*} & 0.66 & 0.03 & 0.03 & 0.78 & 0.03 & 0.03 \\
 {Aligned spin} & {with} & {56789*} & 1.13 & 0.38 & 0.18 & 1.08 & 0.17 & 0.17 \\
\end{tabular}
\caption{\label{tab:Separate}\textbf{Mutual information between intrinsic and extrinsic variables tied to time and phase }: The mutual information $I$ between intrinsic variables
  (represented as either $A\equiv (\tau_0,\tau_3)$ or  $A'=A+\chi$) and extrinsic variables (represented as either
  $B=(d_L,\cos\iota,\phi_{orb},RA,DEC)$ or 
 some of  $C=(t,\psi)$).  After marginalizing over time and/or phase, covariances  have extremely weak correlations between
 extrinsic and intrinsic parameters. 
}
\end{table*}

\section{Comparing parameter estimation to the effective Fisher matrix} \label{sec:results:Compare}

\subsection{Effective Fisher matrix predictions}
\label{sec:sub:RevisedResult}
\abbrvEFM{} calculated an effective Fisher matrix using a specific post-Newtonian approximation that neglected the
quadrupole-monopole~\cite{1998PhRvD..57.5287P}, self-spin~\cite{Mikoczi:2005dn} and 2.5PN
  spin-orbit terms\footnote{But again, we have not included 3PN and 3.5PN SO terms, which were implemented in {\tt lalsimulation} after this work as well underway.}~\cite{Faye:2006gx,Blanchet:2006gy,gw-astro-mergers-approximations-SpinningPNHigherHarmonics}.  These terms have since been added to the
  \texttt{lalsimulation} code, and so they are included in this work.   
Table \ref{tab:Fisher:NoAndAlignedSpin} provides a revised effective Fisher matrix, including their effect.      
Table \ref{tab:AlignedSpin:SensitiveToPN} compares our revised effective Fisher matrix (with the new spin terms) 
 in the aligned-spin case to the effective Fisher matrix computed with the older waveform model.
The waveform model is unchanged for non-spinning binaries.

To provide a benchmark for comparison, we have calculated the KL divergence between the effective Fisher matrices
derived with and without higher harmonics:
\begin{eqnarray}
D_{KL}(\mc,\eta|\text{zero spin}) &=&  0.019 \\
D_{KL}(\mc,\eta|\text{aligned spin}) &=& 0.38 \\
D_{KL}(\mc,\eta, \chi|\text{aligned spin}) &=& 0.46 
\end{eqnarray} 
In other words, the effective Fisher matrix suggests higher harmonics will marginally influence the posterior correlations
between intrinsic parameters, with shape changes comparable  
to the typical fluctuations between noise realizations seen in our study [Table \ref{tab:RunComparisons:ToEachOtherAndTheory}].  
In the non-spinning case,  both noise fluctuations and the influence of higher harmonics are small.  
When including spin as a parameter,  both noise fluctuations and higher harmonics have a larger
impact on the posterior shape.

\begin{table*}[!]
\begin{tabular}{c | ccc|cc|ccc|ccc  }
   Source                                                                          & \multicolumn{5}{|c}{Zero spin}                                                                    & \multicolumn{6}{|c}{Aligned spin}   \\
     \hline
 Harmonics                                                               &\multicolumn{3}{|c|}{no}  &\multicolumn{2}{|c|}{with}   &\multicolumn{3}{|c|}{no}  &\multicolumn{3}{|c}{with}  \\
     \hline
Parameter                                                                   &                         &$M_{\rm c}$ &  $\eta$  &   $M_{\rm c}$   &  $\eta$      &$M_{\rm c}$ &  $\eta$ & $\chi$          &$M_{\rm c}$ &  $\eta$ & $\chi$  \\ 
 \hline
\multirow{3}{*}{$(\hat{\Gamma}_{ij})_{\rm eff}$}  &  $M_{\rm c}$ &5688 &-8900  &6017 & -9611&  6044     & -246.5&-1414  &7073&-603.8&-1718  \\
                                                                                              & $\eta$   & -         &15197    &-        &16928   &       -    &379.9 & 146.8    &           -     &646.2&275.0\\
			                                                                  & $\chi$&   -        & -            & -          & -           &          -    & -       &  354.3     & -          &      -          &448.1      \\
 \hline
\multirow{3}{*}{$c_{ij}$}                                           &$M_{\rm c}$&1.00  &0.957    &1.00& 0.952&          1.00  &-0.950&0.997&               1.00& -0.936&0.995 \\
                                                                                             & $\eta$&-          &1.00     & -      & 1.00&         -         &1.00&-0.957&                   -     & 1.00& -0.949 \\
			                                                                   & $\chi$&-         &-             &      -     & -        &               -       &-       &1.00&                        -   & -        &1.00  \\
\hline
$ \sigma_i \times10^3$			                                       &       & 2.29  &1.40 &             2.11 &1.26&               7.98  &8.94 & 35.5      & 6.44 &6.49&28.5\\	
\hline
$\gamma_i$                                                                           &  &20533&352.34& 22524& 420.43 &6389  &387.9&1.792&    7553&       610.9&  2.798 \\	                                                  
    \end{tabular}
 \caption{\label{tab:Fisher:NoAndAlignedSpin}{\bf Effective fitting parameters}: 
Following the tables in \abbrvEFM{}, this table provides $\hat{\Gamma}_{\rm eff}$, a locally quadratic fit to a specific, idealized ambiguity function (the
``effective Fisher matrix''); the correlation coefficients derived from $\hat{\Gamma}_{\rm eff}$; the eigenvalues of
$\hat{\Gamma}$; and the
one-dimensional covariances $\sigma_a =\sqrt{\Sigma_{aa}}$ for $\Sigma=(20)^2\hat{\Gamma}_{\rm eff}$.
Including many more significant figures than shown above, the data used in our own calculations is available on
request.   Due to the orders-of-magnitude difference between eigenvalues ($\gamma_i$) shown, many significant figures are required to
reproduce our calculations in full.  
}
 \end{table*}

\subsection{Comparing predicted and calculated intrinsic parameter distributions}

As described above, our simulations produce (one-, two-, and three-dimesional) posterior distributions of 
intrinsic parameters that are similar to one another.  
The heavy black solid and dotted curves in  Figures \ref{fig:CompareAllIntrinsic:ZeroSpin} 
and~\ref{fig:CompareAllIntrinsic:AlignedSpin} compare the predictions of the
effective Fisher matrix, scaled to $\rho=20$, to the results of our simulations.  
Table \ref{tab:RunComparisons:ToEachOtherAndTheory} provides a quantitative comparison 
between each two- and three-dimensional
covariance matrix $\Sigma_{ab}$ and the corresponding two- and three-dimensional effective Fisher matrix $K_*$, via
$D_{KL}$ [Eq. (\ref{eq:Dkl})].  
For a sense of scale, any two two-dimensional Monte Carlo posteriors that sample the same
distribution should  differ by less than 
$D_{KL}\simeq 2\times 10^{-3} $ [Eq. (\ref{eq:DklDistinguishCriteria})], while any two independent noise realizations should
have two-dimensional posteriors which differ by less than $D_{KL}\simeq 0.015 $
[Eq. (\ref{eq:DklCriteria:FluctuateAmplitude})].

\ForInternalReference{
\begin{table*}
\begin{tabular}{llr|llllll}
Source & Harmonics & Seed &  $D_{KL}(\mc,\eta)$ & $D_{KL}(\tau_0,\tau_3)$ & $D_{KL}(\mc,\eta^{-2})$
&$D_{KL}(\mc,\eta,a)$ & $D_{KL}(\tau_0,\tau_3,a_ 1)$ & $D_{KL}(\mc,\eta^{-2},a)$ \\ \hline
 {Zero spin} & {no} & - & 0.039 & 0.041 & 0.038 & - & - & -   \\
 {Zero spin} & {no} & {1234} & 0.005 & 0.004 & 0.005 & - & - & -   \\
 {Zero spin} & {no} & {56789} & 0.023 & 0.018 & 0.016 & - & - & -   \\
 {Zero spin} & {with} & - & 0.037 & 0.041 & 0.042 & - & - & -   \\
 {Zero spin} & {with} & {1234} & 2.764 & 2.374 & 2.152 & - & - & -  \\
 {Zero spin} & {with} & {56789} & 0.026 & 0.021 & 0.019 & - & - & - \\
 {Aligned spin} & {no} & - & 0.072 & 0.030 & 0.023 & 2.243 & 2.174 & 2.130  \\
 {Aligned spin} & {no} & {1234} & 0.141 & 0.115 & 0.107 & 2.081 & 2.040 & 2.003  \\
 {Aligned spin} & {no} & {56789} & 0.511 & 0.214 & 0.143 & 2.202 & 1.946 & 1.908  \\
 {Aligned spin} & {with} & - & 0.049 & 0.025 & 0.018 & 2.279 & 2.236 & 2.202 \\
 {Aligned spin} & {with} & {1234} & 0.154 & 0.208 & 0.266 & 2.180 & 2.219 & 2.255  \\
 {Aligned spin} & {with} & {56789} & 0.104 & 0.178 & 0.337 & 1.911 & 2.049 & 2.255  \\
\end{tabular}

\caption{\label{tab:CompareFisherToSimulations}\textbf{Simulations versus the effective Fisher matrix}:  For each simulation, a comparison of that
  simulation's shape with the corresponding effective Fisher matrix provided in Table
  \ref{tab:AlignedSpin:SensitiveToPN}.
As suggested in Section \ref{sec:Methods:Metrics}, the large $D_{KL}$ values seen in this table, particularly in three
dimensions, occur because the correlations identified by the 
effective Fisher matrix and our simulations are misaligned relative to one another.  
}
\end{table*}
}

In the absence of spin, as demonstrated qualitatively via the heavy black curves in Figure
\ref{fig:CompareAllIntrinsic:ZeroSpin} and quantitatively by the  $D_{KL}$ in Table
\ref{tab:RunComparisons:ToEachOtherAndTheory}, 
our effective Fisher matrix predictions agree remarkably well with the simulated results, despite the substantial
simplifications they employ.
By contrast, when spin is included as a parameter, the predictions of the effective Fisher matrix seem to fare more
poorly, particularly in capturing  multiparameter correlations; see  the heavy black curves in Figure
\ref{fig:CompareAllIntrinsic:AlignedSpin}.  
To some extent, as anticipated in Section \ref{sec:Methods:Metrics} our poor performance  reflects poorly chosen
coordinates.  In our coordinates, the confidence intervals are long and non-ellipsoidal, being distorted by significant
changes to the metric across substantial ranges of parameter space,
whereas the effective Fisher matrix contours are always ellipsoidal by construction. 
The large $D_{KL}$ values provided in Table~\ref{tab:RunComparisons:ToEachOtherAndTheory}
for aligned-spin binaries indicate that the effective Fisher matrix and simulations have covariances with
different principal axes. 
Due to the extremely large ratio of
eigenvalues in three dimensions ($\simeq 10^5$), even an extremely small relative misalignment $\theta$ between the predicted
and simulated correlation leads to a large $D_{KL} \simeq 10^5\theta^2$.  To a lesser extent, both the one and
two-dimensional correlations are negatively influenced by the hard cutoff in $\chi$.  Given the sensitivity of $D_{KL}$ to
misalignment, one could argue that the \abbrvEFM{} procedure is performing rather well even for the
aligned-spin case.

\subsection{Geometric parameters}
\label{sec:sub:Geometric}
Including source orientation and distance, the posterior is well-known to have several correlations which cannot be
captured with a locally Gaussian approximation in conventional coordinates.  Nonetheless, the posterior over  geometric parameters can be
well-approximated using a few simple network-independent expressions:
\begin{subequations}
\label{eq:PosteriorDistanceIota:NoHarmonics}
\begin{align}
A(\theta) &\equiv \sqrt{|\Y{-2}_{22}(\theta,0)|^2 +|\Y{-2}_{2,-2}(\theta,0)|^2} \\
\frac{dp(r,\iota)}{dr  d\cos \iota} & \propto  r^2  e^{ - \frac{\rho^2}{2 }\left[1 + \frac{A(\iota)^2}{A(\pi/4)^2 (r/d_0)^2} - 2
  P(\iota,\pi/4)\frac{A(\iota)}{A(\pi/4) (r/d_0)}
  \right]
} 
\end{align}
\end{subequations}
where $P$, an inner product evaluated over all orientations,  is provided by Eq. (28) in \abbrvEFM{}.   
The seemingly-complicated expression appearing in the exponential is nothing more than 
\begin{align}
\frac{1}{2}\left[\qmstateproduct{h}{h} + \qmstateproduct{h'}{h'}  - 2 \qmstateproduct{h}{h'}\right]
\end{align}
evaluated by re-expressing $h'=\rho \hat{h}' A(\iota)/A(\pi/4)$ where  $h'$ is normalized ($\qmstateproduct{\hat{h}'}{\hat{h}'}=1$); performing a similar
replacement for $h$; and  replacing the overlap  $\qmstateproduct{\hat{h}}{\hat{h}'}$ between normalized complex-valued
states by $P$. 
 As illustrated using one contour (the heavy black curve) in  Figure
\ref{fig:CompareGeometry}, this distribution accurately approximates the source distance and orientation posterior in
the absence of higher harmonics.  
This agreement occurs despite considerable differences in the two models:  our simulations use a realistic three-detector configuration and
include all correlations with sky position,  while
\abbrvEFM{}  assume an idealized network and omit  any correlation with sky location.  

Similar network-independent approximations can reproduce features seen in other correlations.  As a concrete example,
using a well-chosen reference frequency we expect and Figure \ref{fig:CompareGeometry} confirms that the posterior in $\psi,\phiref$ is concentrated near hyperplanes
of either constant $\psi_{\pm} \equiv  \psi\pm \phiref$, depending on which harmonic dominates.    Specifically, in the
absence of higher harmonics, the marginalized posterior $p(\psi_+,\psi_-)d \psi_+d\psi_-$, defined by marginalizing all
other degrees of freedom
\begin{align}
\label{def:DistributonVersusPsiPlusMinus}
{}p(\psi_+,\psi_-) d\psi_+d\psi_- \equiv \int_\lambda p(\psi,\phiref,\lambda)d\lambda
\end{align}
can be approximated by a one-dimensional Gaussian distribution depending on either $\psi_+$ or $\psi_-$,   
in the neigborhood of
one such hyperplane.   For the solid curves shown in the top left panel of Figure \ref{fig:CompareGeometry} and the
covariance provided in Table \ref{tab:ParameterErrors}, a good local approximation is given by 
\begin{subequations}
\label{eq:PsiPlus:Marginalized}
\begin{align}
p(\psi_+)       &\propto e^{-\rho^2 \bar{\Gamma}_{++}\delta\psi_+^2/2} \quad \text{for this case} \\
\bar{\Gamma}_{++} &\equiv -\frac{2}{\rho^2} \partial^2_{\psi_+} \ln \int \int_\lambda p(\psi,\phiref,\lambda)d\lambda
\end{align}
\end{subequations}
\ForInternalReference{
Motivated by the general form of the likelihood and the observed periodicities in $\psi_{\pm}$, a better global
approximation  has the form
\begin{eqnarray}
\frac{dp(\phi_{orb},\psi)}{d\phi_{orb}d\psi} &\propto 
e^{
   -\rho^2
   \left[1 -
      \sum_{k=1}^{\infty} a_k \cos 2k (\phi_{orb}+\psi  -(\phi_{orb,in}+\psi_{inj}))
  \right]
}
\end{eqnarray}
where $a_k$ is some  sequence with $\sum_k a_k=1$ and $\sum_k a_k(2 k)^2 = \bar{\Gamma}_{\psi\psi}$.  
}
In the large-signal-amplitude limit, we can calculate $\bar{\Gamma}_{++}$ by using the known functional form of the
waveform emitted along the orbital angular momentum direction $h(t,\hat{z},\lambda)$ on all intrinsic parameters:
\begin{eqnarray}
\bar{\Gamma}_{++} 
&= 4 - \sum_{ab \ne \psi} \Gamma_{\psi a}\Gamma_{\psi b}[\Gamma_{ab}]^{-1}  \nonumber \\
 &=4- 4 \frac{\qmstateproduct{\partial_ah}{h}\qmstateproduct{\partial h}{\partial h}^{-1}_{ab}\qmstateproduct{h}{\partial_b h}}{\rho^2}
\end{eqnarray}
where $[\Gamma_{ab}]^{-1} - = \qmstateproduct{\partial h}{\partial h}^{-1}_{ab}$ is the inverse of the submatrix 
not involved with $\psi_{\pm}$. 
For the post-Newtonian approximation adopted in the text, we find $\bar{\Gamma}_{++} \simeq 0.27-0.28$, which
adequately reproduce the observed widths  in Table \ref{tab:ParameterErrors} and  Figure \ref{fig:CompareGeometry}: 
$\sigma_+\simeq 1/\sqrt{\bar{\Gamma}\rho^2}\simeq 0.095$.
This approximation holds independent of the network's relative sensitivity to the two polarizations.\footnote{To a first
  approximation, the signal can be approximated as circularly polarized; the relative sensitivity to one or another
  linear polarization is irrelevant. }
Finally and similarly, the posterior  probability distribution in $t$ and $\psi_+$ can be approximated by  a  Gaussian.
Our ability to correctly reproduce the source orientation distribution, as well as to correctly model the intrinsic
parameter distribution of nonspinning binaries, suggests that the \abbrvEFM{} approach correctly approximates the
posterior distribution.  %
In particular, for the examples shown here, multidetector physics is not needed to model the posterior distribution to a
zeroth approximation.

\section{Generalizing the results} \label{sec:implications}

\begin{figure}
\includegraphics[width=\columnwidth]{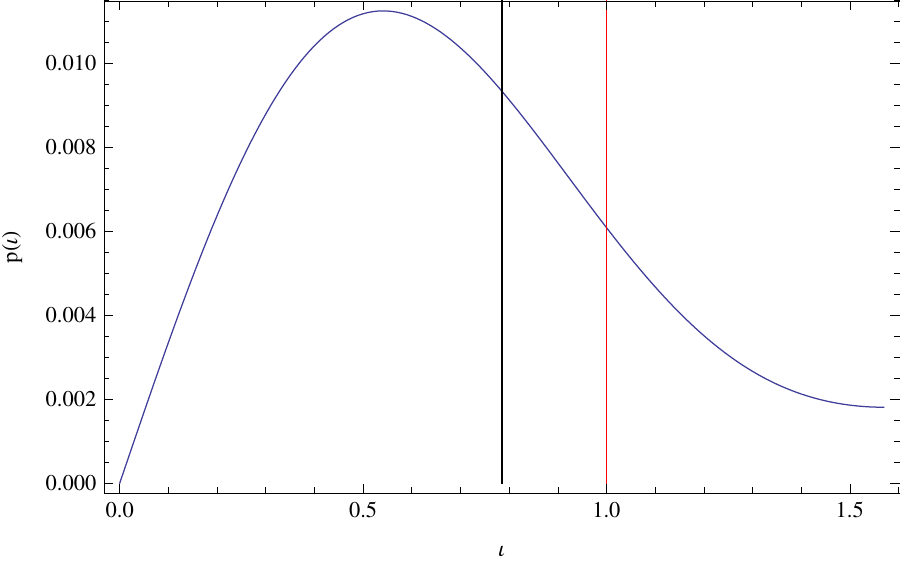}
\caption{\label{fig:InclinationDistribution}\textbf{Reference inclination is typical}: The vertical black line shows the source
  inclination ($\iota=\pi/4$) adopted in this
  work.  For comparison, the curve shows the  relative probability $p(\iota)\propto
  \int d\phi |h(\iota,\phi)|^3\sin \iota $ of different inclinations $\iota$
  for a binary detected by an idealized network with isotropic sensitivity to both polarizations.  
Finally, the red line at $\iota \simeq 1$ indicates the approximate angle above which both the $(2,2)$ and $(2,-2)$
modes influence the signal amplitude by more than a fractional change in   $1/\rho^2$ for $\rho \simeq 20$; for inclinations smaller than the
red line, the signal can be described as nearly circularly polarized for network amplitudes $\rho \lesssim 20$.   Source
inclinations closer to $\iota \simeq 0,\pi$ will be even better described by a single circular polarization than the
systems explored in this work.   By contrast, the $\simeq
18\%$ of all orientations between $\iota \simeq
1$ and $\pi-1$ contain significant contributions from both left- and right-handed emission.  
}
\end{figure}

\begin{figure}
\includegraphics[width=\columnwidth]{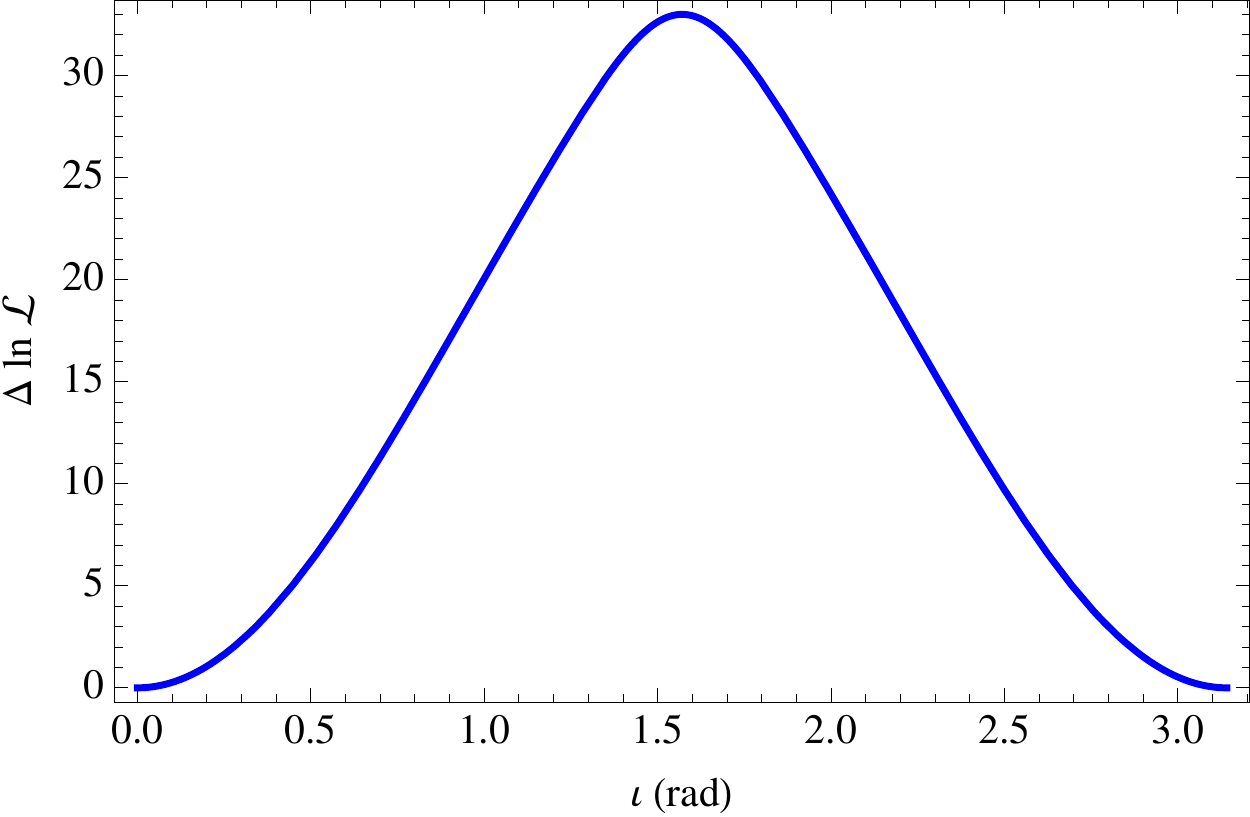}
\caption{\label{fig:IotaScaling} \textbf{Influence on higher harmonics with inclination}: Here we plot 
the expected change in max log-likelihood from the inclusion of higher harmonics
$\Delta \ln {\cal L} = \rho_{0PN}^2 \Delta \ln \rho^2 / 2$ vs inclination ($\iota$), see Eq.~(\ref{eq:fracSNRsq}). 
We assume all other parameters are those of our fiducial non-spinning binary.
From our rule-of-thumb criteria $\Delta \ln {\cal L} \gtrsim 1$, we expect higher harmonics to have
a significant effect on our recovered posterior except when the binary is very nearly face-on.
}
\end{figure}

Though we present a detailed analysis of only two specific systems, certain aspects of our results can be generalized
to make inferences about the parameter estimation prospects for a broader range of binary systems.
In this section, we argue that the binary location and orientation we consider are, in a sense, ``typical'' of 
detectable binaries.

As  gravitational wave emission is strongest along the orbital
angular momentum axis, most nonprecessing binaries will be detected nearly face on and hence be dominated by
circularly-polarized emission.  The inclination distribution for detected, nonprecessing binaries, 
which follows from Eq.~\ref{eq:PosteriorDistanceIota:NoHarmonics}, is plotted in Fig.~\ref{fig:InclinationDistribution}. 
This figure shows that the distribution is strongly biased towards small inclination.  
By contrast, higher harmonics have their greatest impact near the
orbital plane, since the odd harmonics are proportional to $\sin \iota$.  

We expect that higher harmonics can have a significant influence on the posterior
if they changes the max log-likelihood ratio by $\gtrsim 1$. From Eq.~(\ref{eq:def:rhoRecovered}), this is equivalent 
to increasing the square of SNR by $\gtrsim 2$. We have the following formulae for the SNR-squared from 
restricted waveforms, from first-order amplitude-corrected waveforms, 
and the fractional increase from amplitude corrections:
\begin{eqnarray}
\rho_{0PN}^2 &=& ||h_{22}||^2 \left( \left|\Y{-2}_{22}\right|^2 + \left|\Y{-2}_{2-2}\right|^2 \right)\ , \\
\rho_{0.5PN}^2 &=& \rho_{0PN}^2 \left( 1 + \Delta \ln \rho^2 \right) \ , \\
\Delta \ln \rho^2 &=& \frac{1}{\rho_{0PN}^2} \left[ ||h_{21}||^2 \left( \left|\Y{-2}_{21}\right|^2 + \left|\Y{-2}_{2-1}\right|^2 \right) \right. \nonumber\\
	&+& \left. ||h_{33}||^2 \left( \left|\Y{-2}_{33}\right|^2 + \left|\Y{-2}_{3-3}\right|^2 \right) \right. \nonumber\\
	&+& \left. ||h_{31}||^2 \left( \left|\Y{-2}_{31}\right|^2 + \left|\Y{-2}_{3-1}\right|^2 \right) \right] \ .
\end{eqnarray}
Note that $|| h_{\ell m}||^2 \equiv \langle h_{\ell m}| h_{\ell m} \rangle$, we have used the fact that 
$|| h_{\ell -m}||^2 = || h_{\ell m}||^2$, and we have assumed that different $h_{\ell m}$ modes 
are orthogonal.\footnote{This assumption works very well for the masses considered here, 
	but may break down at very high mass, when the modes may be only a few cycles long.}
The fractional increase in SNR can be written as a combination of prefactors, a ratio of frequency moments of the PSD, 
and ratios of $\Y{-2}_{\ell m}$'s to make the dependence on inclination and mass parameters more explicit. 
In particular,
\begin{equation} \label{eq:fracSNRsq}
\Delta \ln \rho^2 = {\cal I} \left[ \frac{1}{9} {\cal Y}(2,1) + \frac{135}{224} {\cal Y}(3,3) + \frac{1}{2016} {\cal Y}(3,1) \right]\ ,
\end{equation}
with
\begin{eqnarray}
{\cal I} &=& (\pi M)^{2/3} \, \delta^2 \frac{\int \frac{f^{-5/3}}{S_n(f)}\, df}{\int \frac{f^{-7/3}}{S_n(f)}\, df}
	\simeq 0.05\ ,  \label{eq:Ifactor}\\
{\cal Y}(\ell,m) &=& \frac{\left|\Y{-2}_{\ell m}(\iota,0)\right|^2 + \left|\Y{-2}_{\ell -m}(\iota,0)\right|^2}{\left|\Y{-2}_{22}(\iota,0)\right|^2 + \left|\Y{-2}_{2-2}(\iota,0)\right|^2}\ . \label{eq:Yratio}
\end{eqnarray}
where $\delta = (m_1-m_2)/M = \sqrt{1 - 4 \eta}$ 
and the approximate numerical value for ${\cal I}$ was computed assuming our initial LIGO and Virgo 
three-detector network. The ${\cal Y}$ functions appearing in Eq.~(\ref{eq:fracSNRsq}) all peak at $\iota = \pi/2$,
are symmetric about that point, and approach zero as $\iota \rightarrow 0, \pi$.

We can write our condition for when higher harmonics to significantly affect the posterior as 
$\Delta \ln {\cal L} = \rho_{0PN}^2 \Delta \ln \rho^2 / 2 \gtrsim 1$. 
Eqs.~(\ref{eq:fracSNRsq})-(\ref{eq:Yratio}) quantify how the influence of higher harmonics 
scales with inclination, mass ratio and total mass, showing that they become less important 
for face-on, nearly equal mass and lower total mass, as is well known.
Higher harmonics can influence our posterior for almost any inclination, 
so long as $\iota$ is in the approximate range $[0.1, \pi - 0.1]$, as illustrated in Fig.~\ref{fig:IotaScaling}.

\begin{figure}
\includegraphics[width=\columnwidth]{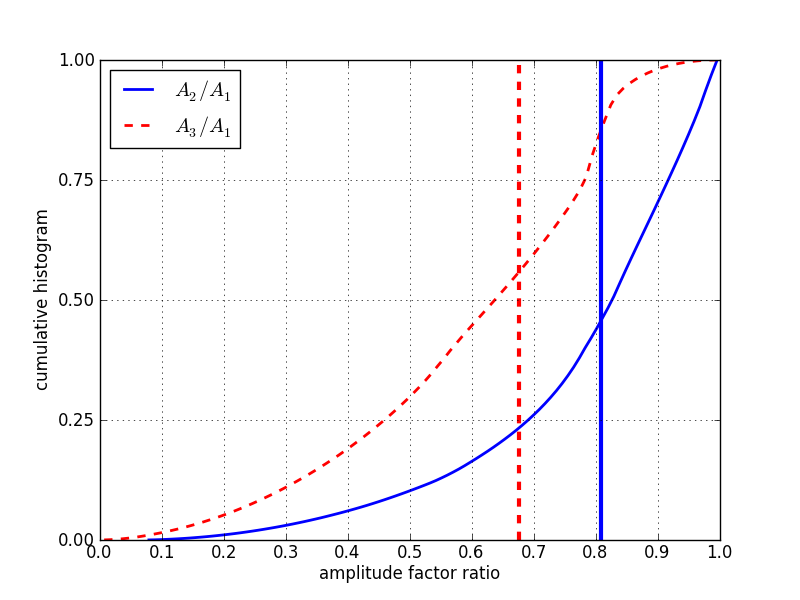}
\caption{\label{fig:ampratio} \textbf{Amplitude factor ratios versus sky location}: We compute the amplitude factor 
(see Eq.~(\ref{eq:ampfactor})) of each LIGO and Virgo detector for a uniformly distributed set of sky locations and
plot a cumulative histogram of the ratios $A_2/A_1$ (solid blue curve) and $A_3/A_1$ (dashed red curve),
which tell us the signal amplitude of the second- and third-best oriented detectors, 
relative to the best oriented detector. The vertical lines denote these ratios for our injected signals.
These results show that comparable signal amplitude in two and three detectors will be common for the LIGO-Virgo
network, and that this network's sensitivity to our injection is typical.
}
\end{figure}

For the sky location of our binary, we intentionally chose a location so that the signal amplitude at each detector site
was of comparable strength.  To demonstrate our source sky location  was representative, 
we distributed  each of $10^6$ points distributed uniformly across the sky and evaluated the optimality of orientation
for each detector in the network at that sky location.  
Specifically, for each sky location, we compute the ``amplitude factor'' at each of the LIGO and Virgo detector sites
\begin{equation} \label{eq:ampfactor}
0 \leq A \equiv \sqrt{ F_+^2 \frac{1}{4} \left( 1 + \cos^2 \iota \right)^2 + F_\times^2 \cos^2 \iota} \leq 1\ .
\end{equation}
Used in the  ``effective distance'', this expression is  the ratio of the observed signal amplitude 
in a detector to the amplitude of the same signal if it were optimally oriented.
The values $A_1$, $A_2$ and $A_3$ denote the amplitude factors of the
first-, second- and third-best oriented detectors. So, for example, $A_2 / A_1 = 0.5$ ($A_3 / A_1 = 0.5$) 
means that the signal was half as loud in the second- (third-) best detector 
as it was in the most favorably oriented one.
In Fig.~(\ref{fig:ampratio}), we plot the cumulative histograms of $A_2/A_1$ and $A_3/A_1$. The vertical lines 
represent the values of these ratios for our injected signals (roughly $0.8$ and $0.7$, respectively).
First, note that the amplitude factor ratios of our injected signals are quite close to the mean of the distribution.
In addition, the second- (third-) best oriented detector will have at least half the signal amplitude of the
best oriented detector for $90\%$ ($70\%$) of the sky.   Comparable angular response $A_k$ in all detectors is the norm; 
a signal in the blind spot of one or more detectors is the exception. 
Note that we have carefully framed this point in terms of an amplitude factor which does not refer to the PSD
of any detector. If, for example, one detector is far less sensitive than the others then one would get the best
parameter estimation performance when both of the sensitive detectors have large amplitude factors, with little
regard for the amplitude factor in the insensitive detector. At any rate, our main point is that the location and orientation
of our fiducial signal is not in the blind spot of any detector in the LIGO-Virgo network, and that this will be the case
for the majority of signals. This is true irrespective of the relative sensitivity of each detector in the network.

\section{Conclusions} \label{sec:conclusions}

In this work we investigate the prospects of parameter estimation for gravitational waves from BH-NS binaries.
We perform full Markov-chain Monte Carlo parameter estimation studies with simulated Gaussian noise
for the three-detector network of the initial LIGO and Virgo interferometers.
Additionally, we predict the performance of these full parameter estimation runs using a
simple analytic estimate based on an effective Fisher matrix method for an idealized detector network.
We generally find that the effective Fisher matrix predictions agree with the full parameter estimation 
results to a reasonable accuracy.

Our primary conclusion is that amplitude corrections do not significantly improve the measurement 
of the masses and spins of the binary. Instead, their main effect is to improve the measurement of the astrophysically
uninteresting source orientation,  by breaking an approximate degeneracy 
between polarization angle and binary orbital phase at some reference point.
By contrast, black hole spin is highly degenerate with all astrophysically interesting parameters (e.g., masses) and
must be included and carefully calibrated, given the significant systematic uncertainties in spinning waveforms
\cite{2013ApJ...766L..14H}. 

We use several analytic tools to study the posterior distributions of our MCMC results and compare them to 
the effective Fisher distribution. 
By computing a prior volume ratio, we argue that higher harmonics principally break the $\psi-\phi_{\rm ref}$
degeneracy, providing marginal improvements in the measurements of other parameters.  
By computing $D_{KL}$, a form of the KL-divergence, we conclude that our MCMC results for different noise 
realizations are self-consistent, and consistent with the effective Fisher matrix predictions.
Using the mutual information, we show that the intrinsic parameters decouple from the extrinsic parameters 
after marginalizing over time and polarization angle.

While the results presented here are limited to a single mass ratio, binary location and orientation, we argue that
they can be generalized to describe parameter estimation performance for a broader class of BH-NS signals.
In particular, we show that the sky location of our signal is ``typical'', and that the LIGO-Virgo detector network would 
have a similar response (in which all three detectors have a comparable signal strength) 
across a very large fraction of the sky. In addition, we show that the binary inclination to line of sight we considered
is typical of detectable signals, and we show approximately but quantitatively how the importance of higher
harmonics will vary with SNR and inclination.

We propose the use of the effective Fisher matrix, as applied in \abbrvEFM,
as a computationally efficient way to predict parameter estimation
performance, to suggest which physical effects are most important to
include in waveform models for various regions of parameter space, and
to provide guidance about the capabilities of future detectors.
Our study also introduces the prior volume $V/V_{\rm prior}$, a powerful global tool to assess when the data constrains
additional parameters.  The application of $V/V_{\rm prior}$ requires considerable care, being derived from the
sometimes difficult-to-compute evidence $Z$.    Any study of subdominant degrees of freedom
in gravitational wave astronomy can adopt the tools presented here ($V/V_{\rm prior}$ and $D_{KL}$) to assess whether new physics is accessible (e.g., modifications of gravity;  spin-orbit
misalignment) and if so precisely what  information those new parameters provide.

Lastly, we make a somewhat technical point about the computational cost of performing parameter estimation 
with amplitude corrected waveforms. To fully and correctly include the effect of amplitude corrections requires 
a higher sampling rate and a lower starting frequency than restricted waveforms, which can significantly increase the
computational cost of using amplitude corrected waveforms in parameter estimation.
Using the same sample rate and starting frequency for amplitude corrected waveforms
as for restricted waveforms will result in aliasing and missing low-frequency portions of the higher harmonics.
However, we find that such effects have a minimal impact on the recovered posterior distributions,
and so one could reasonably perform parameter estimation with amplitude corrected waveforms
for the same computational cost as restricted waveforms.

\ForInternalReference{
\editremark{no finite size effects} -- and this 
  \editremark{some speculation
  about ninja/IMR sampling rates?}

Jump proposals as key ingredient not described here which ensure efficient, physically-motivated exploration of the
parameter space.
}

\appendix
\section{Comparing two distributions' shapes}
\label{ap:Compare:Details}

In the text, we use an extremely simple diagnostic to distinguish between two distributions: their covariance matrix.  
For any two nearly-Gaussian distributions characterized by $K$, so each has the form 
\begin{eqnarray}
p_{orig}(x|\mu,K)&=& \frac{|K|^{1/2}}{(2\pi)^{d/2}} e^{-(x-\mu)K(x-\mu)/2} \nonumber 
\end{eqnarray}
in the neighborhood of some mean, we have
\begin{widetext}
\begin{eqnarray}
D_{KL}&\equiv &\int p_* \ln p_*/p \\
&=& \frac{1}{2}\int p \left[\ln[ |K_*|/|K|]  - (x-\mu)K(x-\mu) + (x-\mu_*)K_*(x-\mu_*)\right] \nonumber\\
&=& \frac{1}{2}\left[\ln[ |K_*|/|K|] + (\mu-\mu_*)K_*(\mu-\mu_*) + \text{Tr}[ (K_*-K)K_*^{-1}]\right]\ .
\label{eq:Gaussian:DKL}
\end{eqnarray}
\end{widetext}
In this work, we parallel-transport both distributions to a common mean; we therefore neglect the middle term in the
above expression.

While the KL divergence has many desirable statistical properties,
for our purposes, the greatest utility of the KL divergence is the ease with which Eq. (\ref{eq:Gaussian:DKL})
can be evaluated and interpreted, allowing us to employ concrete examples to characterize what factors produce a large
$D_{KL}$.  
For example, if $K_*$ and $K$ are related by a rotation $R=\exp ( -i \theta^k L_k)$ through a small angle $\theta_k$,
where $L_k$ are suitable rotation group generators, then
\begin{align}
D_{KL}(K_*,K) 
&= \frac{1}{2}\text{Tr}[K_*^{-1} R K_* R^{-1} -\textbf{1}]  ]] \nonumber \\
 &\simeq \frac{1}{2}\text{Tr}[  - i \theta^k K^{-1}_*[L_k,K_*] 
 \nonumber \\ &
- \frac{1}{2}\theta_k \theta_q K_*^{-1}[L_k,[L_q,K_*]] +
      \ldots ] \nonumber \\
 &\simeq \frac{1}{4}\text{Tr}[ \theta_k \theta_q K_*^{-1}[L_k,[L_q,K_*]] + \ldots ]
\end{align}
because the rotation group generators $L_k$ are traceless.  
Combined with the concrete example provided in the text, this general expression suggests that in many dimensions $D_{KL}$
is extremely sensitive to small misalignments between $K, K_*$, scaling as $\theta^2 \lambda_{+}/\lambda_{-}$ for
$\lambda_{\pm}$ the largest and smallest eigenvalues of $K_*$, respectively.  

\section{Sensitivity to PN model, sampling rate and starting frequency}%
\label{ap:SampleRate}
It is well-known that the agreement between various PN waveform models is not perfect, and that these differences 
can lead to biases in recovered parameters, for example see~\cite{gw-astro-PN-Comparison-AlessandraSathya2009}.
A detailed study of the biases from waveform uncertainty is beyond the scope of this work. However, we do 
offer up an example of the level to which waveform systematics can influence our results. In particular,
our current estimate for the effective Fisher matrix differs from the original \abbrvEFM{} result, because we adopt
an updated model for how the spin influences the orbit.  
Table \ref{tab:AlignedSpin:SensitiveToPN} provides our revised effective Fisher matrix.  Implicitly, this table
illustrates how sensitively our results depend on post-Newtonian order, particularly spin effects.

\begin{table*}[!]
\begin{tabular}{c | cccc|ccc|ccc|ccc  }
  Waveform                                                                          & \multicolumn{7}{|c}{Previous}                                                                    & \multicolumn{6}{|c}{Current}   \\
     \hline
 Harmonics                                                               &\multicolumn{4}{|c|}{no}  &\multicolumn{3}{|c|}{with}   &\multicolumn{3}{|c|}{no}  &\multicolumn{3}{|c}{with}  \\
     \hline
Parameter                                                                   &                         &$M_{\rm c}$ &  $\eta$  &   $\chi$& $M_{\rm c}$   &  $\eta$    &$\chi$  &$M_{\rm c}$ &  $\eta$ & $\chi$          &$M_{\rm c}$ &  $\eta$ & $\chi$  \\ 
 \hline
\multirow{3}{*}{$(\hat{\Gamma}_{ij})_{\rm eff}$}  &  $M_{\rm c}$ &5935 &-1821& -1431  &6598 & -2335 & -1631&   6044     & -246.5&-1414  &7073&-603.8&-1718  \\
                                                                                              & $\eta$   & -         &1381  &559.3  &-        & 1961 & 731.9   &         -    &379.9 & 146.8    &           -     &646.2&275.0   \\
			                                                                  & $\chi$&   -        & -           &364.2  & -          & -         &426.1  &        -    & -       &  354.3     & -          &      -          &448.1      \\
 \hline
\multirow{3}{*}{$c_{ij}$}                                           &$M_{\rm c}$&1.00  &-0.939 &0.995    &1.00& -0.929  & 0.994&         1.00  &-0.950&0.997&               1.00& -0.936&0.995 \\
                                                                                             & $\eta$&-          &1.00     &-0.962 & -      & 1.00&    -0.957&      -         &1.00&-0.957&                   -     & 1.00& -0.949  \\
			                                                                   & $\chi$&-         &-             &1.00   &      -     & -        &    1.00&         -       &-       &1.00&                        -   & -        &1.00   \\
\hline
$ \sigma_i \times10^3$			                                       &       & 8.27   &6.36 &41.9  &             7.14 &5.09& 35.7&              7.98  &8.94 & 35.5      & 6.44 &6.49&28.5\\	
\hline
$\gamma_i$                                                               &                   &6933& 746.1& 1.343      &7992  & 991.5 &1.856   &    6389  &387.9&1.792&    7553&       610.9&  2.798\\	                                                  
    \end{tabular}
\caption{\label{tab:AlignedSpin:SensitiveToPN}
\textbf{Sensitivity of (effective) Fisher matrix to post-Newtonian approximation, SNR=20}: 
The columns labeled ``Previous'' give the  effective Fisher matrix and expected parameter errors for the waveforms 
used in~\cite{gwastro-mergers-HeeSuk-FisherMatrixWithAmplitudeCorrections}.  
The columns labeled ``Current'' give the same results  for waveform which contain additional spin-dependent 
phasing corrections (as explained in Sec.~\ref{sec:sub:RevisedResult}).    
The effective Fisher matrix inevitably depends on the still-uncertain post-Newtonian
  waveform model used.  %
}
\end{table*}

The calculations provided in the main text were performed with $\fSampleRateUsed\unit{Hz}$ sampling. 
Ideally, the sample rate used should be high enough such that the Nyquist frequency is greater than 
the highest frequency obtained by the highest harmonic in the waveform. 
Table \ref{tab1} provides the highest gravitational frequency from the leading-order quadrupole emission.  
When higher harmonics are included, we should
sample at a substantially higher frequency, in direct proportion to the harmonics used.
In particular, the spin-aligned waveform with higher harmonics contains a physical signal up to 
$(5/2)\,1926\, {\rm Hz} \simeq 4815$ Hz, and so $16384$ Hz 
is the lowest power-of-two sampling which will completely avoid any aliasing. 
Unfortunately for \texttt{lalinference\_mcmc}, the computational cost of likelihood evaluations is proportional
to the number of waveform samples, and thus increases linearly with the sampling rate.  

Furthermore, the waveform length scales with the lower frequency as $f_{\rm low}^{-8/3}$ (see Eq.~(\ref{eq:tau0})).  
Therefore, a waveform for which the $5^{\rm th}$ harmonic is present all the way down to $f_{\rm low}$ will be a factor 
$(5/2)^{-8/3} \simeq 11.5$ longer than an equivalent waveform for which we only need the
$2^{nd}$ harmonic down to $f_{\rm low}$.
Therefore, starting at a sufficiently low frequency such that all higher harmonics are entirely in-band can increase
the cost of both waveform generation and likelihood evaluations by more than an order of magnitude!
As described in Sec.~\ref{subsec:waveforms}, we  generate longer
waveforms such that the highest ($5^{\rm th}$) harmonic is entirely in-band.

\begin{figure*}
\includegraphics[width=\columnwidth]{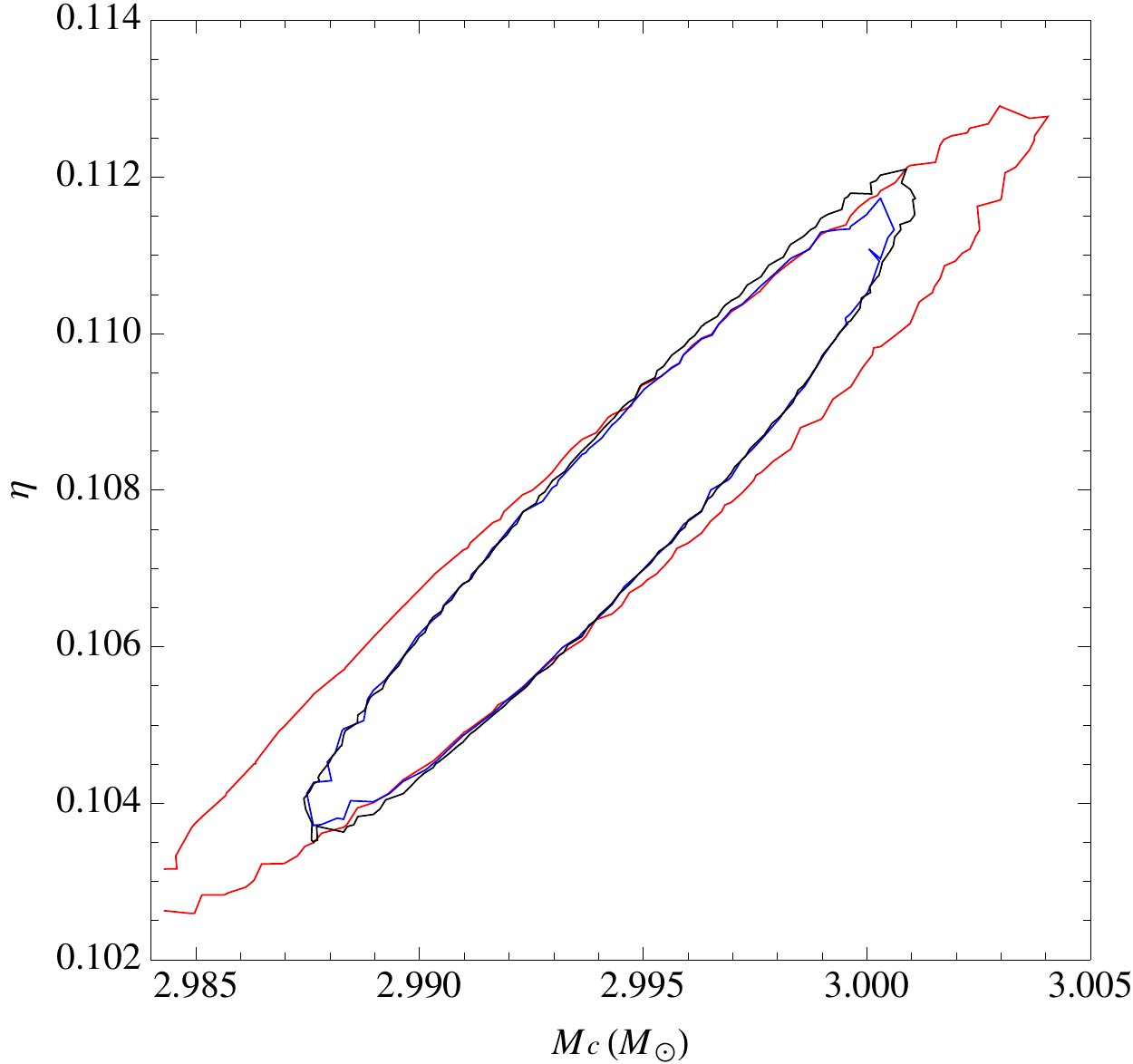}
\includegraphics[width=\columnwidth]{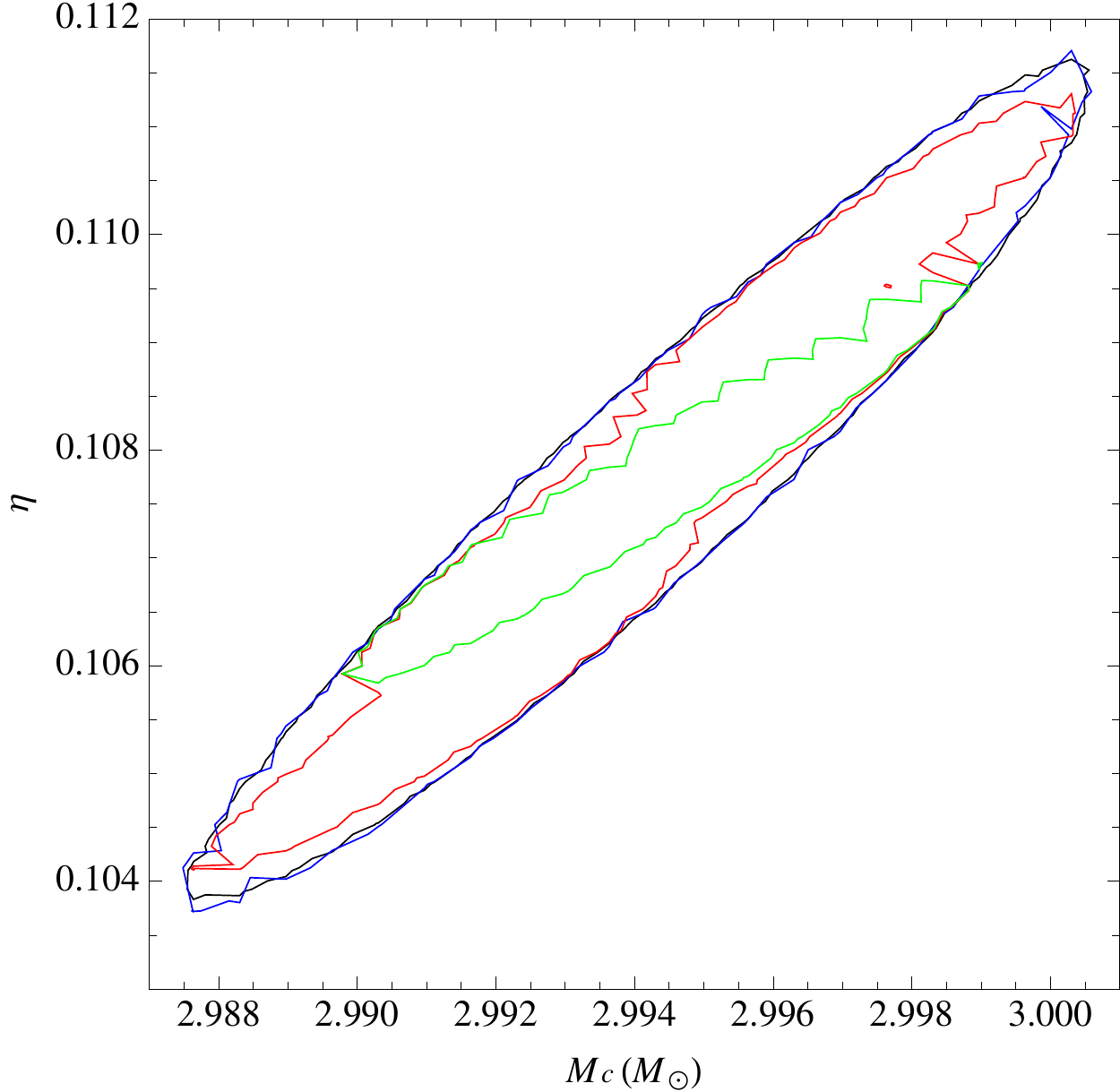}
\caption{\label{fig:SrateDependence}\textbf{Marginalized ambiguity contours (where the normalized overlap is 0.99) of intrinsic parameters with various lower limits and sampling rates  for the zero spin waveforms with higher harmonics.} \emph{Left panel}: The injected signal starts at 12 Hz and the templates at 12 (black),  30 (blue), and 40 Hz (red). No additional information is added below 30 Hz.  \emph{Right panel}: Contours are calculated by using the sampling rate of 8192 (Black), 4096 (blue), 2048 (red), and 1024 Hz (green) with a fixed starting frequency of 30 Hz. $f_{\rm samp} \geq 2 \unit{kHz}$ is required for the zero spin case.}
\end{figure*}

Are such expensive computations necessary to resolve the marginal impact from higher harmonics?   
To estimate how much sampling rate and lower frequency impacts our results, we calculated the local ambiguity function
versus intrinsic parameters for the zero spin binary following \abbrvEFM{}, for several choices of sampling rate and lower frequency limit in Figure~\ref{fig:SrateDependence}. %

For the initial LIGO noise curve, we found that $f_{\rm min}\ge 30 \unit{Hz}$ and $f_{\rm samp}>2 $ \unit{kHz}  was
required for zero spin  templates, with and without higher harmonics. In the case of aligned spin, although the theoretical lower limit of the sampling rate is 16384\unit{kHz} for the higher order waveforms, we found that the local ambiguity function was well recovered by using the sampling rate of 4096\unit{kHz}.
Moreover, as demonstrated in the text, this sampling rate produces good agreement with our theoretical predictions.  
Conversely, to illustrate the pathologies that arise when undersampling a signal, we have performed simulations with
$f_{\rm samp} = 1\unit{kHz}$.  Despite the aliasing high of physical frequencies, the detectors' poor sensitivity to high
frequencies could de facto allow computation at such a low sampling rate.   Despite the relatively small amount of signal power associated with high frequency, aliasing produced
noticable modulations and biases in our posteriors.  No lower sampling rate should be employed, without explicitly
filtering away high frequency content. 

\section{Effective dimension and prior volume versus signal amplitude}
\label{ap:Thermo}
Parallel-tempered markov-chain Monte Carlo has been described at length before, both in general
\cite{mm-stats-MCMC-GeometricLaddear-Neal,mm-stats-MCMC-GeometricLadderChoices-Liu}  and in the context of
ground- and space-based gravitational wave astronomy; see, e.g.,  \cite{2007PhRvD..75f2004R,2008ApJ...688L..61V,2009PhRvD..80f3007L, 2009CQGra..26k4007R} \citeMCMC{} and references therein.
In this appendix, we describe how we reprocessed the multiple MCMC chains to evaluate the effective dimension
($D_{\rm eff}$).  As described in the text, the excellent agreement between our numerical result and the
theoretically-expected value for this quantity gave us additional confidence our simulations had converged.   
Inevitably, we also touch upon an independent method we adopted to evaluate the evidence (thermodynamic integration).
As described in the text, relatively accurate evidence integrals were critical in allowing us to rule
out significant impact of higher harmonics beyond the relatively trivial effect described in the text.

\subsection{Review of parallel tempering and thermodynamic integration}
The \texttt{lalinference\_mcmc} code evolves several parallel  MCMC chains 
simultaneously, each with a likelihood of the form
\begin{eqnarray}
{\cal L}(\beta) \equiv {\cal L}^\beta
\end{eqnarray}
where  $T=1/\beta$ is called the chain's ``temperature''. 
The low-temperature likelihoods resemble our targeted, physical distribution while high-temperature likelihoods
correspond to weaker signal strength and enable efficient exploration of the entire parameter space.   

\begin{figure}
\includegraphics{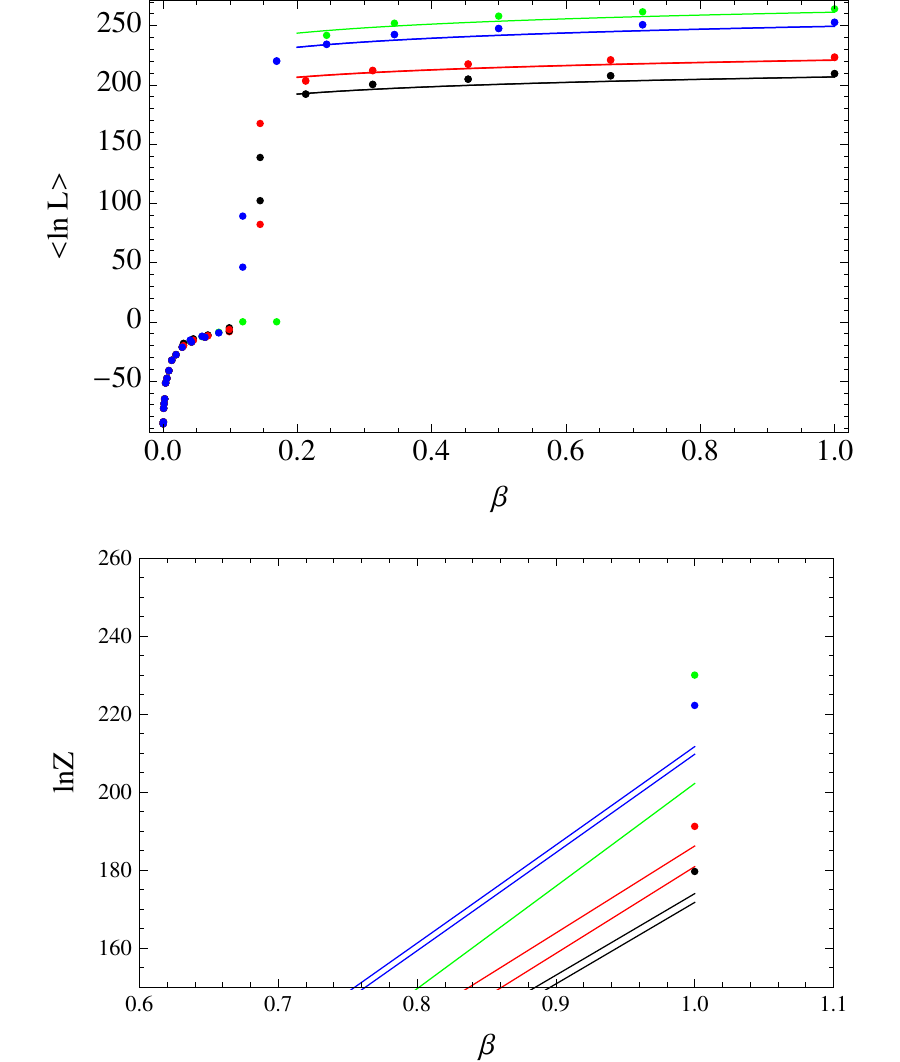}
\caption{\label{fig:ThermoIntegrate}\textbf{Thermodynamic integration }:   Illustration of thermodynamic integration,
  using samples from zero-noise simulations with zero-spin and no harmonics (black); zero spin and with harmonics (red);
  aligned spin and no harmonics (blue); and aligned spin with harmonics (green).  
\emph{Top panel}: Plot of $\left<\ln {\cal L}\right>_\beta$ versus $\beta$, a monotonically increasing function  needed to compute the thermodynamic
integral for the evidence $Z$ [Eq. (\ref{eq:ThermoIntegrate:Z})].   For comparison, a solid line shows $\rho(\beta)^2/2 + d\ln
\rho(\beta)$ where $d$ is the number of parameters (9 or 11) and where $\rho(1)$ is the signal amplitude. 
\emph{Bottom panel}: A plot of the derived evidence $\ln Z(\beta)$ versus $\beta$ derived from the thermodynamic
integral using the temperature chains shown above.  For comparison, the points show evidence derived from direct
integration over the posterior.  
Direct integration evidence is consistently larger.  While the small  difference between these two integrals is consistent with
discretization error in the thermodynamic integral, associated with in the sharp step in $\left<{\cal L}\right>$ near
$\beta \simeq 0.15$, a detailed analysis of the differences between these two evidence calculations is beyond the scope
of this work. 
}
\end{figure}

Theoretically speaking,  parallel-tempered chains allow us to calculate the evidence $Z$ [Eq. (\ref{eq:def:Z:Modified})] and
$V/V_{\rm prior}$  [Eq. (\ref{eq:def:VoverVprior})] using ``thermodynamic integration'' \cite{2011RvMP...83..943V}.
Thermodynamic integration
for the evidence relies on the following relation:
\begin{subequations}
\label{eq:ThermoIntegrate:Z}
\begin{align}
Z(\beta) &\equiv \int  {\cal L}^\beta p(\lambda) d\lambda \\
\frac{d\ln Z}{d\beta} &= \frac{1}{Z} \int p(\lambda) {\cal L}^\beta \ln {\cal L} = \left<\ln {\cal L}\right>_{\beta} \\
\ln Z(\beta) &= \int_0^{\beta} d\beta   \left<\ln {\cal L}\right>_{\beta}
\end{align}
\end{subequations}
where $\left<X\right>_\beta \equiv \int X {\cal L}^\beta p(\lambda)d\lambda/Z$.   
The  averages appearing in each integrand can be calculated from each simulations' samples.  
By definition,  the prior ratio $V/V_{\rm prior}[\beta]$ can be calculated directly from the
temperature-dependent evidence $Z(\beta)$ and the temperature-independent peak likelihoood:
\begin{eqnarray}
\label{eq:ThermoIntegrate:Vmanual}
\ln (V/V_{\rm prior})[\beta] &= \ln Z(\beta) - \beta  \text{max}_\lambda \ln {\cal L} (\{d\})
\end{eqnarray}
Because the recovered signal amplitude $\rho_{\rm rec}\equiv \sqrt{2  \text{max}_\lambda \ln {\cal L} (\{d\})}$  can
differ from the zero-noise signal amplitude $\rho$ by a number of order unity, we always calculate the second maximum
directly from the samples.    
Alternatively, this expression can be rewritten as a thermodynamic integral:
\begin{eqnarray}
\label{eq:ThermoIntegrate:V}
\ln (V/V_{\rm prior})[\beta] &  = \int_0^{\beta} d\beta   \left<\ln {\cal L}/{\cal L}_{\rm max}\right>_{\beta}
\end{eqnarray}

As a concrete example, the top panel of Figure \ref{fig:ThermoIntegrate} shows the functional form of the integrand $\left<\ln {\cal L}\right>_\beta$ appearing
in both thermodynamic expressions for $Z(\beta)$ and $V/V_{\rm prior}$.   
The default temperature chain shown in Figure  \ref{fig:ThermoIntegrate} has a sparse temperature grid; to confirm our
results and control our error,  we have performed followup simulations.  Combining the temperature spacing and
functional forms used here, we expect the thermodynamic evidence  to be accurate to within $\Delta \ln Z \lesssim 5$.
For small $N_{\rm eff}$, as is often the case for slow simulations, this uncertainty is dominated by chain placement and
by slow  convergence of
the average of  $\left< \ln {\cal L}\right>_\beta$ near the critical temperatures where this average changes by $\simeq
\rho^2/2$. 

\subsection{Effective dimension versus amplitude}

\begin{figure}
\includegraphics[width=\columnwidth]{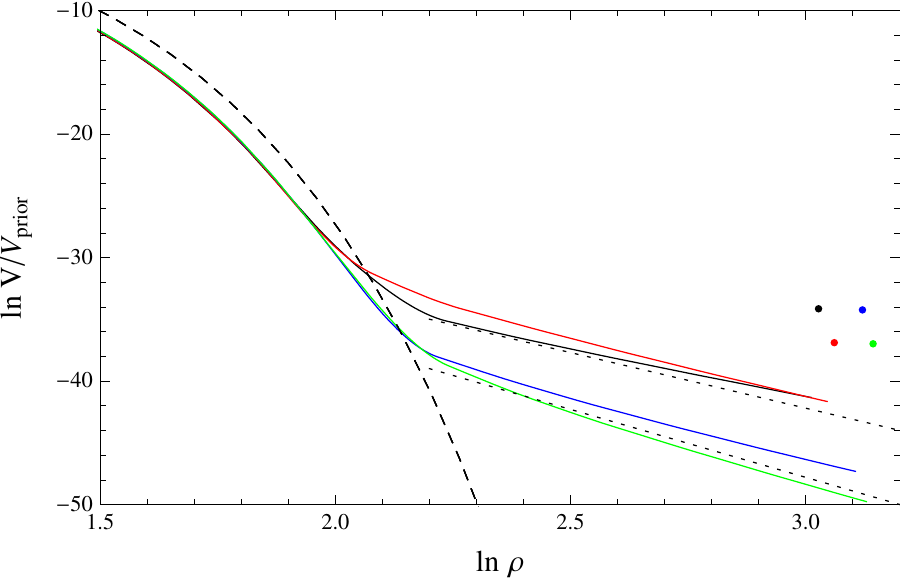}
\caption{\label{fig:VRatioVsAmplitude}\textbf{Prior volume versus amplitude}: 
A plot of $\ln (V/V_{\rm prior})$ versus signal amplitude $\rho$, derived from the signal
amplitude and from $\ln (V/V_{\rm prior})$ versus $\beta$.  For comparison, dotted lines are provided with
a logarithmic slope $-9$ and $-11$, the number of parameters in the nonspinning and spinning model, respectively. 
To illustrate a critical seperatrix between simulation-specific and universal behavior, the dashed line shows the curve  $\ln (V/V_{\rm prior})\simeq -\rho^2/2$
[Eq. (\ref{eq:DeffPredicted})].  This curve corresponds to $Z=1$: the data is equally consistent with the null hypothesis
or the presence of a single signal.  For comparison, the dotted and dashed curves intersect at a network amplitude
$\rho\simeq \exp 2.1$, roughly consistent with a signal amplitude $\lesssim 5.7$ in each detector.  
The colored points show the prior volume calculated using the direct integration evidence, as listed in Table
\ref{tab:RunSummaryAndResults}; as in Figure \ref{fig:ThermoIntegrate}, our approximate thermodynamic evidence integral differs
from direct integration evidence.  
  The vertical scales' absolute units depend on the choice of prior and have been chosen consistently with all other
  calculations described in this work.  
}
\end{figure}

Each parallel-tempered chain represents a different signal amplitude: using our definition 
[Eq. (\ref{eq:def:rhoRecovered})], 
\begin{subequations}
\label{eq:RelateTemperatureToAmplitude}
\begin{align}
\rhoRecovered(\beta) &= \rhoRecovered(1)\sqrt{\beta} \\
\rho(\beta) &= \rho(1)\sqrt{\beta}
\end{align}
\end{subequations}
Also, from first principles we can calculate the \emph{relative} change $\delta \rho/\rho$ that any physical effect will
have on the gravitational wave signal.   As a function of temperature, our simulations lose their ability to discriminate
between coarser and coarser details, until in the limit $\beta \rightarrow 0$ they recover the prior.   
In other words, the low-temperature chains will probe large amplitude and a large  $D_{\rm eff}$ [Eq. (\ref{eq:def:Deff})]; higher-temperature
simulations smooth out fine details using a lower amplitude $\rhoRecovered$ and may have a significantly lower $D_{\rm
  eff}$.  
The prior volume can be calculated by thermodynamic integration both using the likelihood (above) or the effective
dimension  [Eq. (\ref{eq:def:Deff})]:
\begin{align}
\label{eq:ConstrainUsingEffectiveDimension}
d\ln (V/V_{\rm prior}) & = \left<\ln {\cal L}/{\cal L}_{\rm max}\right>_\beta d\beta
= - D_{\rm  eff} d\ln \rho
\end{align}
As a concrete example, Figure \ref{fig:VRatioVsAmplitude} shows the prior volume versus $\rho(\beta)$.  When $\rho(\beta)
>8$, the function $\ln (V/V_{\rm prior})$ is nearly linear versus $\ln \rho$; the slope is $-D_{\rm eff}$.  
Different physical systems have different numbers of parameters, some of which may not be measurable (e.g., $\psi_+$).  
As a result, each curve has a distinct slope at high temperature, set by the number of measurable parameters at $\rho
\simeq 10-20$ [Eq. (\ref{eq:DeffPredicted})].

At sufficiently high temperatures and low amplitudes, the evidence physically should\footnote{As noted above and in
  Footnote \ref{note:Dmax}, the  low-temperature behavior in our simulations reflects the  pathological but conventional
  prior; that choice  should be eliminated in future work.} converge to nearly equal odds -- equivalently,  because the signal cannot be
distinguished from noise.  In this regime,  $V/V_{\rm prior}$
follows the dashed line in Figure \ref{fig:VRatioVsAmplitude}.   Below this threshold, the form of $\ln V/V_{\rm
  prior}$ versus $\ln \rho$ should be universal, set by the definition of the signal amplitude $\rho$ and completely
unrelated to the physical dimension of the problem.  

\begin{acknowledgements}
The authors appreciate helpful discussions with  Madeline Wade, Will Farr, and Tyson Littenburg.
This material is based upon work supported by the National Science Foundation Graduate Research Fellowship under Grant
No. DGE-0824162  and computing resources supported by the National Science Foundation Major Research Instrumentation
Program under Grants No. PHY-0923409 and PHY-1126812.  
ROS and EO are supported by   NSF awards PHY-0970074 and PHY-1307429.  
H.\ S.\ C., C.\ K., and C.\ H.\ L.\ are supported in part by the 
National Research Foundation Grant funded from the Korean 
Government (No.\ NRF-2011-220-C00029) and in part by the KISTI Global 
Science Experimental Data Hub Center (GSDC).
H.\ S.\ C.\ and C.\ H.\ L.\ are supported in part by the BAERI 
Nuclear R \& D program (No.\ M20808740002), Korea.
C.\ K.\ is supported in part by the Research Corporation 
for Scientific Advancement and by a WVEPSCoR Research 
Challenge Grant. This work uses computing resources at the KISTI GSDC.
\end{acknowledgements}
\bibliography{paperexport}

\begin{thebibliography}{74}
\expandafter\ifx\csname natexlab\endcsname\relax\def\natexlab#1{#1}\fi
\expandafter\ifx\csname bibnamefont\endcsname\relax
  \def\bibnamefont#1{#1}\fi
\expandafter\ifx\csname bibfnamefont\endcsname\relax
  \def\bibfnamefont#1{#1}\fi
\expandafter\ifx\csname citenamefont\endcsname\relax
  \def\citenamefont#1{#1}\fi
\expandafter\ifx\csname url\endcsname\relax
  \def\url#1{\texttt{#1}}\fi
\expandafter\ifx\csname urlprefix\endcsname\relax\def\urlprefix{URL }\fi
\providecommand{\bibinfo}[2]{#2}
\providecommand{\eprint}[2][]{\url{#2}}

\bibitem[{\citenamefont{{Abbott et al. (The LIGO Scientific
  Collaboration)}}(2004)}]{gw-detectors-LIGO-original}
\bibinfo{author}{\bibnamefont{{Abbott et al. (The LIGO Scientific
  Collaboration)}}}, \bibinfo{journal}{Nuclear Instruments and Methods in
  Physics Research A} \textbf{\bibinfo{volume}{517}}, \bibinfo{pages}{154}
  (\bibinfo{year}{2004}), \eprint{arXiv:gr-qc/0308043},
  \urlprefix\url{http://arxiv.org/abs/gr-qc/0308043}.

\bibitem[{\citenamefont{{Acernese} et~al.}(2006)\citenamefont{{Acernese},
  {Amico}, {Alshourbagy}, {Antonucci}, {Aoudia}, {Avino}, {Babusci},
  {Ballardin}, {Barone}, {Barsotti} et~al.}}]{gw-detectors-VIRGO-original}
\bibinfo{author}{\bibfnamefont{F.}~\bibnamefont{{Acernese}}},
  \bibinfo{author}{\bibfnamefont{P.}~\bibnamefont{{Amico}}},
  \bibinfo{author}{\bibfnamefont{M.}~\bibnamefont{{Alshourbagy}}},
  \bibinfo{author}{\bibfnamefont{F.}~\bibnamefont{{Antonucci}}},
  \bibinfo{author}{\bibfnamefont{S.}~\bibnamefont{{Aoudia}}},
  \bibinfo{author}{\bibfnamefont{S.}~\bibnamefont{{Avino}}},
  \bibinfo{author}{\bibfnamefont{D.}~\bibnamefont{{Babusci}}},
  \bibinfo{author}{\bibfnamefont{G.}~\bibnamefont{{Ballardin}}},
  \bibinfo{author}{\bibfnamefont{F.}~\bibnamefont{{Barone}}},
  \bibinfo{author}{\bibfnamefont{L.}~\bibnamefont{{Barsotti}}},
  \bibnamefont{et~al.}, \bibinfo{journal}{Classical and Quantum Gravity}
  \textbf{\bibinfo{volume}{23}}, \bibinfo{pages}{S635} (\bibinfo{year}{2006}).

\bibitem[{\citenamefont{{Hannam} et~al.}(2008)\citenamefont{{Hannam}, {Husa},
  {Br{\"u}gmann}, and {Gopakumar}}}]{2008PhRvD..78j4007H}
\bibinfo{author}{\bibfnamefont{M.}~\bibnamefont{{Hannam}}},
  \bibinfo{author}{\bibfnamefont{S.}~\bibnamefont{{Husa}}},
  \bibinfo{author}{\bibfnamefont{B.}~\bibnamefont{{Br{\"u}gmann}}},
  \bibnamefont{and}
  \bibinfo{author}{\bibfnamefont{A.}~\bibnamefont{{Gopakumar}}},
  \bibinfo{journal}{\prd} \textbf{\bibinfo{volume}{78}},
  \bibinfo{pages}{104007} (\bibinfo{year}{2008}).

\bibitem[{\citenamefont{{Buonanno} et~al.}(2009)\citenamefont{{Buonanno},
  {Iyer}, {Ochsner}, {Pan}, and
  {Sathyaprakash}}}]{gw-astro-PN-Comparison-AlessandraSathya2009}
\bibinfo{author}{\bibfnamefont{A.}~\bibnamefont{{Buonanno}}},
  \bibinfo{author}{\bibfnamefont{B.~R.} \bibnamefont{{Iyer}}},
  \bibinfo{author}{\bibfnamefont{E.}~\bibnamefont{{Ochsner}}},
  \bibinfo{author}{\bibfnamefont{Y.}~\bibnamefont{{Pan}}}, \bibnamefont{and}
  \bibinfo{author}{\bibfnamefont{B.~S.} \bibnamefont{{Sathyaprakash}}},
  \bibinfo{journal}{\prd} \textbf{\bibinfo{volume}{80}},
  \bibinfo{pages}{084043} (\bibinfo{year}{2009}).

\bibitem[{\citenamefont{{Buonanno} et~al.}(2003)\citenamefont{{Buonanno},
  {Chen}, and {Vallisneri}}}]{2003PhRvD..67j4025B}
\bibinfo{author}{\bibfnamefont{A.}~\bibnamefont{{Buonanno}}},
  \bibinfo{author}{\bibfnamefont{Y.}~\bibnamefont{{Chen}}}, \bibnamefont{and}
  \bibinfo{author}{\bibfnamefont{M.}~\bibnamefont{{Vallisneri}}},
  \bibinfo{journal}{\prd} \textbf{\bibinfo{volume}{67}},
  \bibinfo{pages}{104025} (\bibinfo{year}{2003}),
  \urlprefix\url{http://arxiv.org/pdf/gr-qc/0211087}.

\bibitem[{\citenamefont{{Buonanno} et~al.}(2004)\citenamefont{{Buonanno},
  {Chen}, {Pan}, and {Vallisneri}}}]{2004PhRvD..70j4003B}
\bibinfo{author}{\bibfnamefont{A.}~\bibnamefont{{Buonanno}}},
  \bibinfo{author}{\bibfnamefont{Y.}~\bibnamefont{{Chen}}},
  \bibinfo{author}{\bibfnamefont{Y.}~\bibnamefont{{Pan}}}, \bibnamefont{and}
  \bibinfo{author}{\bibfnamefont{M.}~\bibnamefont{{Vallisneri}}},
  \bibinfo{journal}{\prd} \textbf{\bibinfo{volume}{70}},
  \bibinfo{pages}{104003} (\bibinfo{year}{2004}).

\bibitem[{\citenamefont{{Arun} et~al.}(2009)\citenamefont{{Arun}, {Buonanno},
  {Faye}, and
  {Ochsner}}}]{gw-astro-mergers-approximations-SpinningPNHigherHarmonics}
\bibinfo{author}{\bibfnamefont{K.~G.} \bibnamefont{{Arun}}},
  \bibinfo{author}{\bibfnamefont{A.}~\bibnamefont{{Buonanno}}},
  \bibinfo{author}{\bibfnamefont{G.}~\bibnamefont{{Faye}}}, \bibnamefont{and}
  \bibinfo{author}{\bibfnamefont{E.}~\bibnamefont{{Ochsner}}},
  \bibinfo{journal}{\prd} \textbf{\bibinfo{volume}{79}},
  \bibinfo{pages}{104023} (\bibinfo{year}{2009}).

\bibitem[{\citenamefont{{Pan} et~al.}(2004)\citenamefont{{Pan}, {Buonanno},
  {Chen}, and {Vallisneri}}}]{BCV:PTF}
\bibinfo{author}{\bibfnamefont{Y.}~\bibnamefont{{Pan}}},
  \bibinfo{author}{\bibfnamefont{A.}~\bibnamefont{{Buonanno}}},
  \bibinfo{author}{\bibfnamefont{Y.}~\bibnamefont{{Chen}}}, \bibnamefont{and}
  \bibinfo{author}{\bibfnamefont{M.}~\bibnamefont{{Vallisneri}}},
  \bibinfo{journal}{\prd} \textbf{\bibinfo{volume}{69}},
  \bibinfo{pages}{104017} (\bibinfo{year}{2004}),
  \urlprefix\url{http://arxiv.org/abs/gr-qc/0310034}.

\bibitem[{\citenamefont{{Damour} et~al.}(2004)\citenamefont{{Damour},
  {Gopakumar}, and {Iyer}}}]{2004PhRvD..70f4028D}
\bibinfo{author}{\bibfnamefont{T.}~\bibnamefont{{Damour}}},
  \bibinfo{author}{\bibfnamefont{A.}~\bibnamefont{{Gopakumar}}},
  \bibnamefont{and} \bibinfo{author}{\bibfnamefont{B.~R.}
  \bibnamefont{{Iyer}}}, \bibinfo{journal}{\prd} \textbf{\bibinfo{volume}{70}},
  \bibinfo{pages}{064028} (\bibinfo{year}{2004}).

\bibitem[{\citenamefont{{Buonanno} et~al.}(2005)\citenamefont{{Buonanno},
  {Chen}, {Pan}, {Tagoshi}, and {Vallisneri}}}]{2005PhRvD..72h4027B}
\bibinfo{author}{\bibfnamefont{A.}~\bibnamefont{{Buonanno}}},
  \bibinfo{author}{\bibfnamefont{Y.}~\bibnamefont{{Chen}}},
  \bibinfo{author}{\bibfnamefont{Y.}~\bibnamefont{{Pan}}},
  \bibinfo{author}{\bibfnamefont{H.}~\bibnamefont{{Tagoshi}}},
  \bibnamefont{and}
  \bibinfo{author}{\bibfnamefont{M.}~\bibnamefont{{Vallisneri}}},
  \bibinfo{journal}{\prd} \textbf{\bibinfo{volume}{72}},
  \bibinfo{pages}{084027} (\bibinfo{year}{2005}).

\bibitem[{\citenamefont{{K{\"o}nigsd{\"o}rffer} and
  {Gopakumar}}(2006)}]{2006PhRvD..73l4012K}
\bibinfo{author}{\bibfnamefont{C.}~\bibnamefont{{K{\"o}nigsd{\"o}rffer}}}
  \bibnamefont{and}
  \bibinfo{author}{\bibfnamefont{A.}~\bibnamefont{{Gopakumar}}},
  \bibinfo{journal}{\prd} \textbf{\bibinfo{volume}{73}},
  \bibinfo{pages}{124012} (\bibinfo{year}{2006}).

\bibitem[{\citenamefont{{Tessmer} and {Gopakumar}}(2007)}]{2007MNRAS.374..721T}
\bibinfo{author}{\bibfnamefont{M.}~\bibnamefont{{Tessmer}}} \bibnamefont{and}
  \bibinfo{author}{\bibfnamefont{A.}~\bibnamefont{{Gopakumar}}},
  \bibinfo{journal}{\mnras} \textbf{\bibinfo{volume}{374}},
  \bibinfo{pages}{721} (\bibinfo{year}{2007}).

\bibitem[{\citenamefont{{Hinder} et~al.}(2010)\citenamefont{{Hinder},
  {Herrmann}, {Laguna}, and {Shoemaker}}}]{gr-astro-eccentric-NR-2008}
\bibinfo{author}{\bibfnamefont{I.}~\bibnamefont{{Hinder}}},
  \bibinfo{author}{\bibfnamefont{F.}~\bibnamefont{{Herrmann}}},
  \bibinfo{author}{\bibfnamefont{P.}~\bibnamefont{{Laguna}}}, \bibnamefont{and}
  \bibinfo{author}{\bibfnamefont{D.}~\bibnamefont{{Shoemaker}}},
  \bibinfo{journal}{\prd} \textbf{\bibinfo{volume}{82}}, \bibinfo{eid}{024033}
  (\bibinfo{year}{2010}), \eprint{0806.1037}.

\bibitem[{\citenamefont{{K{\"o}nigsd{\"o}rffer} and
  {Gopakumar}}(2005)}]{2005PhRvD..71b4039K}
\bibinfo{author}{\bibfnamefont{C.}~\bibnamefont{{K{\"o}nigsd{\"o}rffer}}}
  \bibnamefont{and}
  \bibinfo{author}{\bibfnamefont{A.}~\bibnamefont{{Gopakumar}}},
  \bibinfo{journal}{\prd} \textbf{\bibinfo{volume}{71}},
  \bibinfo{pages}{024039} (\bibinfo{year}{2005}).

\bibitem[{\citenamefont{{J. Aasi et al (The LIGO Scientific Collaboration and
  the Virgo Collaboration)}}(2013)}]{LIGO-CBC-S6-PE}
\bibinfo{author}{\bibnamefont{{J. Aasi et al (The LIGO Scientific Collaboration
  and the Virgo Collaboration)}}}, \bibinfo{journal}{\prd}
  \textbf{\bibinfo{volume}{88}}, \bibinfo{eid}{062001} (\bibinfo{year}{2013}),
  \urlprefix\url{http://arxiv.org/abs/arXiv:1304.1775}.

\bibitem[{\citenamefont{{Del Pozzo} et~al.}(2011)\citenamefont{{Del Pozzo},
  {Veitch}, and {Vecchio}}}]{2011PhRvD..83h2002D}
\bibinfo{author}{\bibfnamefont{W.}~\bibnamefont{{Del Pozzo}}},
  \bibinfo{author}{\bibfnamefont{J.}~\bibnamefont{{Veitch}}}, \bibnamefont{and}
  \bibinfo{author}{\bibfnamefont{A.}~\bibnamefont{{Vecchio}}},
  \bibinfo{journal}{\prd} \textbf{\bibinfo{volume}{83}}, \bibinfo{eid}{082002}
  (\bibinfo{year}{2011}).

\bibitem[{\citenamefont{{Cornish} et~al.}(2011)\citenamefont{{Cornish},
  {Sampson}, {Yunes}, and {Pretorius}}}]{2011PhRvD..84f2003C}
\bibinfo{author}{\bibfnamefont{N.}~\bibnamefont{{Cornish}}},
  \bibinfo{author}{\bibfnamefont{L.}~\bibnamefont{{Sampson}}},
  \bibinfo{author}{\bibfnamefont{N.}~\bibnamefont{{Yunes}}}, \bibnamefont{and}
  \bibinfo{author}{\bibfnamefont{F.}~\bibnamefont{{Pretorius}}},
  \bibinfo{journal}{\prd} \textbf{\bibinfo{volume}{84}}, \bibinfo{eid}{062003}
  (\bibinfo{year}{2011}).

\bibitem[{\citenamefont{{Li} et~al.}(2012)\citenamefont{{Li}, {Del Pozzo},
  {Vitale}, {Van Den Broeck}, {Agathos}, {Veitch}, {Grover}, {Sidery},
  {Sturani}, and {Vecchio}}}]{gr-extensions-tests-Europeans2011}
\bibinfo{author}{\bibfnamefont{T.~G.~F.} \bibnamefont{{Li}}},
  \bibinfo{author}{\bibfnamefont{W.}~\bibnamefont{{Del Pozzo}}},
  \bibinfo{author}{\bibfnamefont{S.}~\bibnamefont{{Vitale}}},
  \bibinfo{author}{\bibfnamefont{C.}~\bibnamefont{{Van Den Broeck}}},
  \bibinfo{author}{\bibfnamefont{M.}~\bibnamefont{{Agathos}}},
  \bibinfo{author}{\bibfnamefont{J.}~\bibnamefont{{Veitch}}},
  \bibinfo{author}{\bibfnamefont{K.}~\bibnamefont{{Grover}}},
  \bibinfo{author}{\bibfnamefont{T.}~\bibnamefont{{Sidery}}},
  \bibinfo{author}{\bibfnamefont{R.}~\bibnamefont{{Sturani}}},
  \bibnamefont{and}
  \bibinfo{author}{\bibfnamefont{A.}~\bibnamefont{{Vecchio}}},
  \bibinfo{journal}{\prd} \textbf{\bibinfo{volume}{85}}, \bibinfo{eid}{082003}
  (\bibinfo{year}{2012}).

\bibitem[{\citenamefont{{Veitch}
  et~al.}(2012{\natexlab{a}})\citenamefont{{Veitch}, {Mandel}, {Aylott},
  {Farr}, {Raymond}, {Rodriguez}, {van der Sluys}, {Kalogera}, and
  {Vecchio}}}]{gwastro-mergers-PE-Aylott-LIGOATest}
\bibinfo{author}{\bibfnamefont{J.}~\bibnamefont{{Veitch}}},
  \bibinfo{author}{\bibfnamefont{I.}~\bibnamefont{{Mandel}}},
  \bibinfo{author}{\bibfnamefont{B.}~\bibnamefont{{Aylott}}},
  \bibinfo{author}{\bibfnamefont{B.}~\bibnamefont{{Farr}}},
  \bibinfo{author}{\bibfnamefont{V.}~\bibnamefont{{Raymond}}},
  \bibinfo{author}{\bibfnamefont{C.}~\bibnamefont{{Rodriguez}}},
  \bibinfo{author}{\bibfnamefont{M.}~\bibnamefont{{van der Sluys}}},
  \bibinfo{author}{\bibfnamefont{V.}~\bibnamefont{{Kalogera}}},
  \bibnamefont{and}
  \bibinfo{author}{\bibfnamefont{A.}~\bibnamefont{{Vecchio}}},
  \bibinfo{journal}{\prd} \textbf{\bibinfo{volume}{85}}, \bibinfo{eid}{104045}
  (\bibinfo{year}{2012}{\natexlab{a}}).

\bibitem[{\citenamefont{{Nissanke} et~al.}(2011)\citenamefont{{Nissanke},
  {Sievers}, {Dalal}, and {Holz}}}]{2011ApJ...739...99N}
\bibinfo{author}{\bibfnamefont{S.}~\bibnamefont{{Nissanke}}},
  \bibinfo{author}{\bibfnamefont{J.}~\bibnamefont{{Sievers}}},
  \bibinfo{author}{\bibfnamefont{N.}~\bibnamefont{{Dalal}}}, \bibnamefont{and}
  \bibinfo{author}{\bibfnamefont{D.}~\bibnamefont{{Holz}}},
  \bibinfo{journal}{\apj} \textbf{\bibinfo{volume}{739}}, \bibinfo{eid}{99}
  (\bibinfo{year}{2011}).

\bibitem[{\citenamefont{{Veitch}
  et~al.}(2012{\natexlab{b}})\citenamefont{{Veitch}, {Mandel}, {Aylott},
  {Farr}, {Raymond}, {Rodriguez}, {van der Sluys}, {Kalogera}, and
  {Vecchio}}}]{2012PhRvD..85j4045V}
\bibinfo{author}{\bibfnamefont{J.}~\bibnamefont{{Veitch}}},
  \bibinfo{author}{\bibfnamefont{I.}~\bibnamefont{{Mandel}}},
  \bibinfo{author}{\bibfnamefont{B.}~\bibnamefont{{Aylott}}},
  \bibinfo{author}{\bibfnamefont{B.}~\bibnamefont{{Farr}}},
  \bibinfo{author}{\bibfnamefont{V.}~\bibnamefont{{Raymond}}},
  \bibinfo{author}{\bibfnamefont{C.}~\bibnamefont{{Rodriguez}}},
  \bibinfo{author}{\bibfnamefont{M.}~\bibnamefont{{van der Sluys}}},
  \bibinfo{author}{\bibfnamefont{V.}~\bibnamefont{{Kalogera}}},
  \bibnamefont{and}
  \bibinfo{author}{\bibfnamefont{A.}~\bibnamefont{{Vecchio}}},
  \bibinfo{journal}{\prd} \textbf{\bibinfo{volume}{85}}, \bibinfo{eid}{104045}
  (\bibinfo{year}{2012}{\natexlab{b}}), \eprint{1201.1195}.

\bibitem[{\citenamefont{{Cho} et~al.}(2012)\citenamefont{{Cho}, {Ochsner},
  {O'Shaughnessy}, {Kim}, and
  {Lee}}}]{gwastro-mergers-HeeSuk-FisherMatrixWithAmplitudeCorrections}
\bibinfo{author}{\bibfnamefont{H.}~\bibnamefont{{Cho}}},
  \bibinfo{author}{\bibfnamefont{E.}~\bibnamefont{{Ochsner}}},
  \bibinfo{author}{\bibfnamefont{R.}~\bibnamefont{{O'Shaughnessy}}},
  \bibinfo{author}{\bibfnamefont{C.}~\bibnamefont{{Kim}}}, \bibnamefont{and}
  \bibinfo{author}{\bibfnamefont{C.}~\bibnamefont{{Lee}}},
  \bibinfo{journal}{Submitted to PRD (arXiv:1209.4494)}
  (\bibinfo{year}{2012}), \urlprefix\url{http://arxiv.org/abs/arXiv:1209.4494}.

\bibitem[{\citenamefont{{Cokelaer}}(2008)}]{2008CQGra..25r4007C}
\bibinfo{author}{\bibfnamefont{T.}~\bibnamefont{{Cokelaer}}},
  \bibinfo{journal}{Classical and Quantum Gravity}
  \textbf{\bibinfo{volume}{25}}, \bibinfo{pages}{184007}
  (\bibinfo{year}{2008}).

\bibitem[{\citenamefont{{Vallisneri}}(2008{\natexlab{a}})}]{2008PhRvD..77d2001%
V}
\bibinfo{author}{\bibfnamefont{M.}~\bibnamefont{{Vallisneri}}},
  \bibinfo{journal}{\prd} \textbf{\bibinfo{volume}{77}}, \bibinfo{eid}{042001}
  (\bibinfo{year}{2008}{\natexlab{a}}), \eprint{arXiv:gr-qc/0703086}.

\bibitem[{\citenamefont{{Vitale} and {Zanolin}}(2010)}]{2010PhRvD..82l4065V}
\bibinfo{author}{\bibfnamefont{S.}~\bibnamefont{{Vitale}}} \bibnamefont{and}
  \bibinfo{author}{\bibfnamefont{M.}~\bibnamefont{{Zanolin}}},
  \bibinfo{journal}{\prd} \textbf{\bibinfo{volume}{82}}, \bibinfo{eid}{124065}
  (\bibinfo{year}{2010}).

\bibitem[{\citenamefont{{Poisson} and {Will}}(1995)}]{1995PhRvD..52..848P}
\bibinfo{author}{\bibfnamefont{E.}~\bibnamefont{{Poisson}}} \bibnamefont{and}
  \bibinfo{author}{\bibfnamefont{C.~M.} \bibnamefont{{Will}}},
  \bibinfo{journal}{\prd} \textbf{\bibinfo{volume}{52}}, \bibinfo{pages}{848}
  (\bibinfo{year}{1995}), \eprint{arXiv:gr-qc/9502040}.

\bibitem[{\citenamefont{{Vallisneri}}(2008{\natexlab{b}})}]{gw-astro-Vallis-Fi%
sher-2007}
\bibinfo{author}{\bibfnamefont{M.}~\bibnamefont{{Vallisneri}}},
  \bibinfo{journal}{\prd} \textbf{\bibinfo{volume}{77}}, \bibinfo{eid}{042001}
  (\bibinfo{year}{2008}{\natexlab{b}}),
  \urlprefix\url{http://arxiv.org/abs/gr-qc/0703086}.

\bibitem[{\citenamefont{{Cutler} and {Flanagan}}(1994)}]{CutlerFlanagan:1994}
\bibinfo{author}{\bibfnamefont{C.}~\bibnamefont{{Cutler}}} \bibnamefont{and}
  \bibinfo{author}{\bibfnamefont{E.~E.} \bibnamefont{{Flanagan}}},
  \bibinfo{journal}{\prd} \textbf{\bibinfo{volume}{49}}, \bibinfo{pages}{2658}
  (\bibinfo{year}{1994}), \eprint{gr-qc/9402014}.

\bibitem[{\citenamefont{{Van Den Broeck} and
  {Sengupta}}(2007)}]{2007CQGra..24..155V}
\bibinfo{author}{\bibfnamefont{C.}~\bibnamefont{{Van Den Broeck}}}
  \bibnamefont{and} \bibinfo{author}{\bibfnamefont{A.~S.}
  \bibnamefont{{Sengupta}}}, \bibinfo{journal}{Classical and Quantum Gravity}
  \textbf{\bibinfo{volume}{24}}, \bibinfo{pages}{155} (\bibinfo{year}{2007}).

\bibitem[{\citenamefont{{Balasubramanian}
  et~al.}(1996)\citenamefont{{Balasubramanian}, {Sathyaprakash}, and
  {Dhurandhar}}}]{1996PhRvD..53.3033B}
\bibinfo{author}{\bibfnamefont{R.}~\bibnamefont{{Balasubramanian}}},
  \bibinfo{author}{\bibfnamefont{B.~S.} \bibnamefont{{Sathyaprakash}}},
  \bibnamefont{and} \bibinfo{author}{\bibfnamefont{S.~V.}
  \bibnamefont{{Dhurandhar}}}, \bibinfo{journal}{\prd}
  \textbf{\bibinfo{volume}{53}}, \bibinfo{pages}{3033} (\bibinfo{year}{1996}),
  \eprint{arXiv:gr-qc/9508011}.

\bibitem[{\citenamefont{{Raymond}}(2012)}]{gw-astro-PE-Raymond}
\bibinfo{author}{\bibfnamefont{V.}~\bibnamefont{{Raymond}}}, Ph.D. thesis
  (\bibinfo{year}{2012}),
  \urlprefix\url{https://gwic.ligo.org/thesisprize/2012/raymond-thesis.pdf}.

\bibitem[{\citenamefont{{Porter} and {Cornish}}(2008)}]{2008PhRvD..78f4005P}
\bibinfo{author}{\bibfnamefont{E.~K.} \bibnamefont{{Porter}}} \bibnamefont{and}
  \bibinfo{author}{\bibfnamefont{N.~J.} \bibnamefont{{Cornish}}},
  \bibinfo{journal}{\prd} \textbf{\bibinfo{volume}{78}},
  \bibinfo{pages}{064005} (\bibinfo{year}{2008}).

\bibitem[{\citenamefont{{Lang} et~al.}(2011)\citenamefont{{Lang}, {Hughes}, and
  {Cornish}}}]{2011PhRvD..84b2002L}
\bibinfo{author}{\bibfnamefont{R.~N.} \bibnamefont{{Lang}}},
  \bibinfo{author}{\bibfnamefont{S.~A.} \bibnamefont{{Hughes}}},
  \bibnamefont{and} \bibinfo{author}{\bibfnamefont{N.~J.}
  \bibnamefont{{Cornish}}}, \bibinfo{journal}{\prd}
  \textbf{\bibinfo{volume}{84}}, \bibinfo{pages}{022002}
  (\bibinfo{year}{2011}).

\bibitem[{\citenamefont{{Lang} and {Hughes}}(2006)}]{2006PhRvD..74l2001L}
\bibinfo{author}{\bibfnamefont{R.~N.} \bibnamefont{{Lang}}} \bibnamefont{and}
  \bibinfo{author}{\bibfnamefont{S.~A.} \bibnamefont{{Hughes}}},
  \bibinfo{journal}{\prd} \textbf{\bibinfo{volume}{74}},
  \bibinfo{pages}{122001} (\bibinfo{year}{2006}).

\bibitem[{\citenamefont{{Klein} et~al.}(2009)\citenamefont{{Klein}, {Jetzer},
  and {Sereno}}}]{2009PhRvD..80f4027K}
\bibinfo{author}{\bibfnamefont{A.}~\bibnamefont{{Klein}}},
  \bibinfo{author}{\bibfnamefont{P.}~\bibnamefont{{Jetzer}}}, \bibnamefont{and}
  \bibinfo{author}{\bibfnamefont{M.}~\bibnamefont{{Sereno}}},
  \bibinfo{journal}{\prd} \textbf{\bibinfo{volume}{80}}, \bibinfo{eid}{064027}
  (\bibinfo{year}{2009}).

\bibitem[{\citenamefont{Nitz et~al.}(2013)\citenamefont{Nitz, Lundgren, Brown,
  Ochsner, Keppel et~al.}}]{Nitz:2013mxa}
\bibinfo{author}{\bibfnamefont{A.~H.} \bibnamefont{Nitz}},
  \bibinfo{author}{\bibfnamefont{A.}~\bibnamefont{Lundgren}},
  \bibinfo{author}{\bibfnamefont{D.~A.} \bibnamefont{Brown}},
  \bibinfo{author}{\bibfnamefont{E.}~\bibnamefont{Ochsner}},
  \bibinfo{author}{\bibfnamefont{D.}~\bibnamefont{Keppel}},
  \bibnamefont{et~al.} (\bibinfo{year}{2013}), \eprint{1307.1757}.

\bibitem[{\citenamefont{Harry et~al.}(2013)\citenamefont{Harry, Nitz, Brown,
  Lundgren, Ochsner et~al.}}]{Harry:2013tca}
\bibinfo{author}{\bibfnamefont{I.}~\bibnamefont{Harry}},
  \bibinfo{author}{\bibfnamefont{A.}~\bibnamefont{Nitz}},
  \bibinfo{author}{\bibfnamefont{D.~A.} \bibnamefont{Brown}},
  \bibinfo{author}{\bibfnamefont{A.}~\bibnamefont{Lundgren}},
  \bibinfo{author}{\bibfnamefont{E.}~\bibnamefont{Ochsner}},
  \bibnamefont{et~al.} (\bibinfo{year}{2013}), \eprint{1307.3562}.

\bibitem[{\citenamefont{{LIGO Scientific Collaboration}}(2013)}]{lal}
\bibinfo{author}{\bibnamefont{{LIGO Scientific Collaboration}}}
  (\bibinfo{publisher}{{LAL source code is available at
  https://www.lsc-group.phys.uwm.edu/daswg/projects/lalsuite.html, version used
  dated June}}, \bibinfo{year}{2013}),
  \urlprefix\url{https://www.lsc-group.phys.uwm.edu/daswg/projects/lalsuite.ht%
ml}.

\bibitem[{\citenamefont{{Kidder}}(1995)}]{1995PhRvD..52..821K}
\bibinfo{author}{\bibfnamefont{L.~E.} \bibnamefont{{Kidder}}},
  \bibinfo{journal}{\prd} \textbf{\bibinfo{volume}{52}}, \bibinfo{pages}{821}
  (\bibinfo{year}{1995}), \eprint{arXiv:gr-qc/9506022}.

\bibitem[{\citenamefont{BohŽ et~al.}(2013)\citenamefont{BohŽ, Marsat, and
  Blanchet}}]{Bohe:2013cla}
\bibinfo{author}{\bibfnamefont{A.}~\bibnamefont{BohŽ}},
  \bibinfo{author}{\bibfnamefont{S.}~\bibnamefont{Marsat}}, \bibnamefont{and}
  \bibinfo{author}{\bibfnamefont{L.}~\bibnamefont{Blanchet}},
  \bibinfo{journal}{Class.Quant.Grav.} \textbf{\bibinfo{volume}{30}},
  \bibinfo{pages}{135009} (\bibinfo{year}{2013}), \eprint{1303.7412}.

\bibitem[{\citenamefont{Buonanno et~al.}(2013)\citenamefont{Buonanno, Faye, and
  Hinderer}}]{Buonanno:2012rv}
\bibinfo{author}{\bibfnamefont{A.}~\bibnamefont{Buonanno}},
  \bibinfo{author}{\bibfnamefont{G.}~\bibnamefont{Faye}}, \bibnamefont{and}
  \bibinfo{author}{\bibfnamefont{T.}~\bibnamefont{Hinderer}},
  \bibinfo{journal}{Phys.Rev.} \textbf{\bibinfo{volume}{D87}},
  \bibinfo{pages}{044009} (\bibinfo{year}{2013}), \eprint{1209.6349}.

\bibitem[{\citenamefont{Will and Wiseman}(1996)}]{WillWiseman:1996}
\bibinfo{author}{\bibfnamefont{C.~M.} \bibnamefont{Will}} \bibnamefont{and}
  \bibinfo{author}{\bibfnamefont{A.~G.} \bibnamefont{Wiseman}},
  \bibinfo{journal}{Phys. Rev. D} \textbf{\bibinfo{volume}{54}},
  \bibinfo{pages}{4813} (\bibinfo{year}{1996}).

\bibitem[{\citenamefont{{Damour} et~al.}(2001)\citenamefont{{Damour}, {Iyer},
  and {Sathyaprakash}}}]{2001PhRvD..63d4023D}
\bibinfo{author}{\bibfnamefont{T.}~\bibnamefont{{Damour}}},
  \bibinfo{author}{\bibfnamefont{B.~R.} \bibnamefont{{Iyer}}},
  \bibnamefont{and} \bibinfo{author}{\bibfnamefont{B.~S.}
  \bibnamefont{{Sathyaprakash}}}, \bibinfo{journal}{\prd}
  \textbf{\bibinfo{volume}{63}}, \bibinfo{eid}{044023} (\bibinfo{year}{2001}),
  \eprint{gr-qc/0010009}.

\bibitem[{\citenamefont{{The LIGO Scientific Collaboration} and {the Virgo
  Collaboration}}(2010)}]{2010arXiv1003.2481T}
\bibinfo{author}{\bibnamefont{{The LIGO Scientific Collaboration}}}
  \bibnamefont{and} \bibinfo{author}{\bibnamefont{{the Virgo Collaboration}}},
  \bibinfo{journal}{ArXiv e-prints}  (\bibinfo{year}{2010}),
  \eprint{1003.2481}.

\bibitem[{\citenamefont{{The LIGO Scientific Collaboration} and {the Virgo
  Collaboration}}(2012)}]{arXiv:1203.2674}
\bibinfo{author}{\bibnamefont{{The LIGO Scientific Collaboration}}}
  \bibnamefont{and} \bibinfo{author}{\bibnamefont{{the Virgo Collaboration}}},
  \bibinfo{journal}{(arXiv:1203.2674)}  (\bibinfo{year}{2012}),
  \urlprefix\url{http://arxiv.org/abs/arXiv:1203.2674}.

\bibitem[{\citenamefont{{Abadie} et~al.}(2010)\citenamefont{{Abadie}, {Abbott},
  {Abbott}, {Abernathy}, {Accadia}, {Acernese}, {Adams}, {Adhikari}, {Ajith},
  {Allen} et~al.}}]{2010CQGra..27q3001A}
\bibinfo{author}{\bibfnamefont{J.}~\bibnamefont{{Abadie}}},
  \bibinfo{author}{\bibfnamefont{B.~P.} \bibnamefont{{Abbott}}},
  \bibinfo{author}{\bibfnamefont{R.}~\bibnamefont{{Abbott}}},
  \bibinfo{author}{\bibfnamefont{M.}~\bibnamefont{{Abernathy}}},
  \bibinfo{author}{\bibfnamefont{T.}~\bibnamefont{{Accadia}}},
  \bibinfo{author}{\bibfnamefont{F.}~\bibnamefont{{Acernese}}},
  \bibinfo{author}{\bibfnamefont{C.}~\bibnamefont{{Adams}}},
  \bibinfo{author}{\bibfnamefont{R.}~\bibnamefont{{Adhikari}}},
  \bibinfo{author}{\bibfnamefont{P.}~\bibnamefont{{Ajith}}},
  \bibinfo{author}{\bibfnamefont{B.}~\bibnamefont{{Allen}}},
  \bibnamefont{et~al.}, \bibinfo{journal}{Classical and Quantum Gravity}
  \textbf{\bibinfo{volume}{27}}, \bibinfo{eid}{173001} (\bibinfo{year}{2010}),
  \eprint{1003.2480}.

\bibitem[{\citenamefont{Owen and Sathyaprakash}(1999)}]{Owen:1998dk}
\bibinfo{author}{\bibfnamefont{B.~J.} \bibnamefont{Owen}} \bibnamefont{and}
  \bibinfo{author}{\bibfnamefont{B.~S.} \bibnamefont{Sathyaprakash}},
  \bibinfo{journal}{\prd} \textbf{\bibinfo{volume}{60}},
  \bibinfo{pages}{022002} (\bibinfo{year}{1999}).

\bibitem[{\citenamefont{Owen}(1996)}]{Owen:1995tm}
\bibinfo{author}{\bibfnamefont{B.~J.} \bibnamefont{Owen}},
  \bibinfo{journal}{\prd} \textbf{\bibinfo{volume}{53}}, \bibinfo{pages}{6749}
  (\bibinfo{year}{1996}), \urlprefix\url{http://arxiv.org/abs/gr-qc/9511032}.

\bibitem[{\citenamefont{{Cokelaer}}(2007)}]{2007PhRvD..76j2004C}
\bibinfo{author}{\bibfnamefont{T.}~\bibnamefont{{Cokelaer}}},
  \bibinfo{journal}{\prd} \textbf{\bibinfo{volume}{76}}, \bibinfo{eid}{102004}
  (\bibinfo{year}{2007}), \eprint{0706.4437}.

\bibitem[{\citenamefont{{Brown}
  et~al.}(2012{\natexlab{a}})\citenamefont{{Brown}, {Harry}, {Lundgren}, and
  {Nitz}}}]{gwastro-mergers-TemplateBank-AlignedSpin-Andy2012}
\bibinfo{author}{\bibfnamefont{D.~A.} \bibnamefont{{Brown}}},
  \bibinfo{author}{\bibfnamefont{I.}~\bibnamefont{{Harry}}},
  \bibinfo{author}{\bibfnamefont{A.}~\bibnamefont{{Lundgren}}},
  \bibnamefont{and} \bibinfo{author}{\bibfnamefont{A.~H.}
  \bibnamefont{{Nitz}}}, \bibinfo{journal}{(arXiv:1207.6406)}
  (\bibinfo{year}{2012}{\natexlab{a}}),
  \urlprefix\url{http://arxiv.org/abs/arXiv:1207.6406}.

\bibitem[{\citenamefont{{Mackay}}(2003)}]{2003itil.book.....M}
\bibinfo{author}{\bibfnamefont{D.~J.~C.} \bibnamefont{{Mackay}}},
  \emph{\bibinfo{title}{{Information Theory, Inference and Learning
  Algorithms}}} (\bibinfo{year}{2003}).

\bibitem[{\citenamefont{{von Toussaint}}(2011)}]{2011RvMP...83..943V}
\bibinfo{author}{\bibfnamefont{U.}~\bibnamefont{{von Toussaint}}},
  \bibinfo{journal}{Reviews of Modern Physics} \textbf{\bibinfo{volume}{83}},
  \bibinfo{pages}{943} (\bibinfo{year}{2011}).

\bibitem[{\citenamefont{{Burnham} and
  {Anderson}}(2002)}]{book-BurnhamAnderson-ModelSelection}
\bibinfo{author}{\bibnamefont{{Burnham}}} \bibnamefont{and}
  \bibinfo{author}{\bibnamefont{{Anderson}}}, \emph{\bibinfo{title}{Model
  selection and multimodel inference}} (\bibinfo{publisher}{Springer},
  \bibinfo{year}{2002}), ISBN \bibinfo{isbn}{978-0387953649}.

\bibitem[{\citenamefont{{Gentle} et~al.}(2004)\citenamefont{{Gentle}, {Hardle},
  and {Mori}}}]{mm-stats-HandbookOfComputationalMethods}
\bibinfo{author}{\bibfnamefont{J.}~\bibnamefont{{Gentle}}},
  \bibinfo{author}{\bibfnamefont{W.}~\bibnamefont{{Hardle}}}, \bibnamefont{and}
  \bibinfo{author}{\bibfnamefont{M.}~\bibnamefont{{Mori}}},
  \emph{\bibinfo{title}{{Handbook of Computational Statistics: Concepts and
  Methods}}} (\bibinfo{publisher}{Berlin: Springer-Verlag},
  \bibinfo{year}{2004}), ISBN \bibinfo{isbn}{3-540-40464-3}.

\bibitem[{\citenamefont{{Skilling}}(2006)}]{mm-stats-NestedSampling-Skilling20%
06}
\bibinfo{author}{\bibfnamefont{J.}~\bibnamefont{{Skilling}}},
  \bibinfo{journal}{Bayesian Analysis} \textbf{\bibinfo{volume}{1}},
  \bibinfo{pages}{833} (\bibinfo{year}{2006}).

\bibitem[{\citenamefont{{Skilling}}(2009)}]{2009AIPC.1193..277S}
\bibinfo{author}{\bibfnamefont{J.}~\bibnamefont{{Skilling}}}, in
  \emph{\bibinfo{booktitle}{American Institute of Physics Conference Series}},
  edited by \bibinfo{editor}{\bibfnamefont{P.~M.} \bibnamefont{{Goggans}}}
  \bibnamefont{and} \bibinfo{editor}{\bibfnamefont{C.-Y.} \bibnamefont{{Chan}}}
  (\bibinfo{year}{2009}), vol. \bibinfo{volume}{1193} of
  \emph{\bibinfo{series}{American Institute of Physics Conference Series}}, pp.
  \bibinfo{pages}{277--291}.

\bibitem[{\citenamefont{{Feroz} et~al.}(2009)\citenamefont{{Feroz}, {Hobson},
  and {Bridges}}}]{2009MNRAS.398.1601F}
\bibinfo{author}{\bibfnamefont{F.}~\bibnamefont{{Feroz}}},
  \bibinfo{author}{\bibfnamefont{M.~P.} \bibnamefont{{Hobson}}},
  \bibnamefont{and}
  \bibinfo{author}{\bibfnamefont{M.}~\bibnamefont{{Bridges}}},
  \bibinfo{journal}{\mnras} \textbf{\bibinfo{volume}{398}},
  \bibinfo{pages}{1601} (\bibinfo{year}{2009}), \eprint{0809.3437}.

\bibitem[{\citenamefont{{Ajith}}(2011)}]{2011PhRvD..84h4037A}
\bibinfo{author}{\bibfnamefont{P.}~\bibnamefont{{Ajith}}},
  \bibinfo{journal}{\prd} \textbf{\bibinfo{volume}{84}}, \bibinfo{eid}{084037}
  (\bibinfo{year}{2011}), \eprint{1107.1267}.

\bibitem[{\citenamefont{{Brown}
  et~al.}(2012{\natexlab{b}})\citenamefont{{Brown}, {Harry}, {Lundgren}, and
  {Nitz}}}]{2012PhRvD..86h4017B}
\bibinfo{author}{\bibfnamefont{D.~A.} \bibnamefont{{Brown}}},
  \bibinfo{author}{\bibfnamefont{I.}~\bibnamefont{{Harry}}},
  \bibinfo{author}{\bibfnamefont{A.}~\bibnamefont{{Lundgren}}},
  \bibnamefont{and} \bibinfo{author}{\bibfnamefont{A.~H.}
  \bibnamefont{{Nitz}}}, \bibinfo{journal}{\prd} \textbf{\bibinfo{volume}{86}},
  \bibinfo{eid}{084017} (\bibinfo{year}{2012}{\natexlab{b}}),
  \eprint{1207.6406}.

\bibitem[{\citenamefont{{O'Shaughnessy}}(2012)}]{PSconstraints3-MassDistributi%
onMethods-NearbyUniverse}
\bibinfo{author}{\bibfnamefont{R.}~\bibnamefont{{O'Shaughnessy}}},
  \bibinfo{journal}{Submitted to PRD (arxiv:1204.3117)}
  (\bibinfo{year}{2012}), \urlprefix\url{http://arxiv.org/abs/1204.3117}.

\bibitem[{\citenamefont{{Nissanke} et~al.}(2010)\citenamefont{{Nissanke},
  {Holz}, {Hughes}, {Dalal}, and
  {Sievers}}}]{gw-astro-ShortGRBSirens-Hughes2009}
\bibinfo{author}{\bibfnamefont{S.}~\bibnamefont{{Nissanke}}},
  \bibinfo{author}{\bibfnamefont{D.~E.} \bibnamefont{{Holz}}},
  \bibinfo{author}{\bibfnamefont{S.~A.} \bibnamefont{{Hughes}}},
  \bibinfo{author}{\bibfnamefont{N.}~\bibnamefont{{Dalal}}}, \bibnamefont{and}
  \bibinfo{author}{\bibfnamefont{J.~L.} \bibnamefont{{Sievers}}},
  \bibinfo{journal}{\apj} \textbf{\bibinfo{volume}{725}}, \bibinfo{pages}{496}
  (\bibinfo{year}{2010}), \eprint{0904.1017}.

\bibitem[{\citenamefont{{LIGO Scientific Collaboration}
  et~al.}(2013)\citenamefont{{LIGO Scientific Collaboration}, {Virgo
  Collaboration}, {Aasi}, {Abadie}, {Abbott}, {Abbott}, {Abbott}, {Abernathy},
  {Accadia}, {Acernese} et~al.}}]{LIGO-2013-WhitePaper-CoordinatedEMObserving}
\bibinfo{author}{\bibnamefont{{LIGO Scientific Collaboration}}},
  \bibinfo{author}{\bibnamefont{{Virgo Collaboration}}},
  \bibinfo{author}{\bibfnamefont{J.}~\bibnamefont{{Aasi}}},
  \bibinfo{author}{\bibfnamefont{J.}~\bibnamefont{{Abadie}}},
  \bibinfo{author}{\bibfnamefont{B.~P.} \bibnamefont{{Abbott}}},
  \bibinfo{author}{\bibfnamefont{R.}~\bibnamefont{{Abbott}}},
  \bibinfo{author}{\bibfnamefont{T.~D.} \bibnamefont{{Abbott}}},
  \bibinfo{author}{\bibfnamefont{M.}~\bibnamefont{{Abernathy}}},
  \bibinfo{author}{\bibfnamefont{T.}~\bibnamefont{{Accadia}}},
  \bibinfo{author}{\bibfnamefont{F.}~\bibnamefont{{Acernese}}},
  \bibnamefont{et~al.}, \bibinfo{journal}{ArXiv e-prints}
  (\bibinfo{year}{2013}), \eprint{1304.0670}.

\bibitem[{\citenamefont{{Nissanke} et~al.}(2013)\citenamefont{{Nissanke},
  {Kasliwal}, and {Georgieva}}}]{2013ApJ...767..124N}
\bibinfo{author}{\bibfnamefont{S.}~\bibnamefont{{Nissanke}}},
  \bibinfo{author}{\bibfnamefont{M.}~\bibnamefont{{Kasliwal}}},
  \bibnamefont{and}
  \bibinfo{author}{\bibfnamefont{A.}~\bibnamefont{{Georgieva}}},
  \bibinfo{journal}{\apj} \textbf{\bibinfo{volume}{767}}, \bibinfo{eid}{124}
  (\bibinfo{year}{2013}), \eprint{1210.6362}.

\bibitem[{\citenamefont{{Poisson}}(1998)}]{1998PhRvD..57.5287P}
\bibinfo{author}{\bibfnamefont{E.}~\bibnamefont{{Poisson}}},
  \bibinfo{journal}{\prd} \textbf{\bibinfo{volume}{57}}, \bibinfo{pages}{5287}
  (\bibinfo{year}{1998}), \eprint{arXiv:gr-qc/9709032}.

\bibitem[{\citenamefont{Mikoczi et~al.}(2005)\citenamefont{Mikoczi, Vasuth, and
  Gergely}}]{Mikoczi:2005dn}
\bibinfo{author}{\bibfnamefont{B.}~\bibnamefont{Mikoczi}},
  \bibinfo{author}{\bibfnamefont{M.}~\bibnamefont{Vasuth}}, \bibnamefont{and}
  \bibinfo{author}{\bibfnamefont{L.~A.} \bibnamefont{Gergely}},
  \bibinfo{journal}{Phys.Rev.} \textbf{\bibinfo{volume}{D71}},
  \bibinfo{pages}{124043} (\bibinfo{year}{2005}), \eprint{astro-ph/0504538}.

\bibitem[{\citenamefont{Faye et~al.}(2006)\citenamefont{Faye, Blanchet, and
  Buonanno}}]{Faye:2006gx}
\bibinfo{author}{\bibfnamefont{G.}~\bibnamefont{Faye}},
  \bibinfo{author}{\bibfnamefont{L.}~\bibnamefont{Blanchet}}, \bibnamefont{and}
  \bibinfo{author}{\bibfnamefont{A.}~\bibnamefont{Buonanno}},
  \bibinfo{journal}{Phys.Rev.} \textbf{\bibinfo{volume}{D74}},
  \bibinfo{pages}{104033} (\bibinfo{year}{2006}), \eprint{gr-qc/0605139}.

\bibitem[{\citenamefont{Blanchet et~al.}(2006)\citenamefont{Blanchet, Buonanno,
  and Faye}}]{Blanchet:2006gy}
\bibinfo{author}{\bibfnamefont{L.}~\bibnamefont{Blanchet}},
  \bibinfo{author}{\bibfnamefont{A.}~\bibnamefont{Buonanno}}, \bibnamefont{and}
  \bibinfo{author}{\bibfnamefont{G.}~\bibnamefont{Faye}},
  \bibinfo{journal}{Phys.Rev.} \textbf{\bibinfo{volume}{D74}},
  \bibinfo{pages}{104034} (\bibinfo{year}{2006}), \eprint{gr-qc/0605140}.

\bibitem[{\citenamefont{{Hannam} et~al.}(2013)\citenamefont{{Hannam}, {Brown},
  {Fairhurst}, {Fryer}, and {Harry}}}]{2013ApJ...766L..14H}
\bibinfo{author}{\bibfnamefont{M.}~\bibnamefont{{Hannam}}},
  \bibinfo{author}{\bibfnamefont{D.~A.} \bibnamefont{{Brown}}},
  \bibinfo{author}{\bibfnamefont{S.}~\bibnamefont{{Fairhurst}}},
  \bibinfo{author}{\bibfnamefont{C.~L.} \bibnamefont{{Fryer}}},
  \bibnamefont{and} \bibinfo{author}{\bibfnamefont{I.~W.}
  \bibnamefont{{Harry}}}, \bibinfo{journal}{\apjl}
  \textbf{\bibinfo{volume}{766}}, \bibinfo{eid}{L14} (\bibinfo{year}{2013}),
  \eprint{1301.5616}.

\bibitem[{\citenamefont{{Neal}}(2001)}]{mm-stats-MCMC-GeometricLaddear-Neal}
\bibinfo{author}{\bibfnamefont{R.}~\bibnamefont{{Neal}}},
  \bibinfo{journal}{Statistics and Computing} \textbf{\bibinfo{volume}{11}},
  \bibinfo{pages}{125} (\bibinfo{year}{2001}).

\bibitem[{\citenamefont{{Liu}}(2001)}]{mm-stats-MCMC-GeometricLadderChoices-Li%
u}
\bibinfo{author}{\bibfnamefont{J.}~\bibnamefont{{Liu}}},
  \emph{\bibinfo{title}{{Monte Carlo Strategies in Scientific Computing}}}
  (\bibinfo{publisher}{New York: Springer}, \bibinfo{year}{2001}).

\bibitem[{\citenamefont{{R{\"o}ver} et~al.}(2007)\citenamefont{{R{\"o}ver},
  {Meyer}, and {Christensen}}}]{2007PhRvD..75f2004R}
\bibinfo{author}{\bibfnamefont{C.}~\bibnamefont{{R{\"o}ver}}},
  \bibinfo{author}{\bibfnamefont{R.}~\bibnamefont{{Meyer}}}, \bibnamefont{and}
  \bibinfo{author}{\bibfnamefont{N.}~\bibnamefont{{Christensen}}},
  \bibinfo{journal}{\prd} \textbf{\bibinfo{volume}{75}}, \bibinfo{eid}{062004}
  (\bibinfo{year}{2007}), \eprint{gr-qc/0609131}.

\bibitem[{\citenamefont{{van der Sluys} et~al.}(2008)\citenamefont{{van der
  Sluys}, {R{\"o}ver}, {Stroeer}, {Raymond}, {Mandel}, {Christensen},
  {Kalogera}, {Meyer}, and {Vecchio}}}]{2008ApJ...688L..61V}
\bibinfo{author}{\bibfnamefont{M.~V.} \bibnamefont{{van der Sluys}}},
  \bibinfo{author}{\bibfnamefont{C.}~\bibnamefont{{R{\"o}ver}}},
  \bibinfo{author}{\bibfnamefont{A.}~\bibnamefont{{Stroeer}}},
  \bibinfo{author}{\bibfnamefont{V.}~\bibnamefont{{Raymond}}},
  \bibinfo{author}{\bibfnamefont{I.}~\bibnamefont{{Mandel}}},
  \bibinfo{author}{\bibfnamefont{N.}~\bibnamefont{{Christensen}}},
  \bibinfo{author}{\bibfnamefont{V.}~\bibnamefont{{Kalogera}}},
  \bibinfo{author}{\bibfnamefont{R.}~\bibnamefont{{Meyer}}}, \bibnamefont{and}
  \bibinfo{author}{\bibfnamefont{A.}~\bibnamefont{{Vecchio}}},
  \bibinfo{journal}{\apjl} \textbf{\bibinfo{volume}{688}}, \bibinfo{pages}{L61}
  (\bibinfo{year}{2008}), \eprint{0710.1897}.

\bibitem[{\citenamefont{{Littenberg} and
  {Cornish}}(2009)}]{2009PhRvD..80f3007L}
\bibinfo{author}{\bibfnamefont{T.~B.} \bibnamefont{{Littenberg}}}
  \bibnamefont{and} \bibinfo{author}{\bibfnamefont{N.~J.}
  \bibnamefont{{Cornish}}}, \bibinfo{journal}{\prd}
  \textbf{\bibinfo{volume}{80}}, \bibinfo{eid}{063007} (\bibinfo{year}{2009}),
  \eprint{0902.0368}.

\bibitem[{\citenamefont{{Raymond} et~al.}(2009)\citenamefont{{Raymond}, {van
  der Sluys}, {Mandel}, {Kalogera}, {R{\"o}ver}, and
  {Christensen}}}]{2009CQGra..26k4007R}
\bibinfo{author}{\bibfnamefont{V.}~\bibnamefont{{Raymond}}},
  \bibinfo{author}{\bibfnamefont{M.~V.} \bibnamefont{{van der Sluys}}},
  \bibinfo{author}{\bibfnamefont{I.}~\bibnamefont{{Mandel}}},
  \bibinfo{author}{\bibfnamefont{V.}~\bibnamefont{{Kalogera}}},
  \bibinfo{author}{\bibfnamefont{C.}~\bibnamefont{{R{\"o}ver}}},
  \bibnamefont{and}
  \bibinfo{author}{\bibfnamefont{N.}~\bibnamefont{{Christensen}}},
  \bibinfo{journal}{Classical and Quantum Gravity}
  \textbf{\bibinfo{volume}{26}}, \bibinfo{eid}{114007} (\bibinfo{year}{2009}),
  \eprint{0812.4302}.

\end{thebibliography}

\end{document}